\documentclass[pdftex,twocolumn,epjc3]{svjour3}
\usepackage{url}
\usepackage{float}
\usepackage{xcolor}
\usepackage{amsmath}
\usepackage{algorithm}
\usepackage{algorithmic}
\usepackage{tikz}
\usepackage{tikz-feynman}
\usepackage[T1]{fontenc}
\usepackage{microtype}
\usepackage{cite}
\usepackage[hidelinks]{hyperref}

\hypersetup{pdfborder={0 0 0}}

\newcommand\blfootnote[1]{%
  \begingroup
  \renewcommand\thefootnote{}\footnote{#1}%
  \addtocounter{footnote}{-1}%
  \endgroup
}

\journalname{Eur. Phys. J. C}

\begin{document}

\title{Event-by-event Comparison between Machine-Learning-- and Transfer-Matrix--based Unfolding Methods}

\author{\mbox{Mathias Backes\textsuperscript{1}$^{,a}$, Anja Butter\textsuperscript{2,3}$^{,b}$, Monica Dunford\textsuperscript{1}$^{,c}$, Bogdan Malaescu\textsuperscript{2}$^{,d}$}}

\institute{Kirchhoff-Institut f\"{u}r Physik, Universität Heidelberg, Germany 
\and 
LPNHE, Sorbonne Universit\'e, Universit\'e Paris Cit\'e, CNRS/IN2P3, Paris, France
\and
Institut für Theoretische Physik, Universität Heidelberg, Germany}

\date{}

\maketitle

\abstract{
The unfolding of detector effects is a key aspect of comparing experimental data with theoretical predictions.
In recent years, different Machine-Learning methods have been developed to provide novel features, e.g.\ high dimensionality or a probabilistic single-event unfolding based on generative neural networks.
Traditionally, many analyses unfold detector effects using transfer-matrix--based algorithms, which are well established in low-dimensional unfolding.
They yield an unfolded distribution of the total spectrum, together with its covariance matrix.
This paper proposes a method to obtain probabilistic single-event unfolded distributions, together with their uncertainties and correlations, for the transfer-matrix--based unfolding.
The algorithm is first validated on a toy model and then applied to pseudo-data for the $pp\rightarrow Z\gamma \gamma$ process. 
In both examples the performance is compared to the Machine-Learning--based single-event unfolding using an iterative approach with conditional invertible neural networks (IcINN).
}

\blfootnote{$^a$ mathias.backes@kip.uni-heidelberg.de}
\blfootnote{$^b$ anja.butter@lpnhe.in2p3.fr}
\blfootnote{$^c$ monica.dunford@kip.uni-heidelberg.de}
\blfootnote{$^d$ malaescu@in2p3.fr}

\section{Introduction}

Unfolding of detector effects has become one of the standard procedures in particle physics to enable quantitative comparisons between experimental data and theoretical predictions, as well as between data sets collected by different experiments. 

Most common unfolding methods correct statistically~(in the sense of an estimator) distributions parameterized as binned histograms.
They employ response matrices connecting truth and reconstructed quantities, built using Monte Carlo simulations of the detector effects.
A (pseudo-)inversion of the response matrix allows to convert measured distributions into unfolded ones, while possibly involving regularisation procedures too.
Such ``matrix--based'' algorithms include e.g.\ a Simple Inversion~\cite{Cowan:1998ji}, Singular Value Decomposition (SVD)~\cite{Hocker:1995kb}, TUnfold~\cite{Schmitt:2012kp}, Iterative Bayesian Unfolding (IBU)~\cite{DAgostini:1994fjx,DAgostini:2010hil}, Iterative Dynamically Stabilised (IDS) unfolding~\cite{Malaescu:2009dm,Malaescu:2011yg}, as well as other preceding iterative methods~\cite{Richardson:1972,Lucy:1974yx,Shepp:1982,Kondor:1982ah,Multhei:1985qs,Multhei:1986ps}.

More recently, Machine-Learning~(ML) techniques~\cite{Rene:2006,Avdica:2006,Gagunashvili:2010zw,Hosseini:2016,Glazov:2017vni,Adler_2017,ardizzone2018analyzing,Andreassen:2019cjw,ardizzone2019guided,Bellagente:2019uyp,Bellagente:2020piv,Komiske:2021vym,Regadio:2021,Wong:2021zvv,Backes:2022vmn} have enabled the development of a new class of unfolding methods in the context of particle physics.
While allowing to perform a multidimensional unfolding~(i.e.\ to simultaneously correct numerous observables for detector effects), such methods also enable an unbinned treatment of the data~(see e.g.\ Ref.~\cite{Arratia:2021otl}).

One of the ML techniques for unfolding is provided by conditional Invertible Neural Networks (cINN)~\cite{ardizzone2018analyzing,ardizzone2019guided,Bellagente:2019uyp,Bellagente:2020piv}. 
In contrast to many traditional methods the cINN operates on individual events, where the term 'event' refers to the detector response to particle interactions, e.g. at a particle collider, a fixed target experiment or in the context of astrophysical measurements like cosmic rays. 
The unfolding input is not a histogram of a reconstructed distribution, but rather one or several quantities reconstructed in a given event. 
The unfolding of individual events then leads to the notion of event-by-event unfolding, meaning that the overall unfolded distribution is obtained by combining the unfolding results of many individual events. 
Indeed, these methods yield an unfolded distribution for each quantity considered in a given event of an experimental data set.
An iterative version of this approach (IcINN)~\cite{Backes:2022vmn} allows to mitigate the dependency of the unfolding result on the truth distributions used in the Monte Carlo simulation, which are being systematically improved.

While such iterations are also employed in the Omnifold approach~\cite{Andreassen:2019cjw,Komiske:2021vym}, the type of output that this method provides is fundamentally different: these are Monte Carlo events with modified weights compared to the original input sample.
Therefore, in this method one cannot immediately identify a one-to-one correspondence between an input data event and an output Monte Carlo event, which prevents the inclusion of this method in the studies discussed in this paper.
More recently, when our study was close to completion, a new method employing a technique called Schr\"{o}dinger Bridge (SBUnfold) has been proposed~\cite{Diefenbacher:2023wec}.
It benefits from features of both Omnifold and cINN.
SBUnfold maps one set of events on the reconstructed level into another one at the unfolded level, which also makes it challenging to be included in the current study.~\footnote{A generalization of the methodology discussed in this paper, enabling the application to Omnifold and SBUnfold, is left for later studies.}

In former studies, comparisons of the performance of ML unfolding methods and of matrix--based methods~\cite{Arratia:2021otl,Brenner:2019lmf,Baron:2021vvl}, have been performed at the level of the unfolded binned distributions of the full sample of events.
Our goal here is not to compare or challenge the performance of various methods~\footnote{Such comparisons are a delicate exercises that need to be done with care. Even for studies that are a-priori dedicated to such exercise~\cite{Brenner:2019lmf,Baron:2021vvl}, one could challenge e.g.\  the completeness of the ensemble of regularisation parameters that are being studied for various methods, in the specific context of the tests that are being performed, and/or the criteria that are used for selecting their nominal values.}, but rather to propose a new way of extracting the results of matrix--based unfolding and of performing detailed comparisons with the outcome of ML--based methods.

In this paper we generalize the event-by-event approach to matrix--based unfolding methods, allowing hence to preserve more information from the original data in the unfolded output.
Similarly to the ML case~\cite{Arratia:2021otl}, doing so facilitates deriving secondary quantities out of the joint distribution of the unfolded ones, for arbitrary sub-samples of events.
This enables more detailed event-by-event comparisons with ML approaches providing posterior distributions for individual data events. In particular it allows to scrutinize the performance and test the compatibility of the results of the various methods in some restricted phase-space regions.
Furthermore, with this approach one can keep track, event-by-event, of both the unfolded- and the reconstructed-level quantities, which further enhances the amount of preserved information.
This allows e.g.\ to apply selection cuts on reconstructed-level observables or on the time of the data taking, even after the unfolding is performed.

More details on the matrix-- and ML--based unfolding methods employed in this study are recalled in Section~\ref{s:unfolding_methods}.
Section~\ref{s:single} introduces a generalized approach to obtain a single-event unfolding with matrix--based algorithms and a comparison with IcINN is performed in the context of a simple pseudo-data model.
A detailed treatment of the statistical uncertainties with the corresponding correlations is also presented.
A comparison between single-event unfolding with IcINN and IBU in a more complex example is shown in Section~\ref{s:compare_IcINN_IBU}.
The dataset used for this comparison is an Effective Field Theory (EFT) simulation of a $pp\rightarrow Z\gamma\gamma$ final state.
A summary of the studies is presented in Section~\ref{s:Conclusions}.

\section{Unfolding Algorithms}
\label{s:unfolding_methods}

In this section we provide a brief description of the transfer-matrix-- and ML--based unfolding algorithms employed in the current study.

\subsection{Matrix--based Algorithms}
\label{s:matrix_based_unfolding}

Let $f(t)$ be the true underlying function which determines the event distribution.
Instead of observing $f(t)$ directly, only a measured distribution of events $g(r)$ is observed, which is the result of convoluting the true function with the detector response function $R(r|t)$:
\begin{align}
    g(r) = \int R(r|t) f(t) \; \mathrm{d}t.
    \label{eq:unfolding_folding}
\end{align}
The goal of unfolding is to obtain a precise estimate for $f(t)$. 
This can be accomplished by constructing a (pseudo-)inverse $R'(t|r)$ of the detector response function and applying it to the measured distribution.

In a matrix--based unfolding the distributions $f(t)$ and $g(r)$ are converted into histograms $t_j$ and $r_i$, for which a binning is introduced.
The size of this binning is already an important choice, as it effectively induces a regularization in the unfolding procedure.
Too small bins lead to large bin-to-bin statistical (anti-)correlations, while too large bins do not resolve interesting structures.
Eq.~\eqref{eq:unfolding_folding} can then be rewritten as
\begin{align}
    r_i = \sum_j R_{ij} \cdot t_j ,
    \label{Eq:binnedFolding}
\end{align}
with the detector response matrix~(generally also called resolution matrix) $R_{ij} \equiv P(r_i|t_j)$ being the conditional probability for the measured quantity to be in the reconstructed-level bin $i$, given that the true value was in the bin $j$.
Such matrix is typically constructed from a Monte Carlo simulation of the detector effects, filling the corresponding event entries in the true/reconstructed plane and normalising by the number of counts in each truth-level bin, such that $\sum_i R_{ij} = 1$.~\footnote{While in a full unfolding algorithm detector acceptance and efficiency effects have to be corrected for, this aspect factorizes with the main purpose of this paper. Indeed, such corrections are equally relevant (and can be applied in a similar way) for ML-- and transfer-matrix--based methods.}
In order to unfold a measured distribution, a simple matrix-inversion of $R_{ij}$ is the easiest approach, but there are several flaws. 
In general, this induces important fluctuations in the unfolding result, as well as large statistical uncertainties and strong bin-to-bin (anti-)correlations. 
This phenomenon is called the \textit{high frequency problem}.
As a consequence, any further use of the unfolded data in subsequent studies requires a precise knowledge of the corresponding covariance matrix, which is difficult to achieve in such configuration~(see Appendix E of Ref.~\cite{Davier:2023cyp}).
In addition, it is possible that unphysical negative entries appear, since a true inversion of the detector response matrix needs to contain negative entries. 
To overcome these problems it is necessary to introduce some extra regularization, which will always result in a bias (e.g.\ towards the Monte Carlo simulation). 
This trade-off between a small bias and acceptable statistical uncertainties is managed through the strength of the regularization.

The \textbf{Singular Value Decomposition (SVD)} \cite{Hocker:1995kb} is one of the most-used unfolding algorithms. 
The solution to the simple matrix inversion is the maximum likelihood solution for data featuring independent Poissonian fluctuations in each bin. 
In this case the log-likelihood function is 
\begin{align}
    \ln L(\mathbf{\hat{t}}) = \sum_i \biggl( 
    r_i \ln(\textstyle\sum_j R_{ij}\hat{t}_j) 
    - \textstyle\sum_j R_{ij}\hat{t}_j
    - \ln(r_i!) \biggr),
\end{align}
with the measured distribution $r_i$, the response matrix $R_{ij}$, and the estimator of the bin means of the true histogram $\hat{t}_j$.
To avoid the high frequency problem, the Tikhonov regularization term is added to the maximum likelihood in order to achieve smoothness of the resulting unfolded distribution. The new maximum likelihood estimator reads
%
\begin{align}
    \phi = \ln L(\mathbf{\hat{t}}) - \tau  \sum_j \left[ (w_{j+1}-w_j)-(w_{j}-w_{j-1}) \right]^{2} ,  
\end{align}
regularizing the ratio of the estimator and the Monte Carlo truth-level distribution $(\rm{i.e.\ } w_j = \hat{t}_j/\tilde{t}_j)$ with a parameter $\tau$ that determines the strength of this regularization.
The actual implementation of this regularized log-likelihood is efficiently executed using the algorithm by H\"{o}cker and Kartvelishvili~\cite{Hocker:1995kb}, which employs a rescaling of the response matrix, as well as its name-giving \textit{singular value decomposition} (SVD).

The \textbf{Iterative Bayesian Unfolding (IBU)}~\cite{DAgostini:1994fjx,DAgostini:2010hil} provides another possibility to obtain an unfolded distribution with a reasonable balance between bias and statistical uncertainties. 
Given the detector response matrix introduced in Eq.~\ref{Eq:binnedFolding}, the posterior distribution calculated with Bayes' theorem can be written as
\begin{align}
    R'_{ji} = P(t_j|r_i) &= \frac{P(r_i|t_j) P(t_j)}{P(r_i)} \nonumber \\\
    &=  \frac{P(r_i|t_j) P(t_j)}{\sum_k P(r_i|t_k)P(t_k)},
    \label{eq:R'formula}
\end{align}
with
\begin{align}
    P(t_j) = \frac{t_j}{\sum_k t_k} \qquad \text{and} \qquad P(r_i) = \frac{r_i}{\sum_l r_l}.
\end{align}
The posterior constructed with the Monte Carlo simulation $R'_{ji}=P(\tilde{t}_j|\tilde{r}_i)$ can be used to propagate events from the reconstructed level to truth level. 
The unfolded distribution $u_j$ is hence obtained to be:
\begin{align}
    u_j = \sum_i R'_{ji} r_i.
    \label{eq:IBU_unfold}
\end{align}
If the experimental data distributions ($t_j$, $r_i$) and the Monte Carlo simulations ($\tilde{t}_j$, $\tilde{r}_i$) were identical, the unfolding result would be perfect~(provided that the detector effects are also well simulated). 
However, in an example where the experimental spectra and the simulated ones are very different, the resulting unfolded distribution $u_j$ can carry a strong bias towards the Monte Carlo simulation. 
This \textit{model dependency} of the unfolding procedure can be reduced by iterating the calculation several times, with the unfolded distribution being used as a new prior for the truth distribution in simulation.
The crucial parameter in Iterative Bayesian Unfolding is the number of iterations $N$, which provides a regularisation. 
While biases decrease with higher number of iterations, statistical uncertainties increase. 
For an infinite number of iterations the unfolding result converges towards the maximum likelihood solution (see Ref.~\cite{Zech:2011sjd} and references therein).
This is matched by the result of the simple matrix inversion~\cite{Cowan:2002in} if no negative Poisson rate estimates are present in any bin.
In addition, an explicit regularisation can be introduced, e.g.\ by performing fits of the unfolded distribution at each iteration~\cite{DAgostini:1994fjx}.
However, in the studies performed in this paper, only the regularisation based on the number of iterations will be employed for the IBU method.

The \textbf{Iterative Dynamically Stabilised (IDS)} \cite{Malaescu:2009dm,Malaescu:2011yg} method is based on the same principles as IBU, with however a very different regularisation procedure implemented for IDS.
In the case of identical initial and final binning the unfolded distribution is given by \cite{Malaescu:2009dm}
\begin{align}
    u_j = \quad \frac{N_D}{N_{\mathrm{MC}}} \cdot &\tilde{t}_j 
     + \sum_i \biggl[ f(|\Delta r_i |, \hat{\sigma}(r_i) , \lambda_U) \cdot \Delta r_i \cdot R'_{ji} \nonumber \\\
    & + (1-f(|\Delta r_i |, \hat{\sigma}(r_i) , \lambda_U)) \Delta r_i \delta_{ij} \biggr],
    \label{eq:IDS_unfold}
\end{align}
with 
\begin{align}
    \Delta r_i &= r_i -\frac{N_D}{N_{MC}} \tilde{r}_i \, ,   \\\
    \hat{\sigma}(r_i) &= \sqrt{\sigma^2(r_i) + \biggl( \frac{N_D}{N_{MC}} \biggr)^2 \sigma^2(\tilde{r}_i)} \, ,
\end{align}
the total number of Monte Carlo Events $N_{MC}$, the total number of data events $N_D$ and the regularization function
\begin{align}
    f(\Delta x, \sigma, \lambda) := 1-e^{-\bigl(\frac{\Delta x}{\lambda \sigma}\bigr)^2}.
\end{align}
The main idea of the regularization function $f$ is to quantify the information about the significance of an absolute deviation between two spectra in a given bin ($\Delta x$), with respect to its corresponding uncertainty ($\sigma$). 
The free parameter $\lambda$ denotes the strength of the regularization: small $\lambda$ lead to higher values of the function, hence deviations are interpreted as more significant; a large $\lambda$ has the opposite effect.

The first term on the right-hand side of Eq.~\eqref{eq:IDS_unfold} is the Monte Carlo truth component multiplied by a normalization factor. 
The second part (in square brackets) treats the structures that are only present in data and are not included by the simulation. 
Only a fraction $f$ of these events is unfolded using the pseudo-inverse $R'_{ji}$, built employing the current posterior. 
The rest of the events are kept in their respective bin. 
Therefore, reducing the unfolding parameter $\lambda_U$ enhances the number of events which are unfolded using the pseudo-inverse.
It is worth noting that this regularisation procedure acts locally, on each individual bin of the distributions, while the methods employed for SVD and (sometimes) for IBU induce global constraints on the shapes of distributions.
These differences among the approaches have consequences for the level of bin-to-bin statistical correlations, as well as for the induced bias.

In addition to the calculation of the unfolded distribution the IDS algorithm updates the detector response matrix $R_{ij}$ iteratively as well, employing the same regularisation function, with a parameter $\lambda_M$.~\footnote{There is also an extended algorithm that allows e.g.\ to treat fluctuations introduced through the background subtractions. More information about this can be found in Ref.~\cite{Malaescu:2009dm}.}

\subsection{IcINN Unfolding}
\label{s:IcINN_method}

Apart from using matrix--based unfolding algorithms, it is possible to obtain a (pseudo-) inversion of the detector response with ML algorithms. 
Besides algorithms like OmniFold \cite{Andreassen:2019cjw}, which use discriminative neural networks, there are also possibilities to use generative neural networks for unfolding, employing e.g.\ conditional Invertible Neural Networks (cINN) \cite{ardizzone2018analyzing,ardizzone2019guided}. 

cINNs are normalizing flows, which are trained with Monte Carlo events to establish a bijective connection between the truth-level information and a multidimensional Gaussian latent space, while being conditionalised on the detector-level information. 
During training, paired samples of detector- and particle-level events are passed through the network to the latent space.

The correct calibration of the particle-level distributions is guaranteed by optimizing the probability of the network parameters $p(\theta|\tilde{t},\tilde{r})$. The loss function of the cINN encodes therefore the negative log likelihood
\begin{align}
    \mathcal{L}
    = &-\langle\log p(z(\tilde{t})|\theta,\tilde{r})\rangle_{\tilde{t}\sim f, \tilde{r}\sim g}
    -\langle\log \left|\frac{\mathrm{d}z}{\mathrm{d}\tilde{t}\:} \right|\rangle_{\tilde{t}\sim f, \tilde{r}\sim g}
    \nonumber \\\
    &-\lambda \, \theta^2
    + \text{const.} \, ,
    \label{cINN_loss}
\end{align}
using the latent space variable $z$.
Once the training has converged we can sample from the latent space under the condition of a specific detector-level event to generate a distribution of particle-level events.
This preserves the statistical nature of the cINN--based unfolding, which is one of the key features of these methods. 
Further details about the cINN and its application in unfolding can be found in Ref.~\cite{Bellagente:2020piv}.

\begin{figure*}[htbp]
    \centering
    \includegraphics[width=\linewidth]{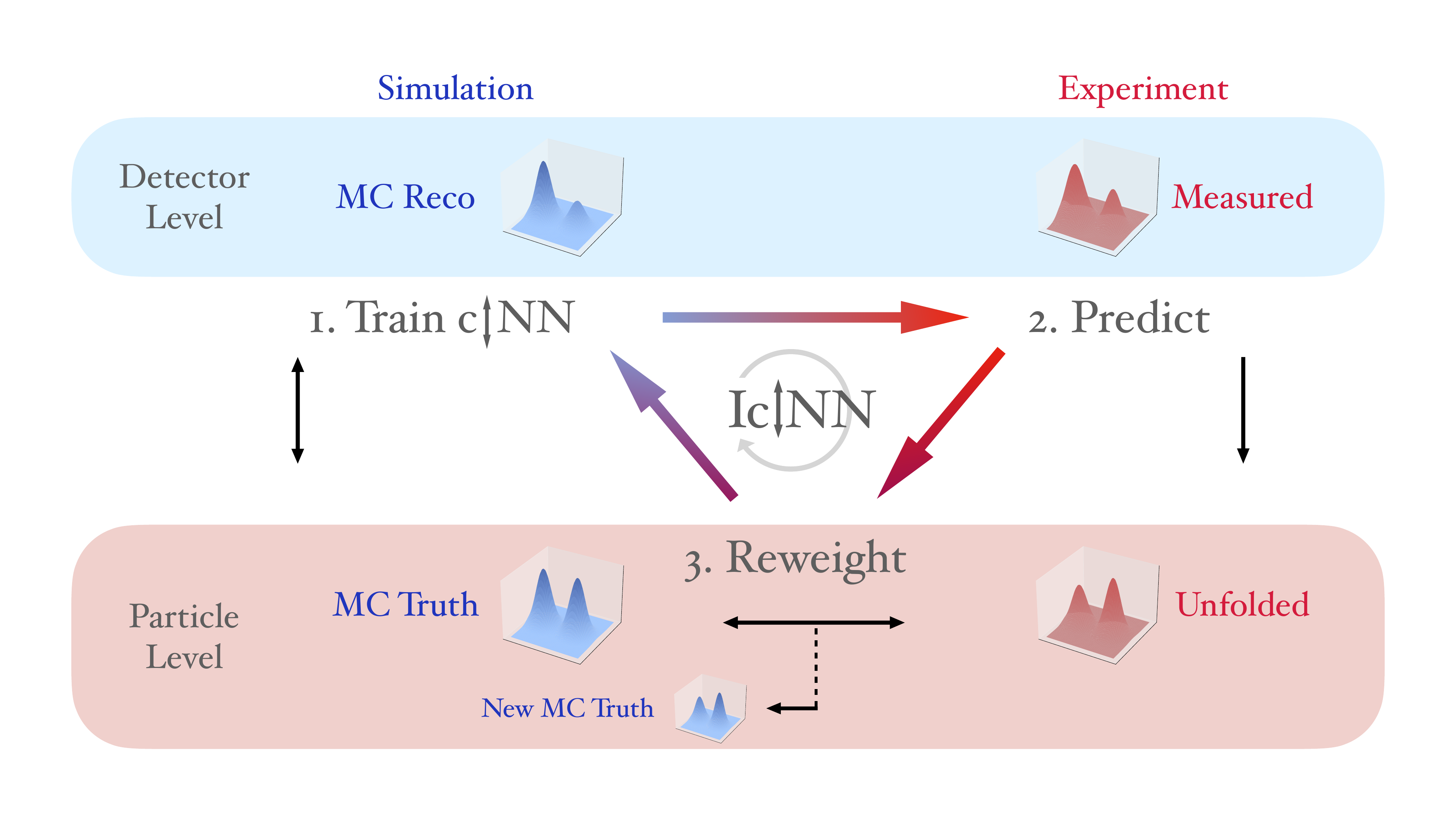}
    \caption{The main steps of the IcINN algorithm. Figure reproduced from Ref.\cite{Backes:2022vmn}.}
    \label{fig:IcINN}
\end{figure*}

In general, the cINN is able to learn a posterior distribution $R'(t|r)$, but, as for IBU and IDS in the matrix--based unfolding, the learned expression will depend on the event distribution of the training sample, i.e.\ the Monte Carlo simulation. 
This model dependency results again in a bias of the unfolded distribution towards the Monte Carlo simulation, which can be reduced by adapting an iterative setup. 
The corresponding IcINN algorithm \cite{Backes:2022vmn}, visualised in Figure \ref{fig:IcINN}, works as following:
\begin{itemize}
    \item The first two steps are identical to the standard cINN setup: we first train the cINN on the simulated sample $(\tilde{t}, \tilde{r})$ and apply it then to the measured data $r$, i.e.\ we sample in the latent space $z$ under the condition of a measured event to obtain an unfolded distribution $u$.
    \item In a third step we train a classifier to learn the ratio between the phase space densities of the unfolded distribution $u$ and the truth-level prior distribution $t$. We then reweight the simulation on particle-level to match the unfolded distribution. Since each event of the simulation on particle-level is connected to one event of the simulation on detector-level, the event weights are transferred from particle- to detector-level.
    \item Employing the new reweighted simulation sample, we then repeat all the steps of training-unfolding-reweighting until the algorithm has converged~(i.e. until the result is stable w.r.t. extra iterations).
\end{itemize}
It has been demonstrated that this setup iteratively reduces the bias towards the Monte Carlo simulation. 
One important feature of the IcINN unfolding is the possibility to predict an unfolded probability \textit{distribution} for a single measured event. 
This is obtained by sampling multiple points in the Gaussian latent space and mapping them to truth level, while fixing the measured event as the condition. 
The ability to unfold one single event in such a probabilistic way has up until now been unique to cINN unfolding.

In Ref.~\cite{Backes:2022vmn} it has already been demonstrated, that the IcINN produces single-event unfolded distributions that agree with the predictions of an analytically solvable toy model. 
Nevertheless, it is worth performing a detailed comparison with the results of the matrix--based unfolding algorithms.
The technical implementation of the IcINN unfolding used in this project is the same as in Appendix A of Ref.~\cite{Backes:2022vmn}.

\section{Matrix--based Unfolding of Single Events}
\label{s:single}

The main difference between ML--based and matrix--based unfolding algorithms is the representation of the data: event-by-event or as a histogram with an introduced binning.
Up to this point it was not possible to validate the single-event unfolded distributions of the IcINN unfolding other than with analytically solvable toy models.
It is hence desirable to construct single-event unfolded distributions using the matrix--based algorithms.
This enables more detailed comparisons, event-by-event, in restricted phase-space regions among various matrix--based methods, as well as with IcINN.
In the following we will develop a transfer-matrix--based method to approximate single-event unfolding.
The approach considers the limit where the additional event has no impact on the unfolding algorithm, so that one can obtain an unfolded distribution by applying the unmodifed unfolding matrix to a single event.
An alternative approach, discussed in~\ref{App:WeightingApproach}, observes the effect of a single additional event by multiplying the event counts of one bin in the measured detector-level distribution with a weighting factor.

\subsection{Analytic Toy Model}
\label{s:analytic}

In order to validate the single-event unfolding for the matrix--based unfolding it is useful to adapt an analytically solvable, Gaussian toy model which gives predictions for the single-event unfolded distributions as well.
Such a toy model can be constructed by explicitly defining a detector response function combined with a truth-level Monte Carlo and data distribution.
In this case, the truth-level Monte Carlo is the assumed prior for the unfolding.

\begin{figure*}[!t]
    \centering
   \includegraphics[width=0.47\textwidth]{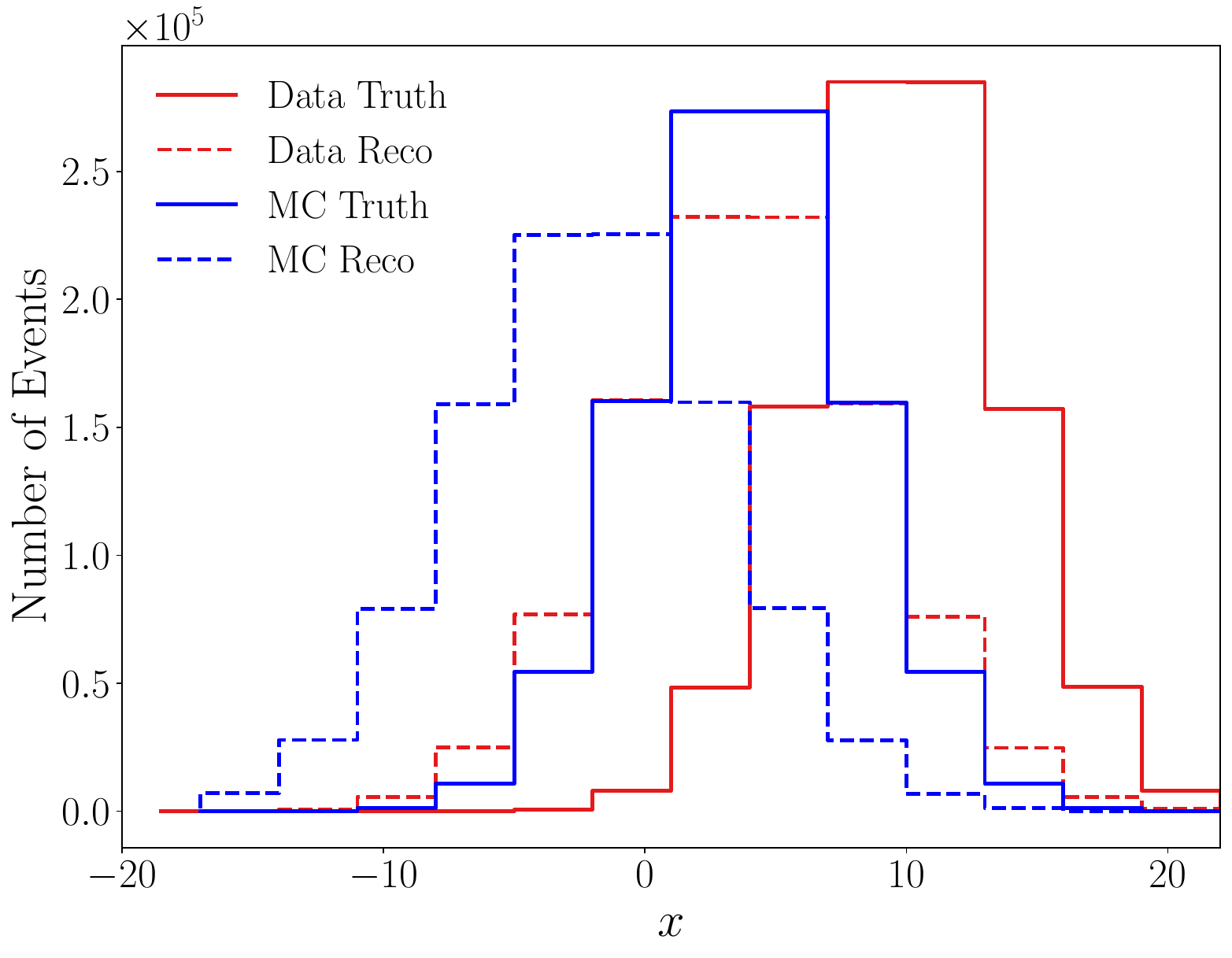}
   \hspace{0.1cm}
   \includegraphics[width=0.47\textwidth]{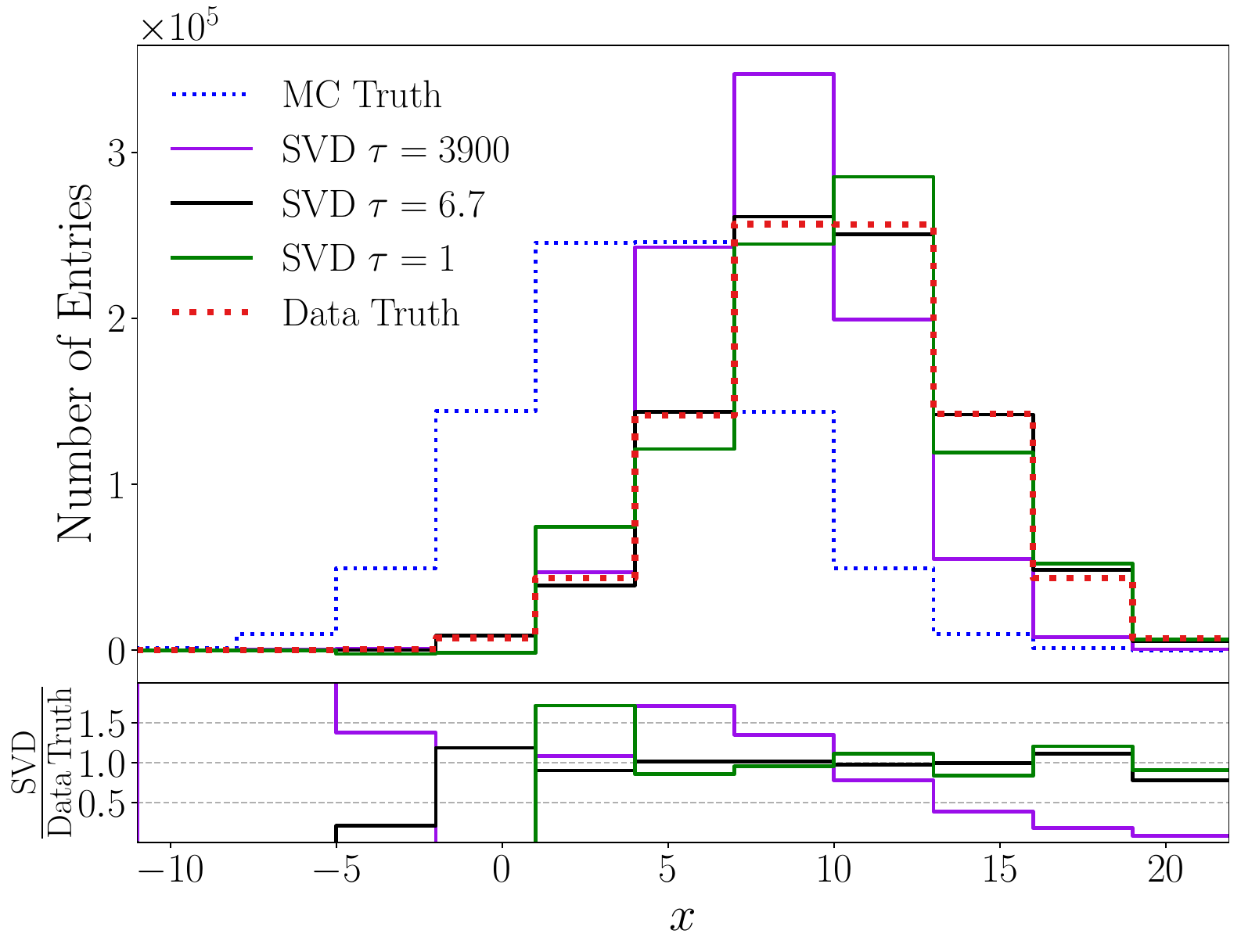}\\
   \includegraphics[width=0.47\textwidth]{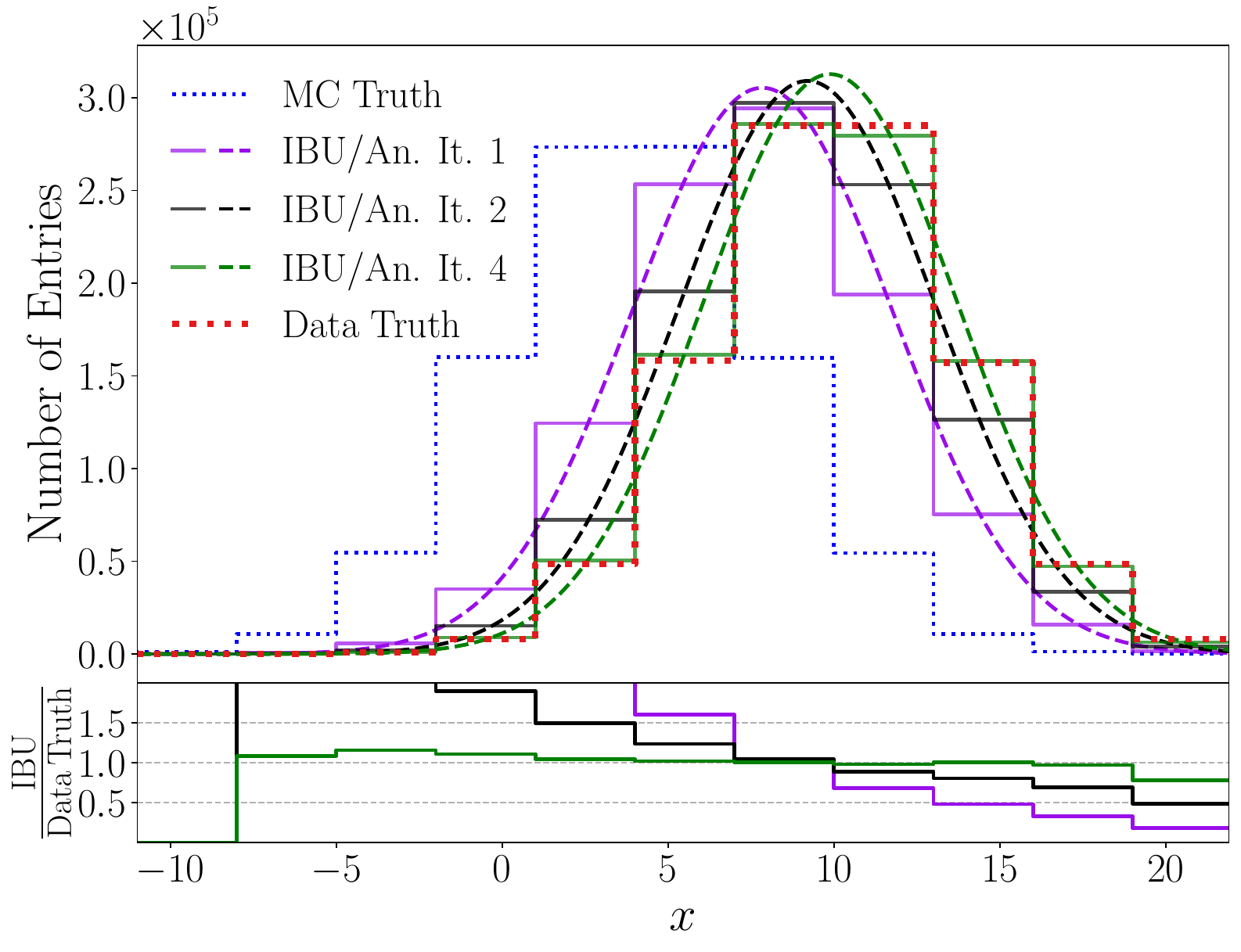}
   \hspace{0.1cm}
   \includegraphics[width=0.47\textwidth]{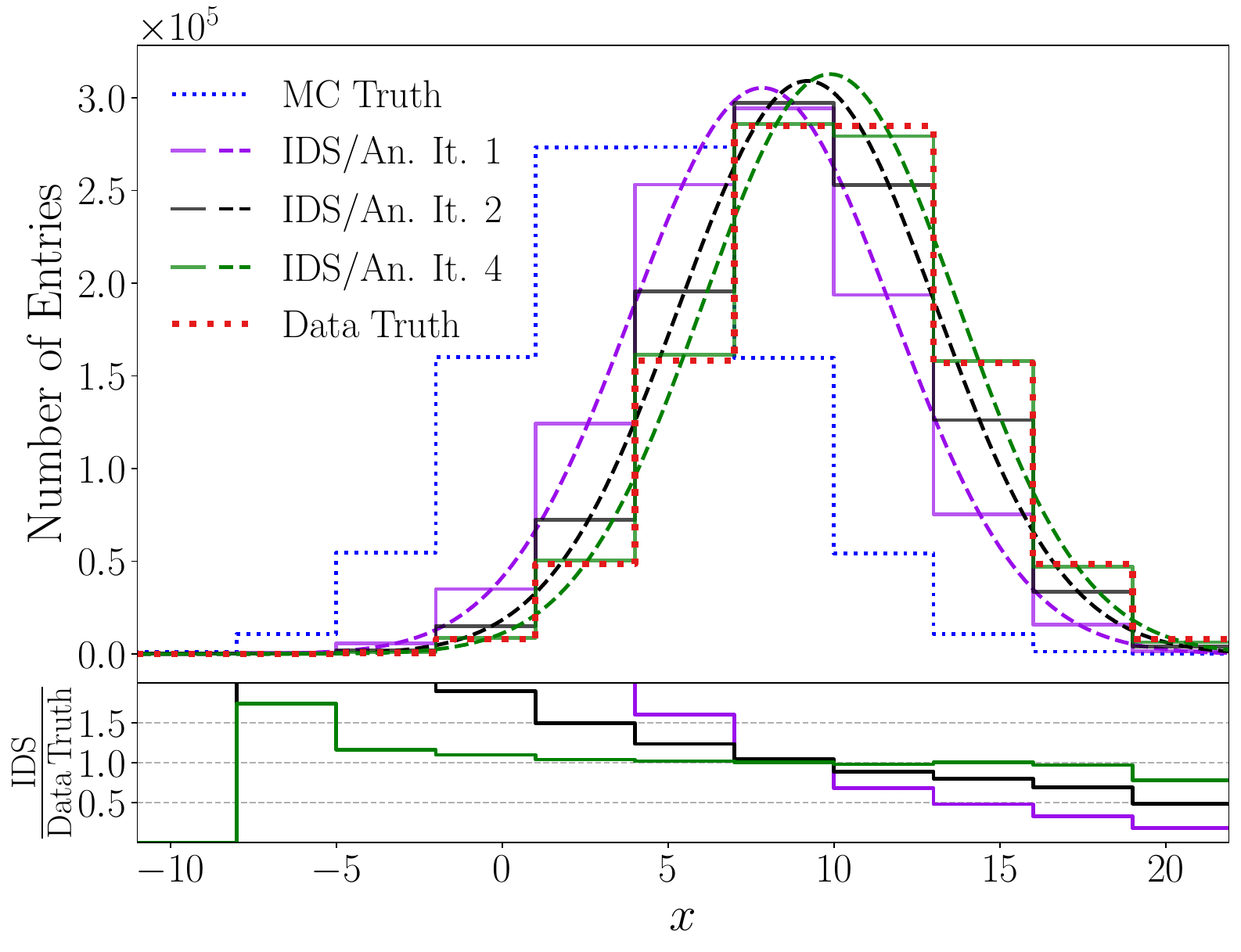}
    \caption{The analytically solvable toy model with several applied unfolding algorithms.
    The upper left plot shows the Monte Carlo distribution (blue) and the pseudo-data (red).
    A dashed histogram indicates detector-level information.
    In the upper right the unfolded distributions (purple, black, green) obtained with the SVD algorithm using different values for $\tau$ are shown.
    The ratio with the experimental truth-level data~(dotted red) shows, that for $\tau=6.7$ the unfolded distribution tends to show the smallest deviation from the truth-level data.
    The lower plots show an unfolding of the toy model with iterative matrix--based algorithms. 
    The results obtained employing various numbers of iterations are shown for IBU~(left) and IDS~(right).
    The unfolded distributions (purple, black and green solid line) are in good agreement with the analytic expectations~(respective dashed lines) in the region of high statistics. The ratios with the experimental truth-level data~(dotted red) show that the bias towards the truth-level Monte Carlo (dotted blue) is iteratively reduced.
    }
    \label{f:unfolding_applied_to_toy}
\end{figure*}

Following Ref.~\cite{Backes:2022vmn}, we choose the Monte Carlo distribution at truth-level to be a Gaussian with $f_{\text{MC}}(t)=G(t; \mu_{MC,t}=4, \sigma_{MC,t}=4)$, while the truth-level data distribution is parameterized as $f_\text{Data}(t)=G(t; \mu_\text{Data,t} = 10, \sigma_\text{Data,t} = 3.8)$.
The detector response function is a Gaussian smearing with $\mu_\text{smear}=-6$, $\sigma_\text{smear}=3$, inducing a relatively important amount of event migrations and therefore making this toy exercise interesting from the unfolding performance point of view.
The respective detector level distributions can be calculated using Eq.~\eqref{eq:unfolding_folding}.
The resulting convolution of two Gaussian functions is also a Gaussian function with a mean value and variance equal to the sum of the means and variances of the convoluted Gaussians.  
The complete set of binned distributions is shown in the upper left plot of Figure~\ref{f:unfolding_applied_to_toy}.
By applying Bayes' theorem it is possible to analytically construct the posterior and obtain the first unfolding result, namely a Gaussian distribution with 
\begin{align}
    \mu_{u,1} 
    &= \frac{(\mu_{\text{Data,r}}-\mu_\text{smear}) \, \sigma_{\text{MC,t}}^2 
    + \mu_{\text{MC,t}} \, \sigma_\text{smear}^2}{\sigma_{\text{MC,t}}^2  + \sigma_\text{smear}^2 } ,\nonumber \\\ 
    \sigma_{u,1} 
    &= \frac{\sigma_{\text{MC,t}} \sqrt{\sigma_{\text{MC,t}}^2 \, \sigma_{\text{Data,r}}^2+ \sigma_{\text{MC,t}}^2 \, \sigma_\text{smear}^2 + \sigma_\text{smear}^4} }{\sigma_{\text{MC,t}}^2+\sigma_\text{smear}^2},
    \label{eq:analytic_result}
\end{align}
using the truth-level Monte Carlo parameters ($\mu_\text{MC,t}$, $\sigma_\text{MC,t}$) and the detector-level data parameters ($\mu_{\text{Data,r}}$, $\sigma_{\text{Data,r}}$).
To obtain the parameters of the distribution after the second iteration we simply recalculate Eq.~\eqref{eq:analytic_result} replacing $( \mu_{\text{MC,t}}, \sigma_{\text{MC,t}} )$ with $(\mu_{u,1}, \sigma_{u,1} )$.
Then one can proceed similarly for the third iteration etc.
By design it is also possible to predict analytically how the single-event unfolded distributions should look like. 
Expressing a single event $r_m$ as a delta distribution leads again to a Gaussian distribution~\cite{Backes:2022vmn} with parameters 
\begin{align}
    \mu_{\text{single}}&= \frac{\sigma_\text{smear}^2 \, \mu_\text{MC,t} - \sigma_\text{MC,t}^2(\mu_\text{smear}-r_m)}{\sigma_\text{smear}^2+\sigma_\text{MC,t}^2}, 
    \nonumber \\\
    \sigma_{\text{single}}^2&= \frac{\sigma_\text{smear}^2\sigma_\text{MC,t}^2}{\sigma_\text{smear}^2+\sigma_\text{MC,t}^2}.
\end{align}
As above, this is the result obtained after one iteration and one can then perform further iterations.

At this point the classical matrix--based unfolding algorithms can be applied to the toy example described above.
Especially for the iterative algorithms it is interesting to compare the unfolded distribution after each iteration with the analytic prediction. 
Here this is rather a closure check for the implementation of the algorithms, since the Iterative Bayesian Unfolding executes the exact same steps used for the analytic derivation with a binned phase space.

The SVD unfolding with a Tikhonov regularization depends on the regularizing parameter $\tau$.
If $\tau$ is too low the regularization is too weak to suppress the high frequency problem. If $\tau$ is too high the unfolded distribution will have a bias towards the truth-level Monte Carlo simulation.
The unfolded distributions for several $\tau$ values are shown in the upper right plot in Figure~\ref{f:unfolding_applied_to_toy}.
An unfolding result close to the truth-level expectation is obtained using $\tau=6.7$, demonstrating the necessity of a reasonable balance between the regularization~(necessary to control the amplitude of the statistical variance) and the bias that it induces.

The iterative IBU and IDS algorithms show good agreement with the analytic prediction after each iteration.
The results up to iteration $N_{\mathrm{iter}}=4$ are also shown in Figure \ref{f:unfolding_applied_to_toy}. 
The parameters for Iterative Dynamic Stabilising are set to $\lambda_U=\lambda_M=0.5$, corresponding to a weak additional regularisation.
Due to this rather conservative choice it is not surprising that the IDS algorithm approaches nearly the same result as the IBU algorithm. The biases associated to their results are iteratively reduced, down to a small level achieved with $N_{\mathrm{iter}}=4$.

\subsection{Applying the Unfolding Matrix to One Reconstructed Event}
\label{Sec:OneEvUnfolding}

Here we propose an approach allowing to employ matrix--based unfolding methods for single events.
It consists in unfolding one single data bin at a time~(i.e.\ the bin containing the event to be unfolded) and normalizing the resulting unfolded distribution such that its integral corresponds to a single-event~(i.e. normalizing the distribution to unit area).

The approach to obtain single-event unfolded distributions is applied here for the iterative matrix--based unfolding algorithms.
This is done without employing any regularisation based on a global constraint of the shape of the unfolded distribution. 
The approach consist in three steps:
\begin{enumerate}
    \item Perform the whole unfolding algorithm with the full distributions in each iteration.
    \item Extract the unfolding matrix (i.e.\ the current pseudo-inverted response matrix) in each iteration.
    \item Apply the unfolding matrix to a histogram with one single event, following the procedure defined in the unfolding algorithm.
    This is equivalent to extracting one column of the unfolding matrix.~\footnote{This approach could be employed for a simple matrix inversion, however with the other caveats of that method, related to the large variances and (anti-)correlations, as mentioned above. It could also be generalized for being applied to the SVD method with Tikhonov regularization, by employing a pseudo-inversion~(see e.g. Section 4.5 from Ref.~\cite{aster2018parameter}). Such developments are left for later studies. Another approach applicable for the SVD method is described in~\ref{App:WeightingApproach}.}
\end{enumerate}
The multiplication of the unfolding matrix with the measured spectrum is a linear operation, hence it makes no difference whether the measured events are summed up to the measured distribution before or after this multiplication.
Nevertheless, the unfolding matrix still needs to be derived using the full histogram.

For IBU this method is implemented in a straight-forward way, by using the unfolding matrix and extracting one column.
The IDS algorithm has a more involved way of deriving the unfolded distribution, which has to be represented in the derivation of the single-event unfolded distribution.

To get a single-event unfolded distribution with the IDS algorithm it is necessary to adjust Eq.~\eqref{eq:IDS_unfold}.
\begin{figure*}[t]
	\centering
	\includegraphics[width=0.45\linewidth]{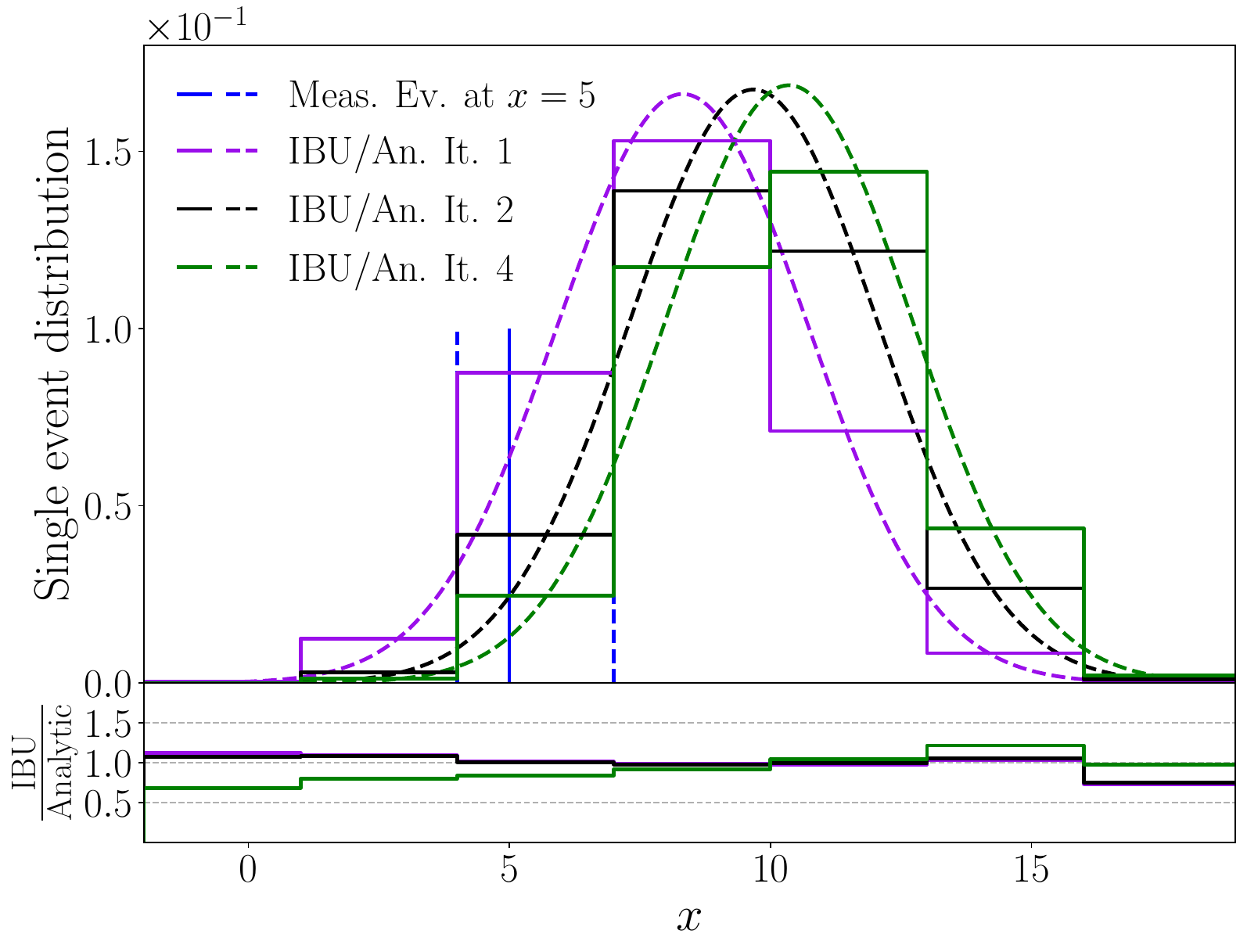}
    \hspace{0.5cm}
    \includegraphics[width=0.45\linewidth]{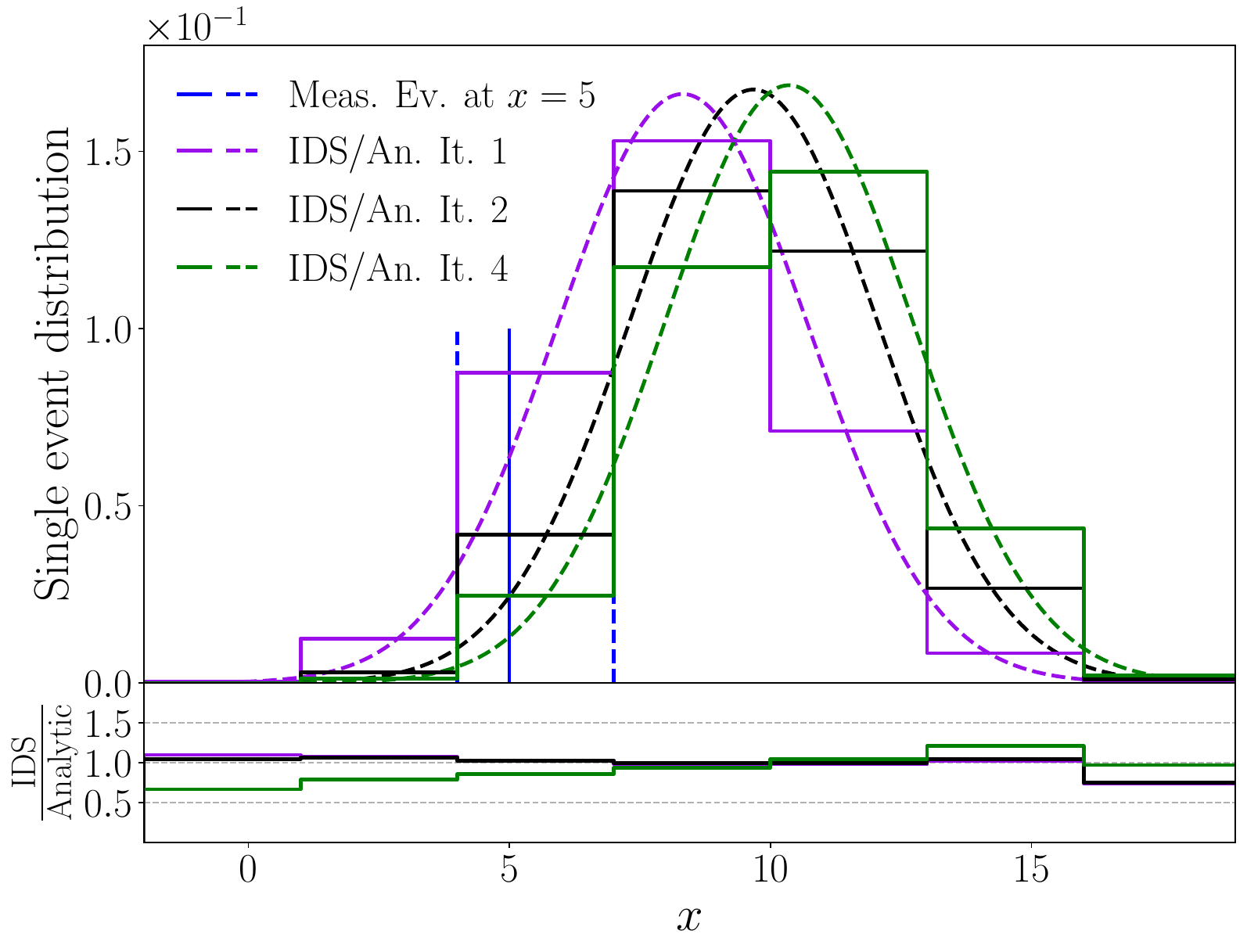}\\
    \includegraphics[width=0.45\linewidth]{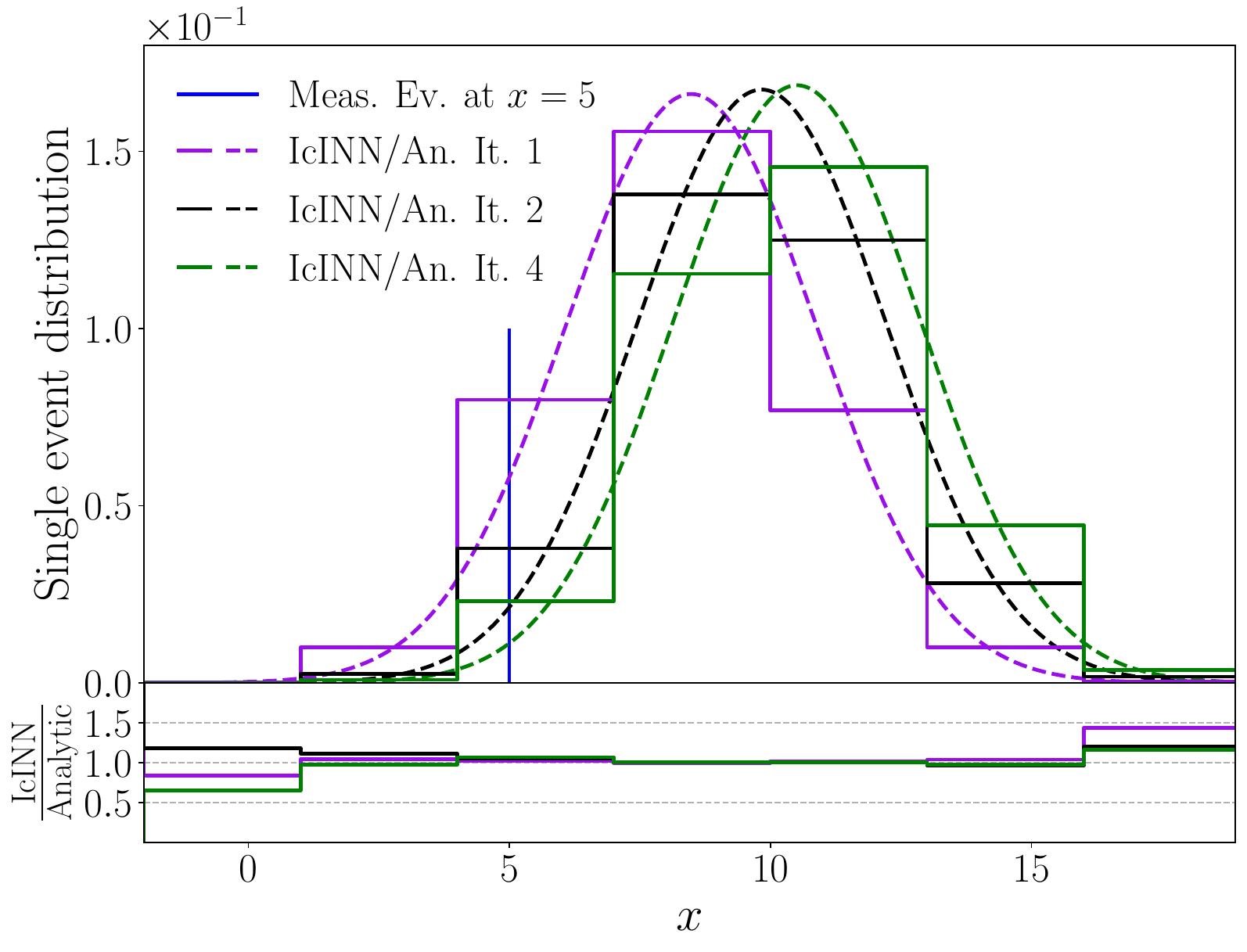}
    \hspace{0.5cm}
    \includegraphics[width=0.45\linewidth]{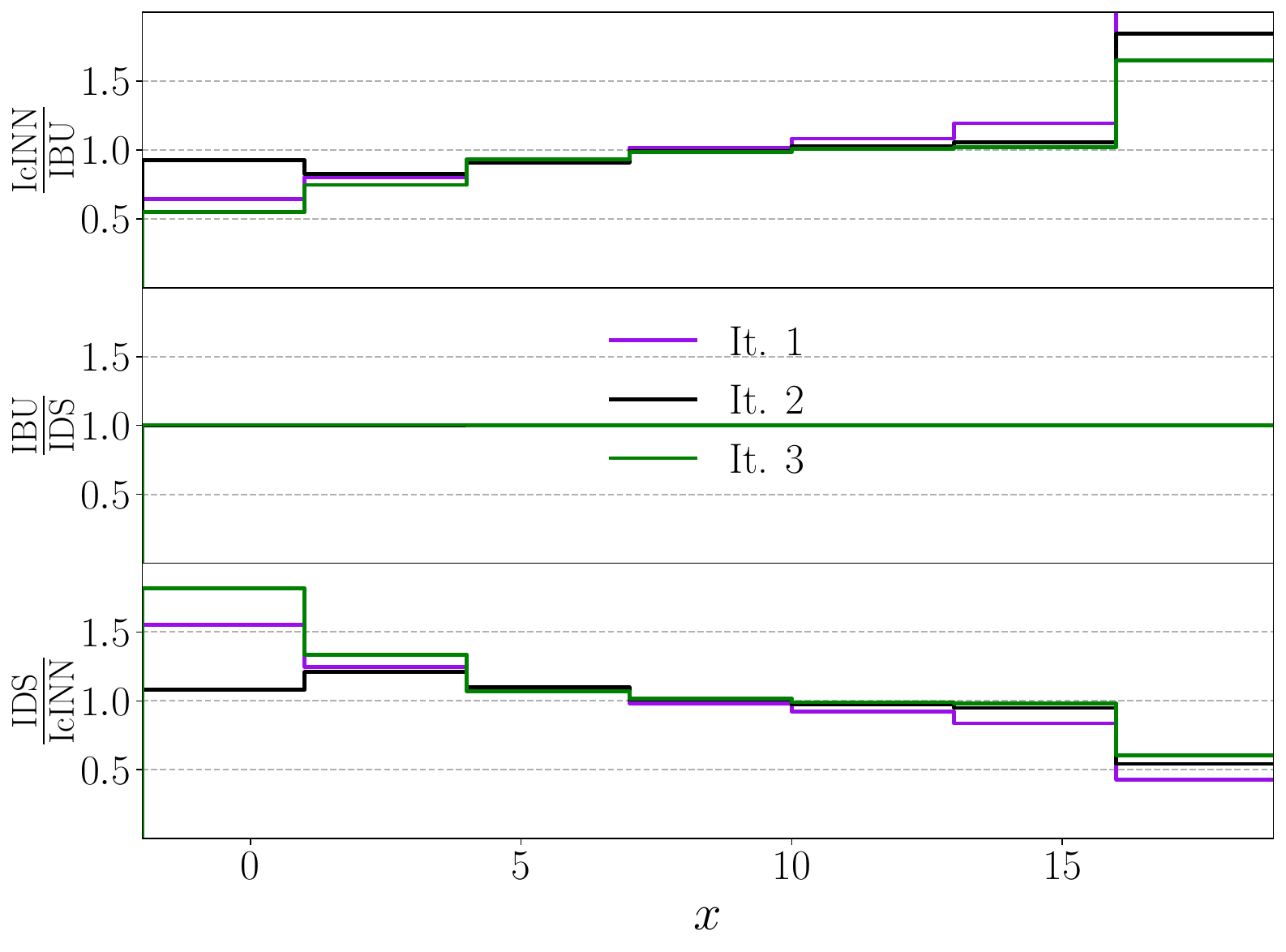}\\
        
	\caption{Single-event unfolded distributions for the matrix--based IBU~(top left), the matrix--based IDS~(top right) and IcINN~(bottom left). The reconstructed single event is represented by the blue vertical line. The unfolding results~(solid histograms) follow closely the analytic expectations for each iteration~(dashed curves). The lower right plot shows a relative comparison between the results of the different algorithms. Results for the iterations $1$, $2$ and $4$~(see main text) are shown in purple, black and green respectively.
    }  
	\label{f:single_true_iterative}
\end{figure*}
Using the fact that unfolding the detector-level Monte Carlo $\tilde{r}_i$ with the unfolding matrix $\Tilde{R}_{ji}$ yields the truth Monte Carlo $\tilde{t}_j$, Eq.~\eqref{eq:IDS_unfold} can be re-written as
\begin{align}
    u_j =& \sum_{i} \biggl[ \left( \frac{N_D}{N_{MC}}\cdot \tilde{r}_i + f(|\Delta r_i|, \hat{\sigma}(r_i), \lambda_U) \cdot \Delta r_i \right) \Tilde{R}_{ji} \nonumber\\\
    & + (1-f(|\Delta r_i|, \hat{\sigma} (r_i), \lambda_U)) \Delta r_i \delta_{ij} \biggr].
    \label{eq:single_unfolding_step_2}
\end{align}  
The summation over the index $i$ runs over all the bins in the detector-level data.
To get a proper {\it single-event} distribution of a detector-level event, the sum over the bins $i$ is dropped and the fixed bin value $i_s$ is used instead.
In addition, the number of entries in the detector-level bin $i_s$ has to be accounted for.
This can be achieved by dividing Eq.~\eqref{eq:single_unfolding_step_2} with the number of entries in the data bin at reconstruction level $r_{i_s}$.
An event in bin $i_s$ of the detector-level data is consequently unfolded to a single-event distribution $e_j(i_s)$ calculated as
\begin{align}
    e_j(i_s) =\quad & \frac{N_D}{N_{MC}}\cdot \frac{\tilde{r}_{i_s}}{r_{i_s}} \cdot \Tilde{R}_{ji_s} \nonumber \\\
     + &\frac{\Delta r_{i_s}}{r_{i_s}} \cdot \biggl[ f(|\Delta r_{i_s}|, \hat{\sigma} (r_{i_s}), \lambda_U) \cdot \Tilde{R}_{ji_s} \nonumber \\\
     + &(1-f(|\Delta r_{i_s}|, \hat{\sigma}(r_{i_s}), \lambda_U)) \cdot \delta_{ji_s} \biggr].
    \label{eq:single_IDS_single_Unfold}
\end{align}
This distribution has indeed an integral normalized to unity, i.e.\ $\sum_j e_j(i_s) =1$.
In conclusion, it is possible to obtain the single-event unfolded distributions by performing a usual IDS unfolding and extracting all necessary variables of Eq.~\eqref{eq:single_IDS_single_Unfold} to construct the single-event distribution for each event of the detector-level data.

The toy model results for the single-event unfolded distributions for IBU and IDS after each iteration are shown together with their analytic prediction in Figure~\ref{f:single_true_iterative}.
It can be noticed that no negative entries are present in these distributions in contrast to other methods~(see~\ref{App:WeightingApproach}).
This is expected, because here the iterative procedure is identical to the one for the unfolding of the full data distribution and the pseudo-inverse of the response matrix does not contain negative entries, even for multiple iterations. In addition, the distributions match the analytic predictions in the regions of high statistics, both for IBU and IDS. 
With this method, the individual single-event distributions sum up to the full unfolded distribution exactly. 
This approach enables the comparison between single-event distributions obtained either with matrix--based methods or with an iterative ML--based unfolding algorithm.
In the ratio plots of Figure~\ref{f:single_true_iterative} we see a good agreement of the unfolding methods (transfer-matrix--based as well as IcINN) in the bulk of the distributions for this example.
It is not surprising that the results of IcINN and the transfer-matrix--based unfolding are similar, since all of them rely on a Monte Carlo-based pseudo-inversion of the detector response and need to be applied iteratively in order to reduce the bias.

\begin{figure*}[t]
    \centering
    \includegraphics[width=0.47\textwidth]{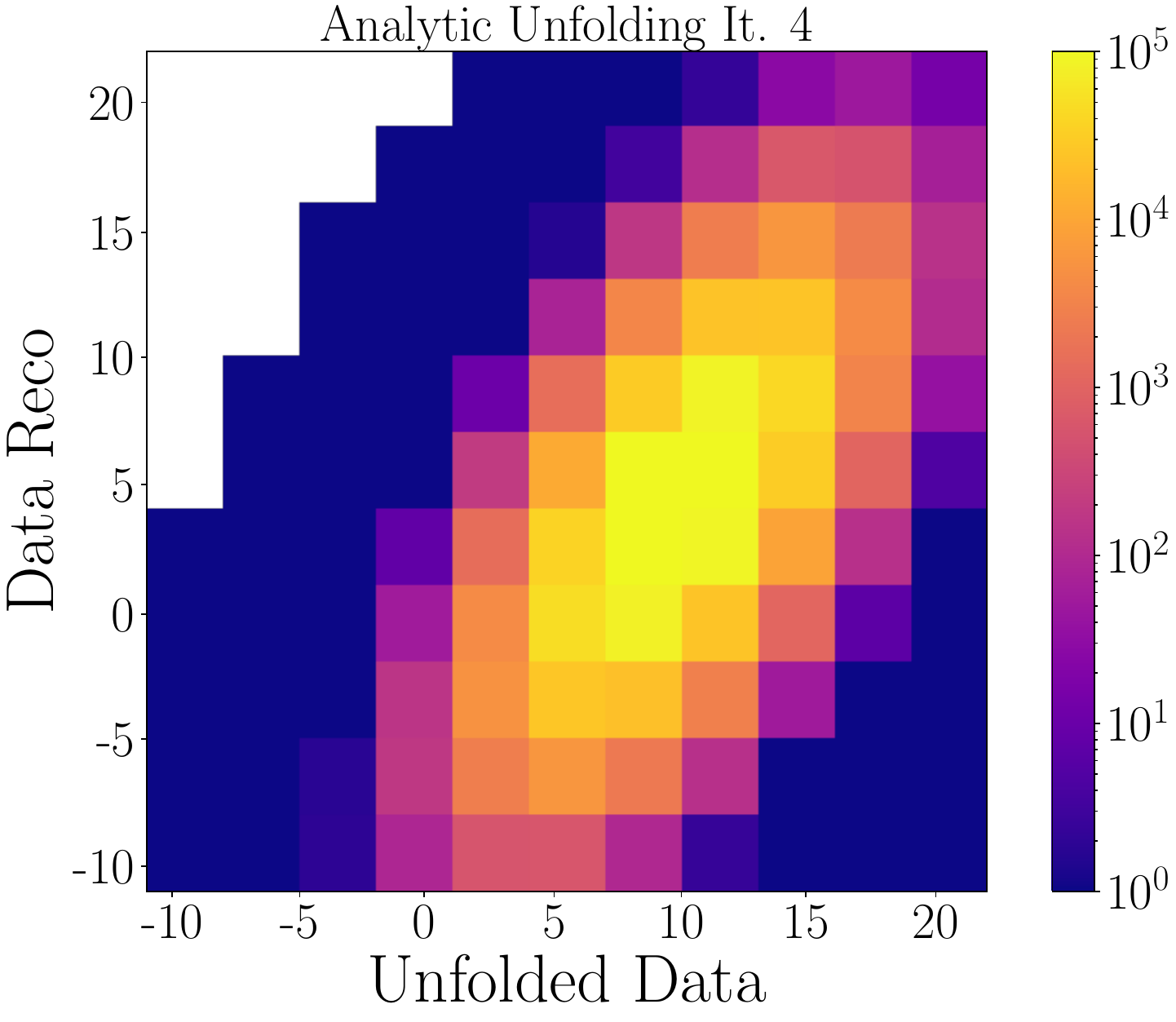}
    \hspace{0.1cm}
    \includegraphics[width=0.47\textwidth]{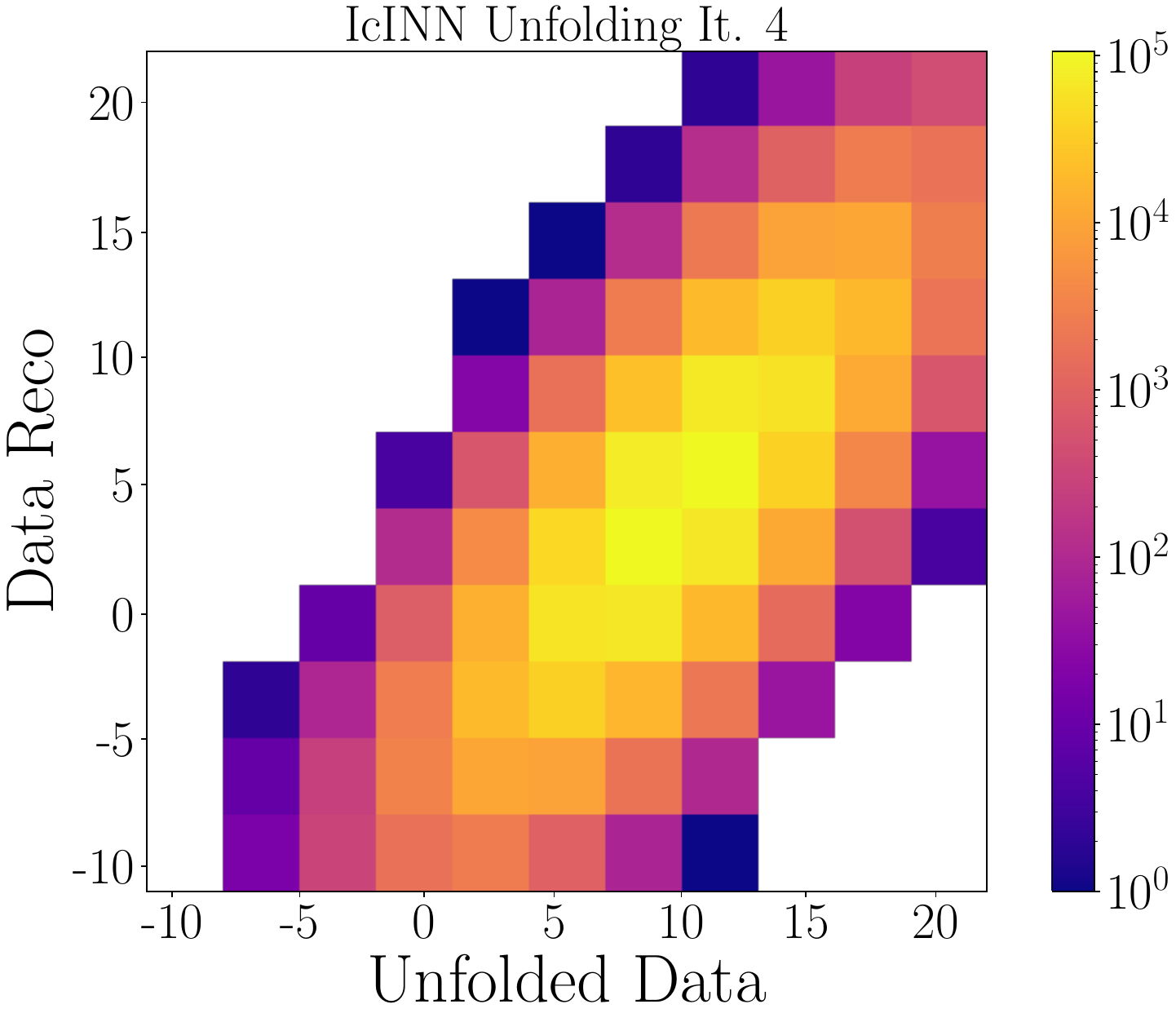}\\
    \vspace{0.3cm}
    \includegraphics[width=0.47\textwidth]{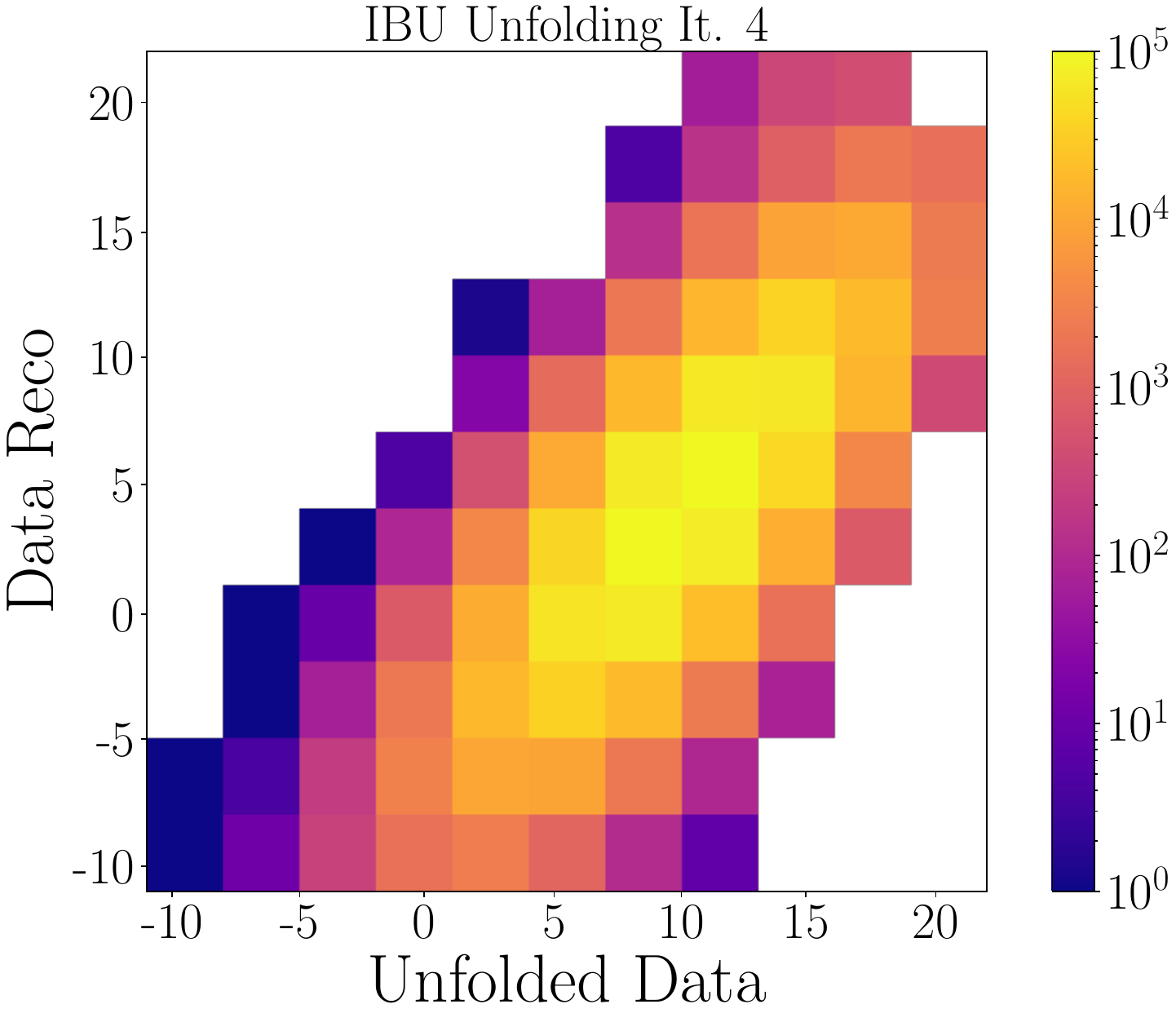}
    \hspace{0.1cm}
    \includegraphics[width=0.47\textwidth]{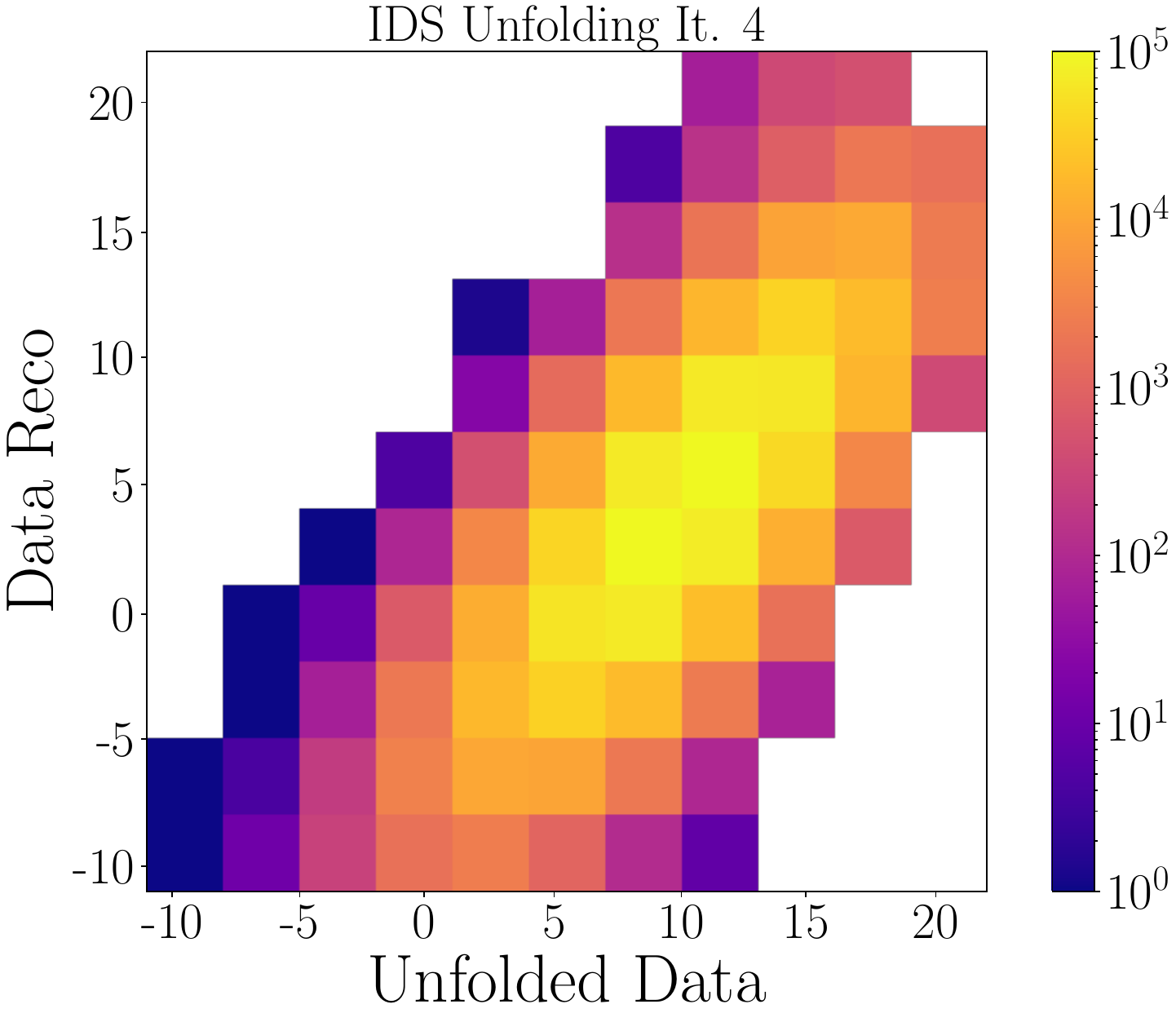}
    \caption{Correlation distributions between detector- and unfolded-level data.
    An analytic prediction for the toy model~(top left) and the results for IcINN~(top right), IBU~(bottom left) and IDS~(bottom right) are shown for $4$ iterations.
    The distributions are similar for all the considered iterative unfolding algorithms.
    The corresponding plots with an additional normalization according to Eq.~\ref{eq:R'formula} (i.e.\ the posterior distribution $R'_{ji}$) are given in~\ref{App:Normalized_correlation_distribution}. }
    \label{f:2D_measured_unfolded}
\end{figure*}

Another benefit from these explicit single-event unfolded distributions is the possibility to visualise the unfolding process itself in a two-dimensional plot containing the detector-level data as well as the unfolded data (Figure~\ref{f:2D_measured_unfolded}).
It is clearly visible that IcINN, IBU and IDS all implement a similar unfolding process.
Furthermore, preserving the correlation between the reconstructed- and the unfolded-level quantities on the event-by-event basis enables the possibility of adjusting the event selection at reconstructed level, even after the unfolding is performed.
This feature can be especially useful, e.g.\ for understanding the potential impact of some detector effects on the physics results.

As a side note, while providing an unfolded distribution for each data event, this procedure yields the same result for all the data events corresponding to the same reconstructed-level bin.
This is caused by the loss of information about the exact measured value for each data event, occurring when the binning is introduced.
The bin size can be decreased only to a certain point, until large bin-to-bin (anti-)correlations appear, in a regime where the unfolding problem is ill-posed and the systematic uncertainties associated to its solution are hard to evaluate reliably.
However, a possible refinement of this study could be achieved by interpolating between the unfolded distributions corresponding to several neighbouring bins on the reconstructed level.   
This shall allow to derive an unfolded distribution that better corresponds to the considered event, by using the actual value of the quantity being unfolded, instead of just the bin that it fits into.
Such an improvement is expected to be most relevant in presence of large and fast-varying gradients in the distributions, large bins and/or of fast-varying detector smearing effects.
This type of developments are left for a follow-up study.

\subsection{Uncertainties when Unfolding a Single Reconstructed Event}

The evaluation of uncertainties for the unfolding result is non-trivial due to non-zero off-diagonal entries in the unfolding matrix which introduce correlations between bins.
A robust way for doing the uncertainty propagation is \textit{bootstrapping}.
There are other possibilities for individual unfolding methods (e.g.\ in the case of IBU~\cite{DAgostini:1994fjx}), but bootstrapping generalizes well (and consistently) to the various unfolding methods considered here.
The idea is to unfold a measured distribution several times while giving a Poissonian weight $\mathcal{P}(1)$ either to each event of the detector-level data distribution or each (truth, reco)-pair of an event in the Monte Carlo simulation.
This approach accounts for statistical fluctuations in the measured spectrum, as well as in the detector response.
These {\it fluctuations} are performed several times to calculate the covariance matrix using the standard formula:
\begin{align}
    \mathrm{cov}_{ij} = \frac{1}{N_{f}-1} \sum_{n=1}^{N_{f}} (u_i^{(n)} - \overline{u}_i)(u_j^{(n)} -\overline{u}_j),
    \label{eq:full_covariance}
\end{align}
with the number of pseudo-experiments~(sometimes called ``toys'') $N_{f}$ and the mean value:
\begin{align}
    \overline{u}_i = \frac{1}{N_{f}} \sum_{n=1}^{N_{f}} u_i^{(n)} \, .
\end{align}
The index $(n)$ accounts for the dependency of the current unfolded distribution on the fluctuations of the current toy model.
The standard deviation in each bin and the correlation matrix can be calculated as:
\begin{align}
    \sigma_j = \sqrt{\mathrm{cov}_{jj}}, \qquad \qquad \mathrm{corr}_{ij} = \frac{\mathrm{cov_{ij}}}{\sigma_i \, \sigma_j}.
\end{align}
The next step is to evaluate the covariance matrices of the single-event distributions. 
For validation, it shall be possible to derive the covariance matrix of the full unfolded distribution using the covariances of the single-event unfolded distributions.
To derive the single-event covariances the following notations are used: 
\begin{itemize}
    \item $e(i_s)$ = distribution resulting from the unfolding of a single event in bin $i_s$ of the detector-level data, with the integral normalised to the event weight,
    \item $e^{(n)}(i_s)$ = single-event unfolded distribution, when Poisson fluctuations are applied to the Monte Carlo (data), as indicated with the number of the pseudo-experiment $(n)$; the integral of the distribution is normalised to the (fluctuated) event weight,
    \item $e_{k}^{(n)}(i_s)$ = $k$'th bin of the $e^{(n)}(i_s)$ distribution,
    \item $\overline{e_k(i_s)} = \frac{1}{N_f}\sum_{n'=1}^{N_f} e_k^{(n')}(i_s)$ is the mean value of $e_{k}^{(n')}(i_s)$ over all fluctuated pseudo-experiments,
    \item $\mathrm{cov}_{kl} (e(i_1),e(i_2))$ = covariance between the shapes of two single-event unfolded distributions~(each of which is normalized to unit integral), where $i_1$ and $i_2$ are either different or identical.
\end{itemize}
It is possible to evaluate the covariance between two single-event unfolded distributions of the measurement, normalized to unit integral, as
\begin{align}
\mathrm{cov}_{kl}^{\mathrm{Data/MC}}(e(i_1), e(i_2)) =\quad & \frac{1}{N_{f}-1}  \sum_{n=1}^{N_{\mathrm{f}}} \cdot \nonumber \\  
\Bigl(e_{k}^{(n)}(i_1) -\overline{e_{k}(i_1)}\Bigr)  
    \cdot &\Bigl(e_{l}^{(n)}(i_2) -\overline{e_{l}(i_2)}\Bigr),
    \label{eq:single_covShape}
\end{align}
where the Data or MC exponent indicates the source of the $N_{\mathrm{f}}$ fluctuations.
It has been checked, both analytically and numerically, that the sum of the single-event covariance matrices allows to reproduce the covariance matrix of the full unfolded distribution.
In order to do so, it is necessary to distinguish between the uncertainty resulting from the response matrix fluctuation and the uncertainty resulting from the detector-level data fluctuation, which can also be factorised in terms of fluctuations of the normalization and shape of the distribution~(see e.g. Ref.~\cite{Casadei:2015hia} for a related discussion on such factorisation).
A more involved discussion of the analytic treatment can be found in~\ref{a:single_event_uncertainty}.

We apply this procedure to obtain the covariances for the analytic toy model example.
Since the results of IBU and IDS are very similar, only the ones obtained for the former are displayed in the following.

\begin{figure*}[p!]
    \centering
    \includegraphics[width=0.47\textwidth]{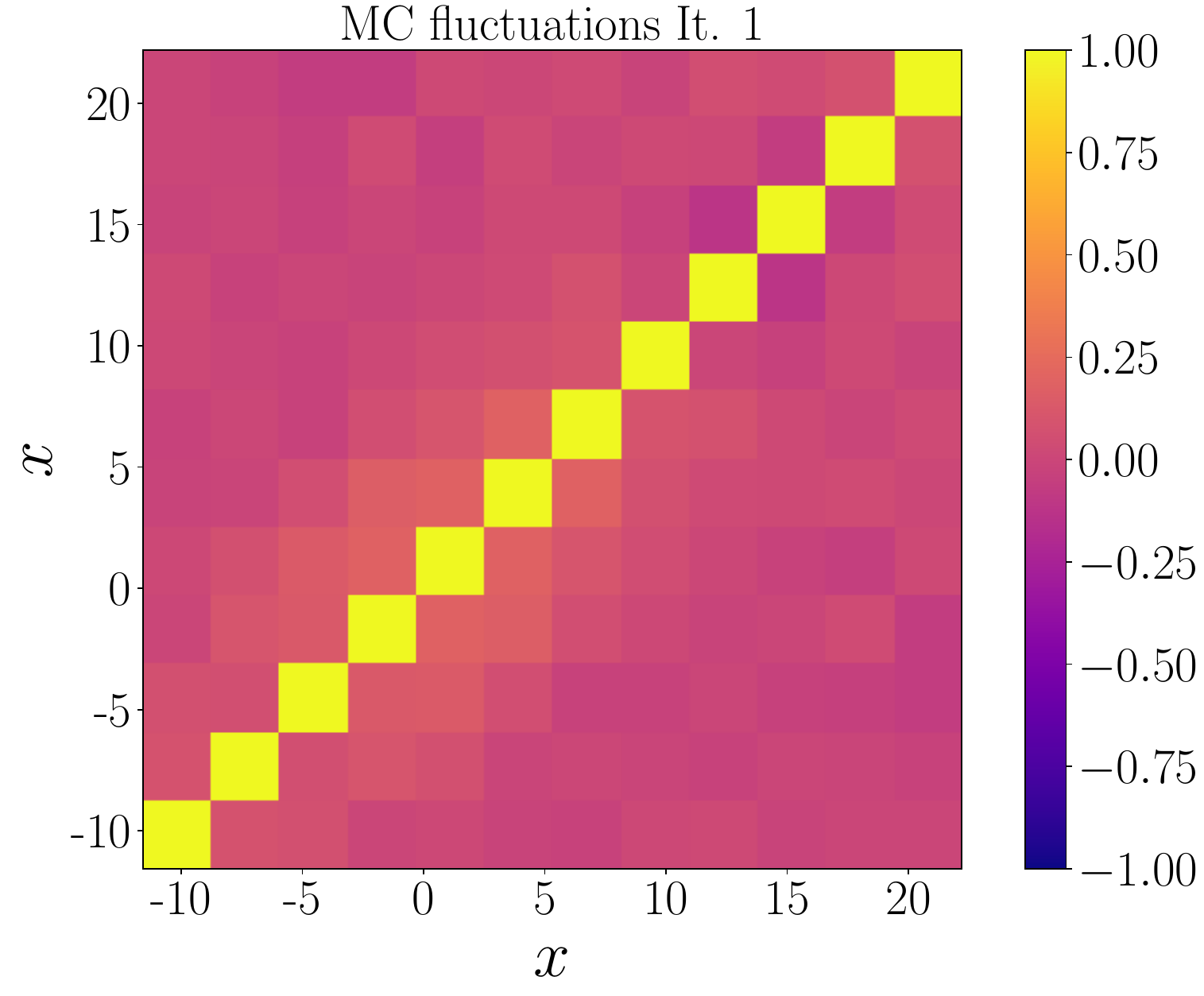}
    \hspace{0.1cm}
    \includegraphics[width=0.47\textwidth]{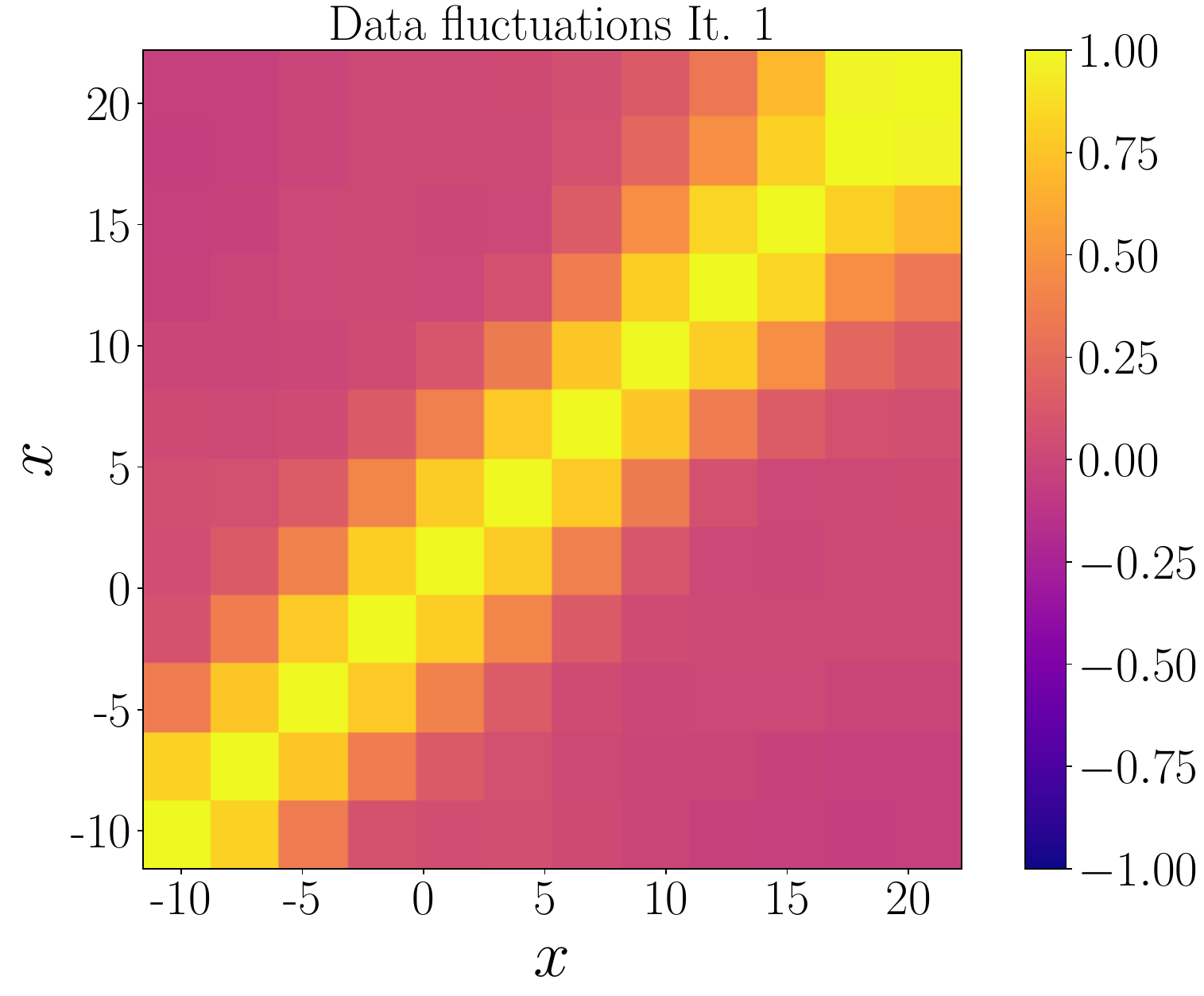}\\
    \vspace{0.3cm}
 \includegraphics[width=0.47\textwidth]{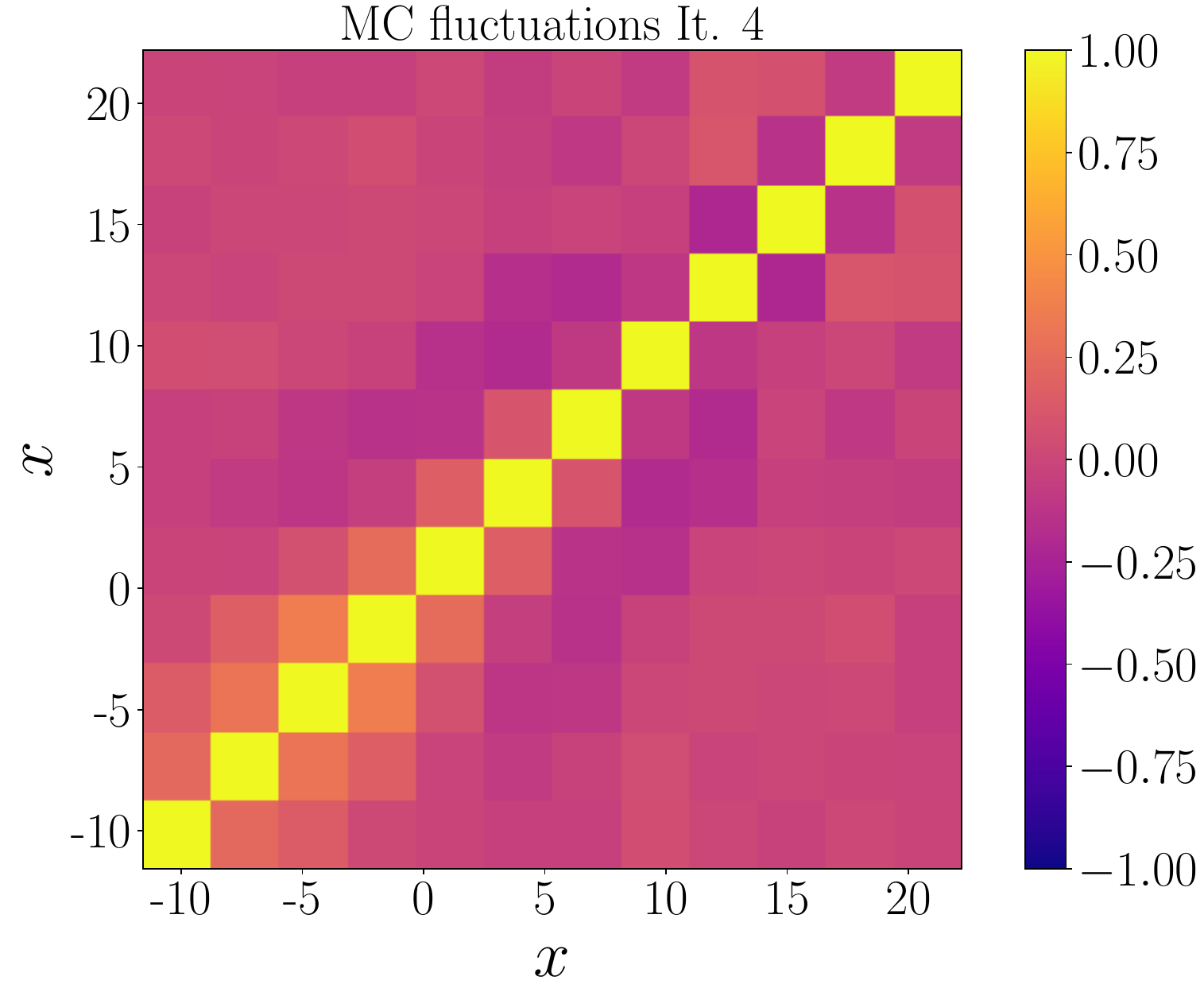}
    \hspace{0.1cm}
    \includegraphics[width=0.47\textwidth]{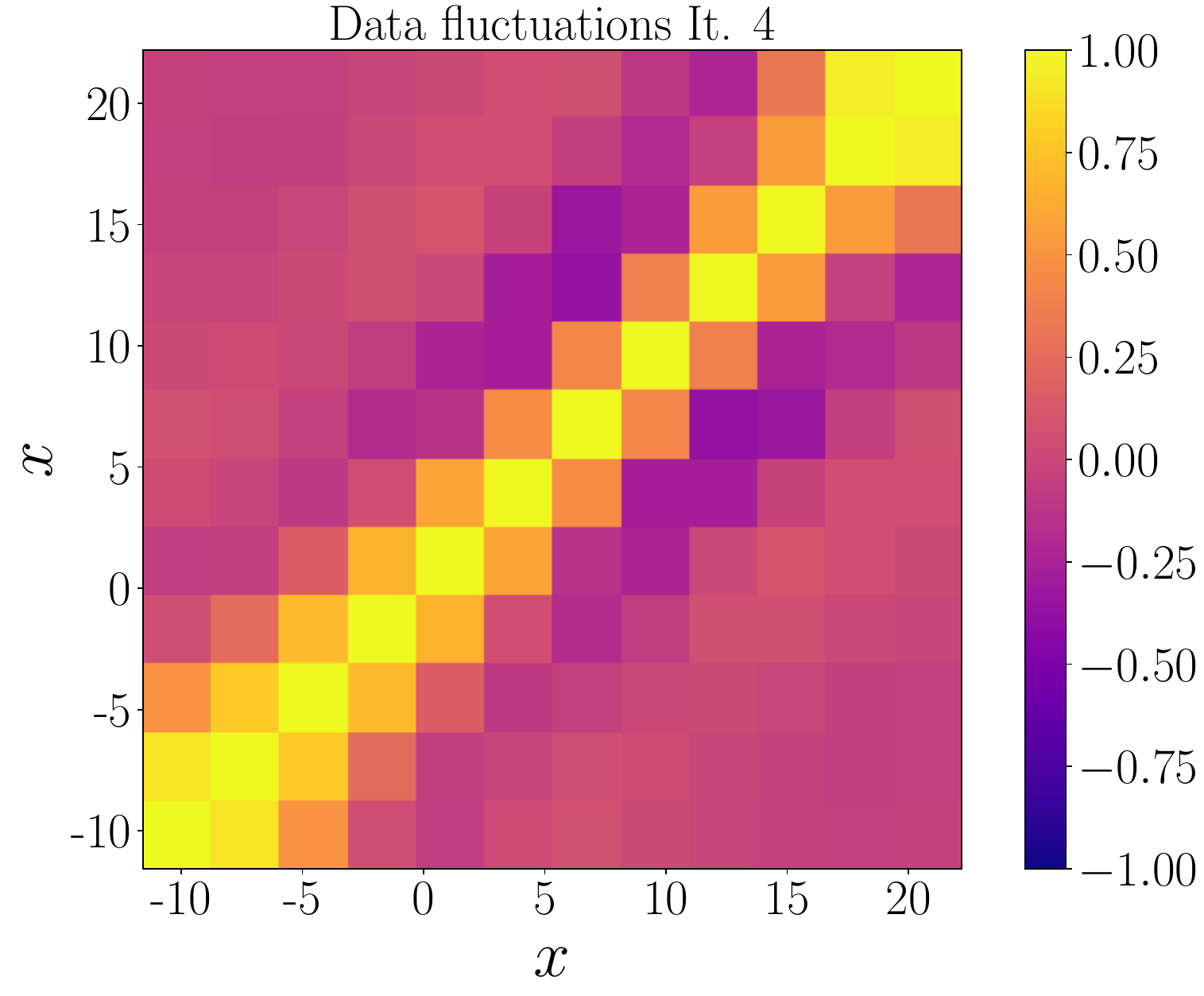}\\
    \vspace{0.3cm}
    \includegraphics[width=0.47\textwidth]{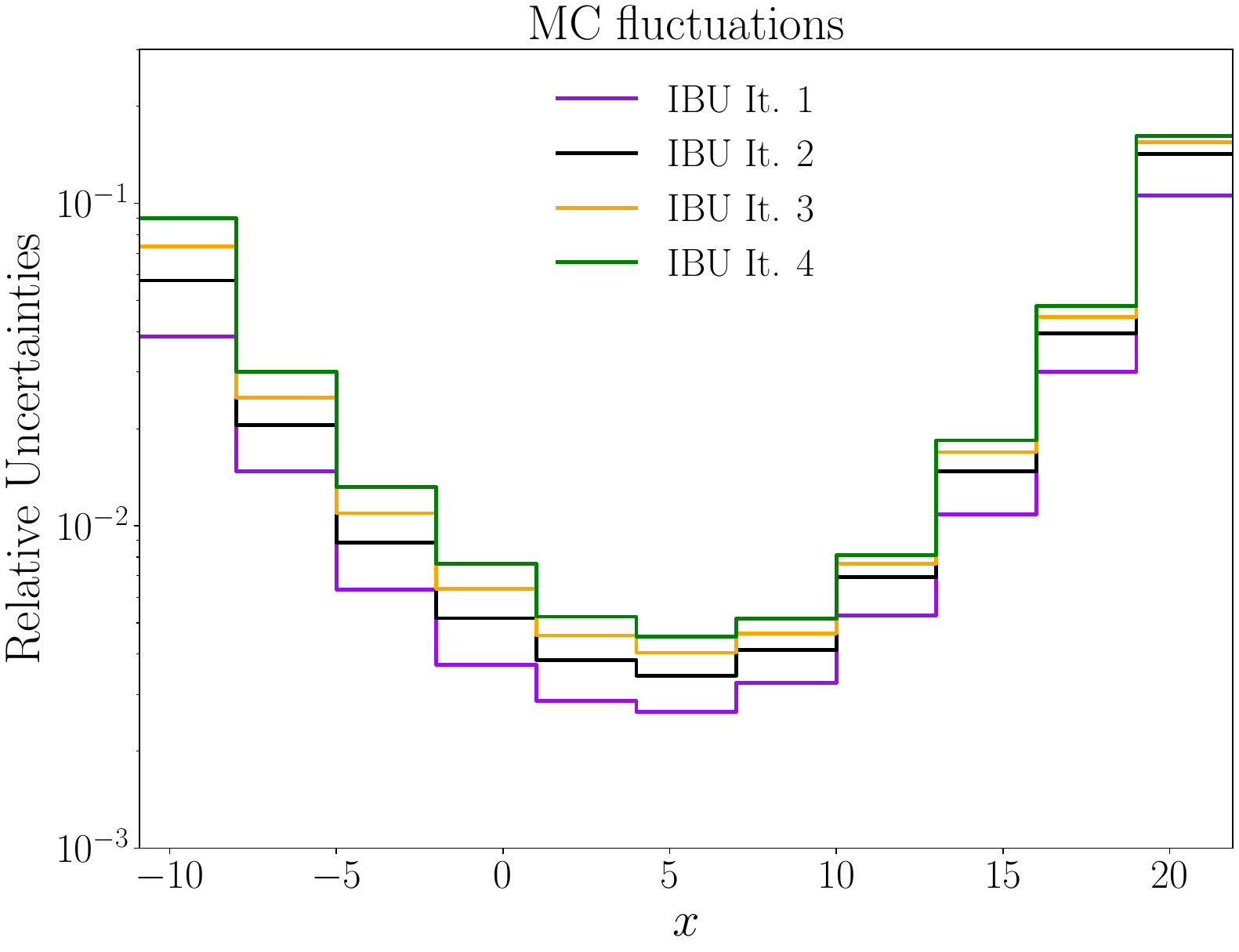}
    \hspace{0.1cm}
    \includegraphics[width=0.47\textwidth]{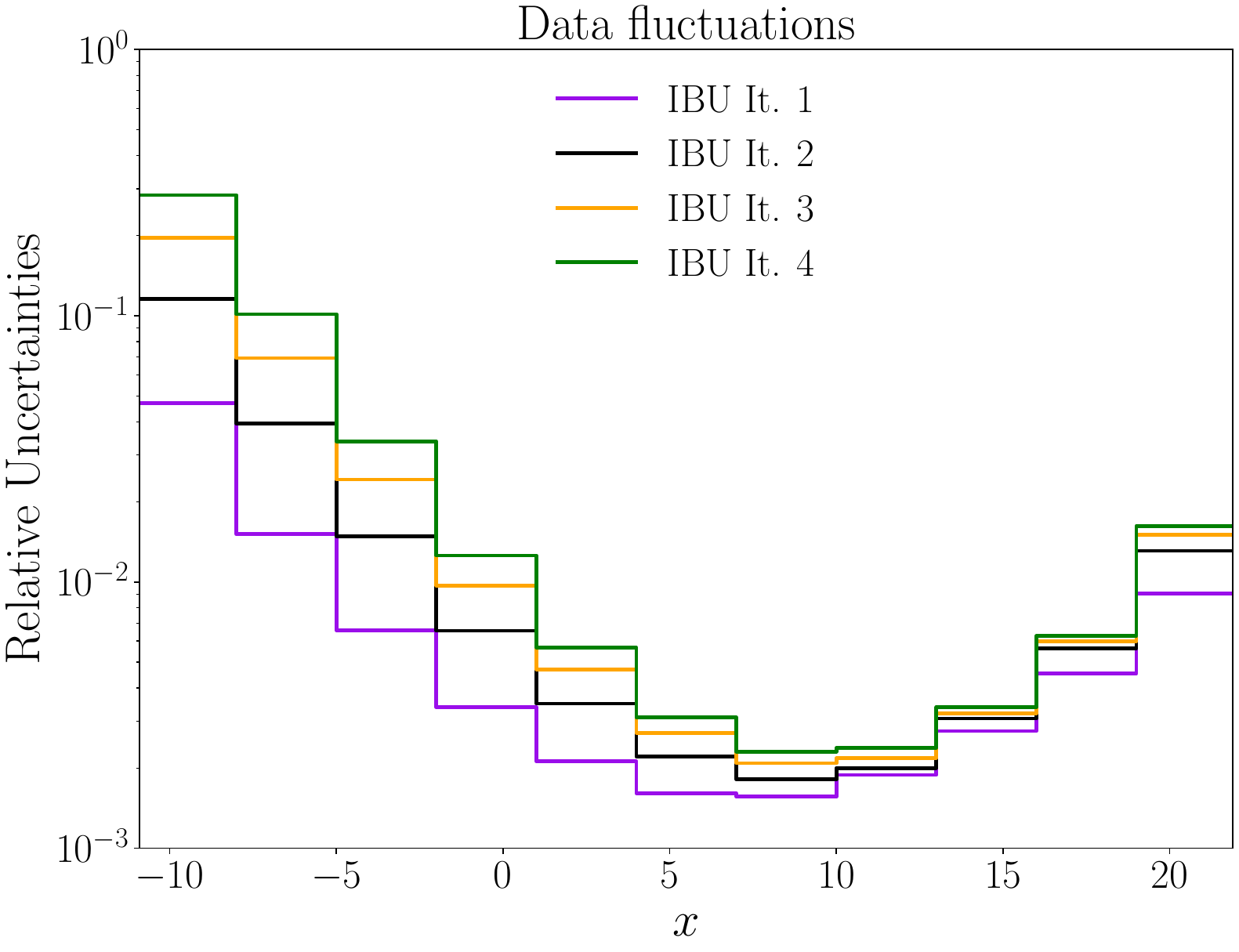}
    \caption{Correlations and amplitude of the statistical uncertainties for the full unfolded distribution, obtained for the analytic example using IBU.
    On the left hand side fluctuations of the response matrix are implemented, on the right fluctuations of the detector-level data.
    The top row shows the correlation matrices after one iteration, the middle row after four iterations.
    The bottom row shows the evolution of the relative uncertainties over several iterations.}
    \label{f:single_covariances_full}
\end{figure*}

Figure~\ref{f:single_covariances_full} shows the correlations of the full unfolded distribution, after one and after four iterations, for the statistical uncertainties from Poissonian fluctuations of the detector response and for Poissonian fluctuations of the detector-level data.~\footnote{Such separate propagation of the statistical uncertainties, originating from the finite amount of either the Monte Carlo simulations or of the data to be unfolded, is often implemented in experimental analyses~(see e.g. Refs.~\cite{ATLAS:2014riz,ATLAS:2019hxz}).} 
In addition, the evolution of the relative statistical uncertainties over several iterations is shown.
The number of events in the simulation is set to $500\,000$ for both the pseudo-data and the Monte Carlo transfer-matrix. The number of fluctuated pseudo-experiments is $N_{f}=1000$.

The plots show a somewhat different behavior for the two kinds of fluctuations.
The fluctuations of the response matrix, i.e.\ the fluctuation of the weights of Monte Carlo simulation events, lead to a correlation matrix which is nearly diagonal after one iteration, showing that the bins are only weakly correlated.
Still, the anti-correlations, visible already at this step, are a consequence of the preservation of the number of events in the unfolding process.
The (anti-)correlations increase somewhat with the larger correction of the smearing effects, achieved using four iterations.

The fluctuation of the detector-level data component introduces important correlations with neighboring bins, already at the first iteration.
This is due to the smearing of the detector response, which was determined in Section \ref{s:analytic} by $\sigma_s = 3$.
A fluctuation of a detector-level bin is smeared out to neighboring bins by the unfolding matrix, in a correlated way as well. 
This introduces strong statistical correlations to neighboring bins. 

At this point it is interesting to look at the differences between the correlation matrices obtained with detector-level data fluctuations, using one and respectively four iterations: while there are anti-correlations (i.e.\ negative entries) after four iterations, the first iteration leads to purely positive correlations, which has been verified numerically.
This can be easily understood recalling the fact that the first pseudo-inversion of the response matrix is unaffected by the detector-level data fluctuations.
If a bin is fluctuated up~(down) statistically, this will induce upward~(downward) fluctuations coherently in all the bins of the unfolded distribution. 
During the second iteration, the fluctuations of the detector-level data also impact the pseudo-inverted response matrix.
This introduces anti-correlations, because the total number of data events is preserved by the unfolding.
In consequence the pseudo-inverted response matrix needs to be normalized along the truth-level axis, in every reconstructed-level bin~(i.e.\ in every column of the matrix).
This normalization causes every upward change of a bin to be connected to a downward change of the other bins of the column, hence the resulting anti-correlation.
This behavior is clearly visible in the correlation matrices in Figure~\ref{f:single_covariances_full}.

The amplitude of the statistical uncertainties can be derived from the square-root of the diagonal elements of the covariance matrix.
As can be seen in the bottom panels of Figure~\ref{f:single_covariances_full}, the expected behavior is apparent: the higher the number of iterations, the larger the uncertainty.
As explained previously, the balance between bias~(see Figure~\ref{f:unfolding_applied_to_toy}) and statistical uncertainties is an important criterion when choosing the number of iterations.

\begin{figure*}[p!]
    \centering
    \includegraphics[width=0.35\textwidth]{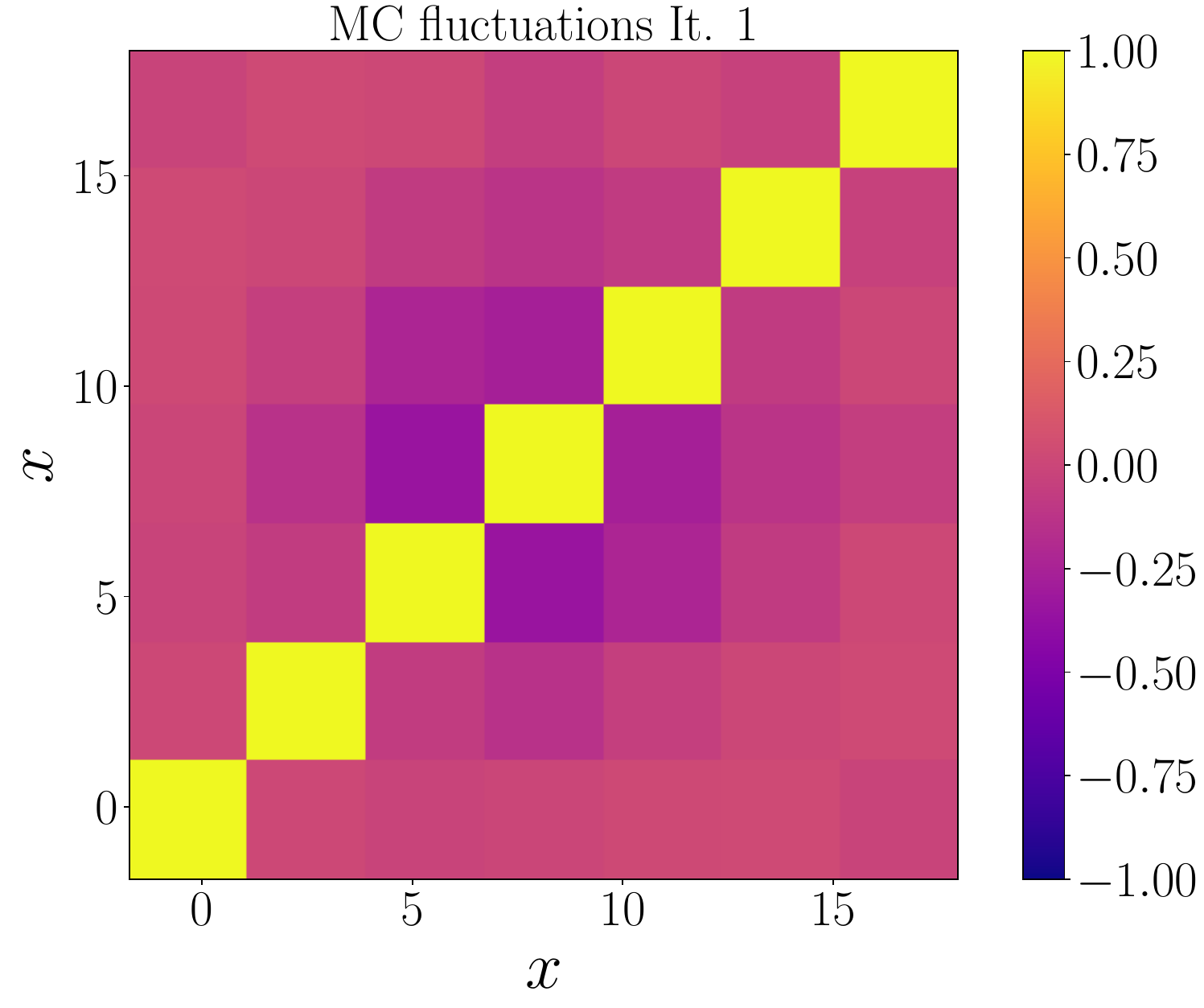}
    \hspace{0.1cm}
    \includegraphics[width=0.35\textwidth]{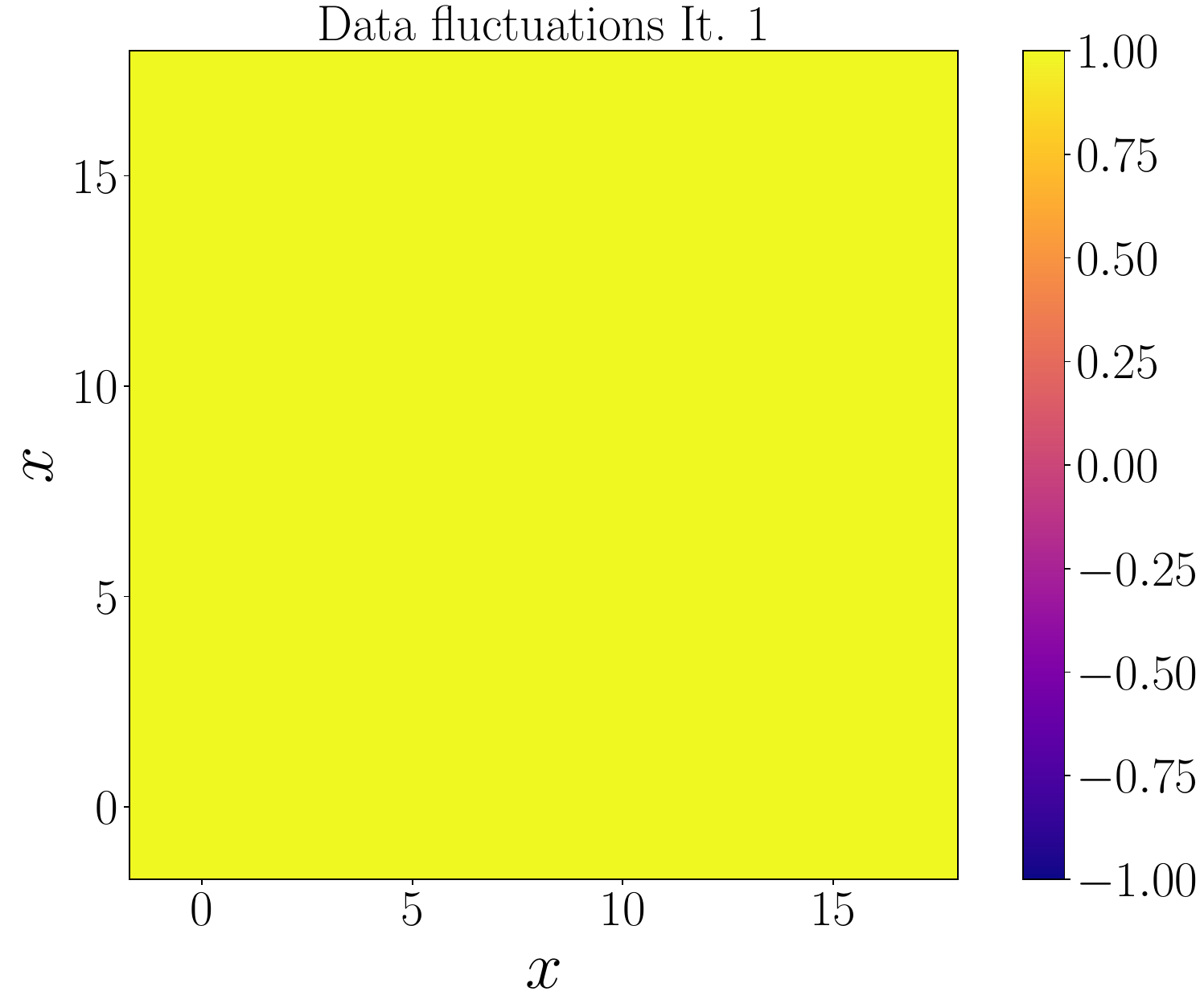}\\
    \vspace{0.3cm}
    \includegraphics[width=0.35\textwidth]{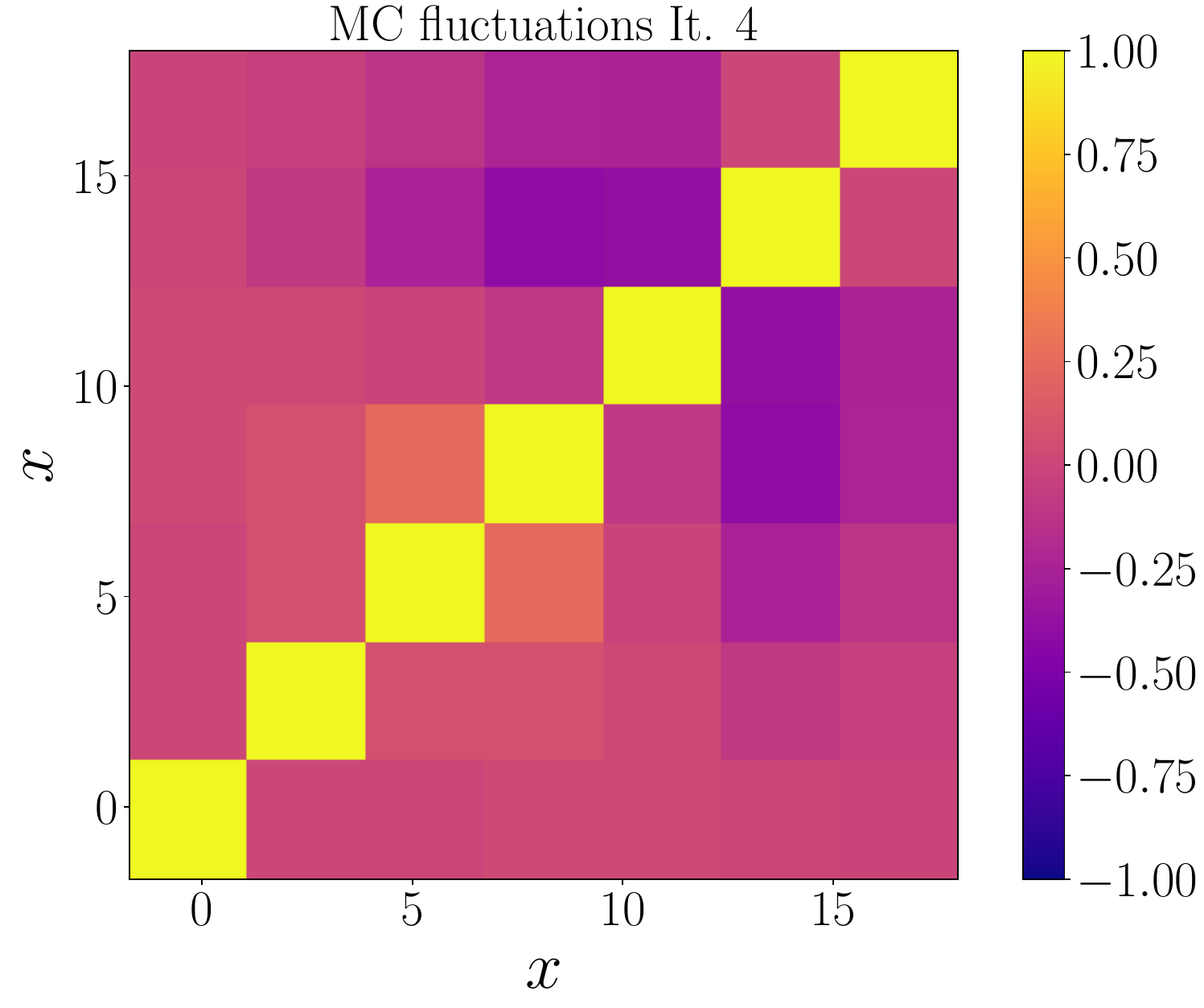}
    \hspace{0.1cm}
    \includegraphics[width=0.35\textwidth]{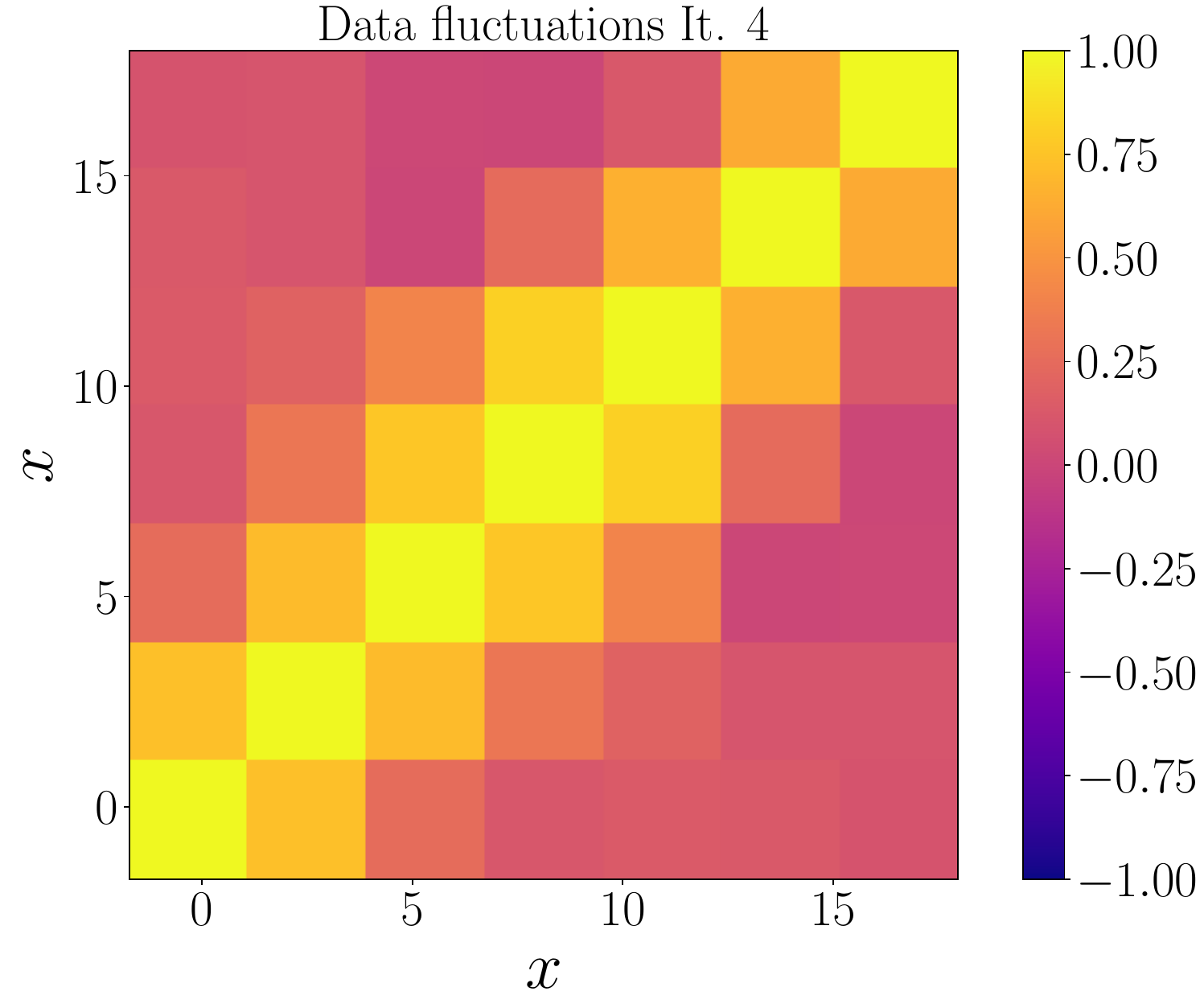}\\
    \vspace{0.3cm}
    \includegraphics[width=0.39\textwidth]{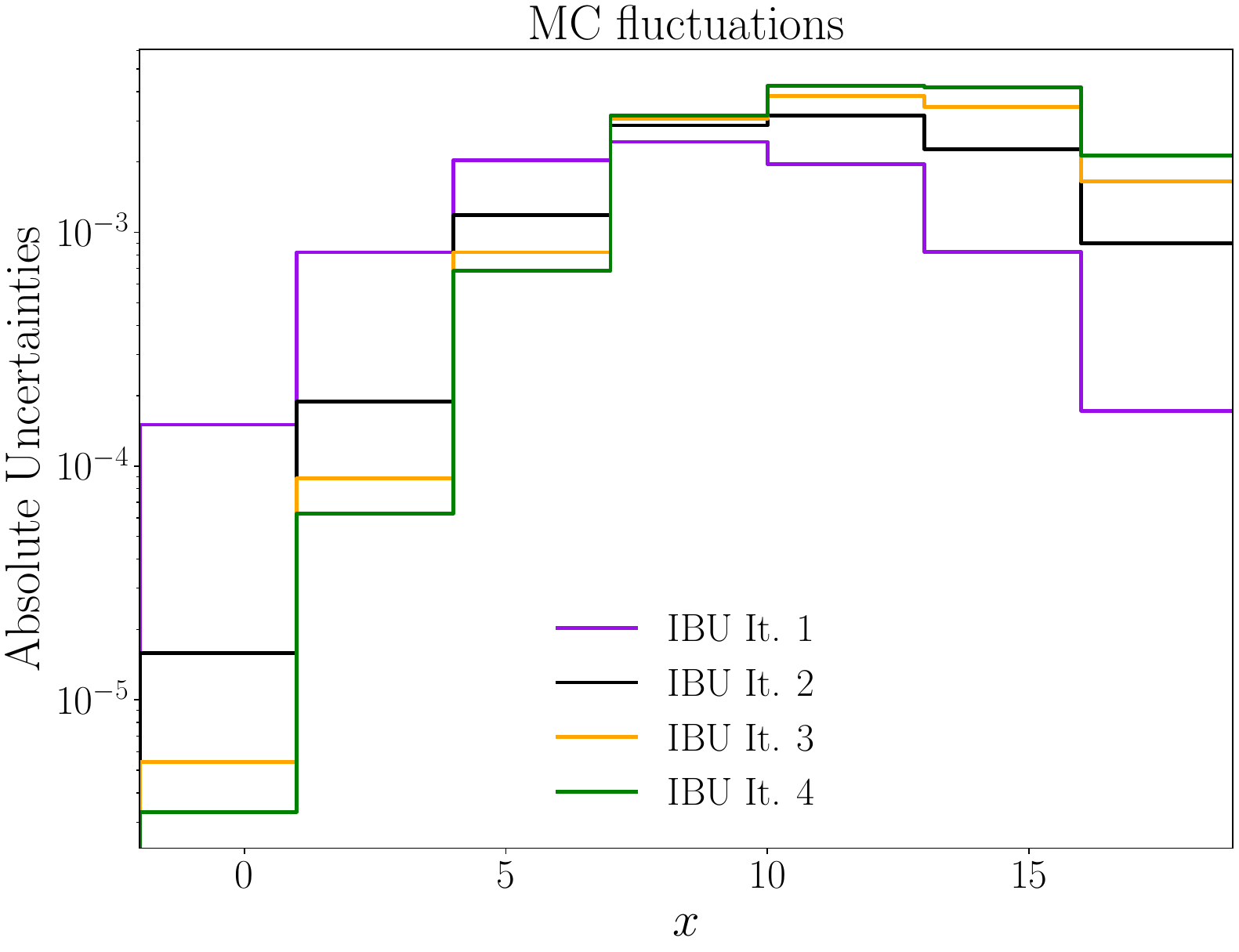}
    \hspace{0.1cm}
    \includegraphics[width=0.39\textwidth]{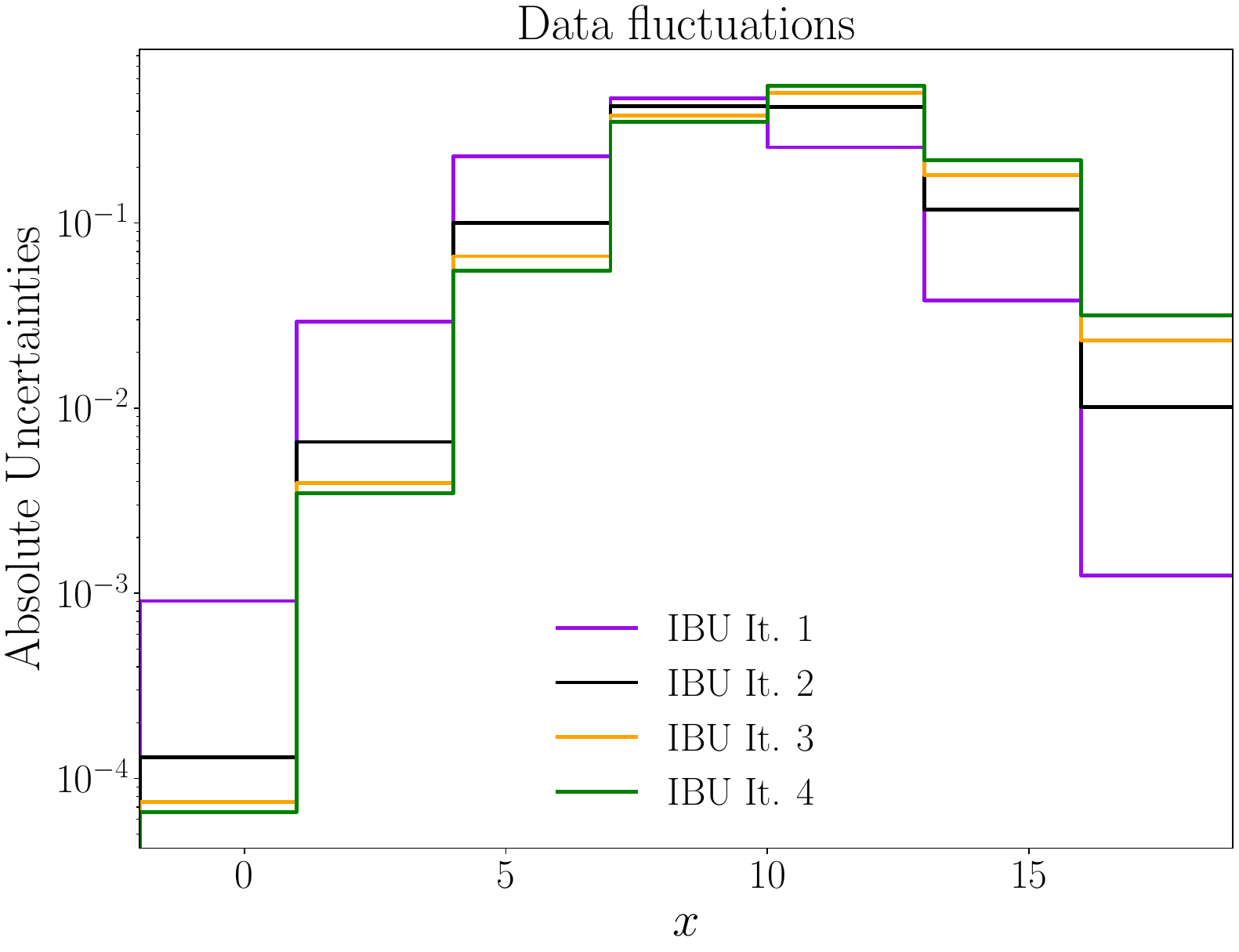}\\
    \vspace{0.3cm}
    \includegraphics[width=0.35\textwidth]{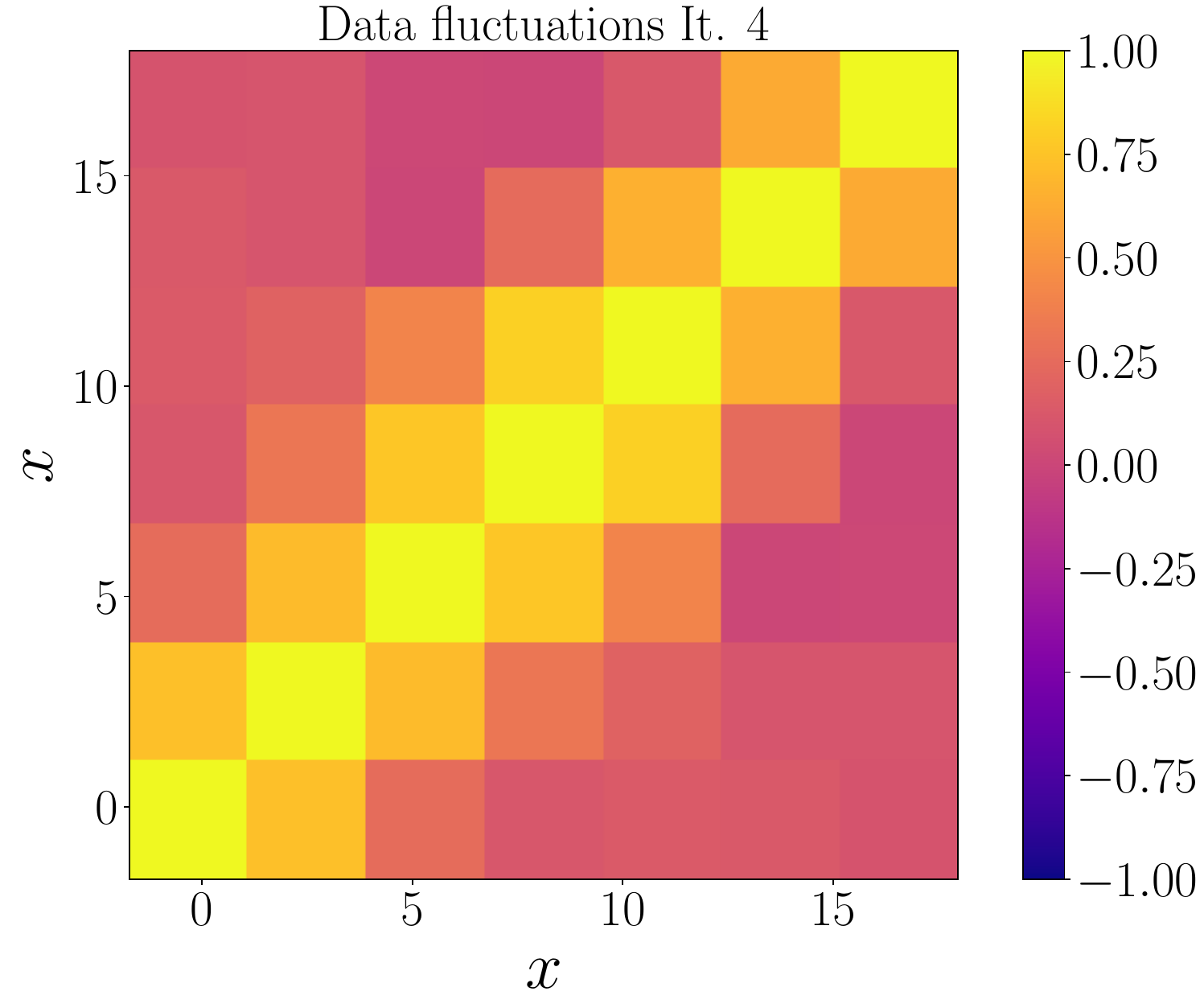}
    \hspace{0.1cm}
    \includegraphics[width=0.39\textwidth]{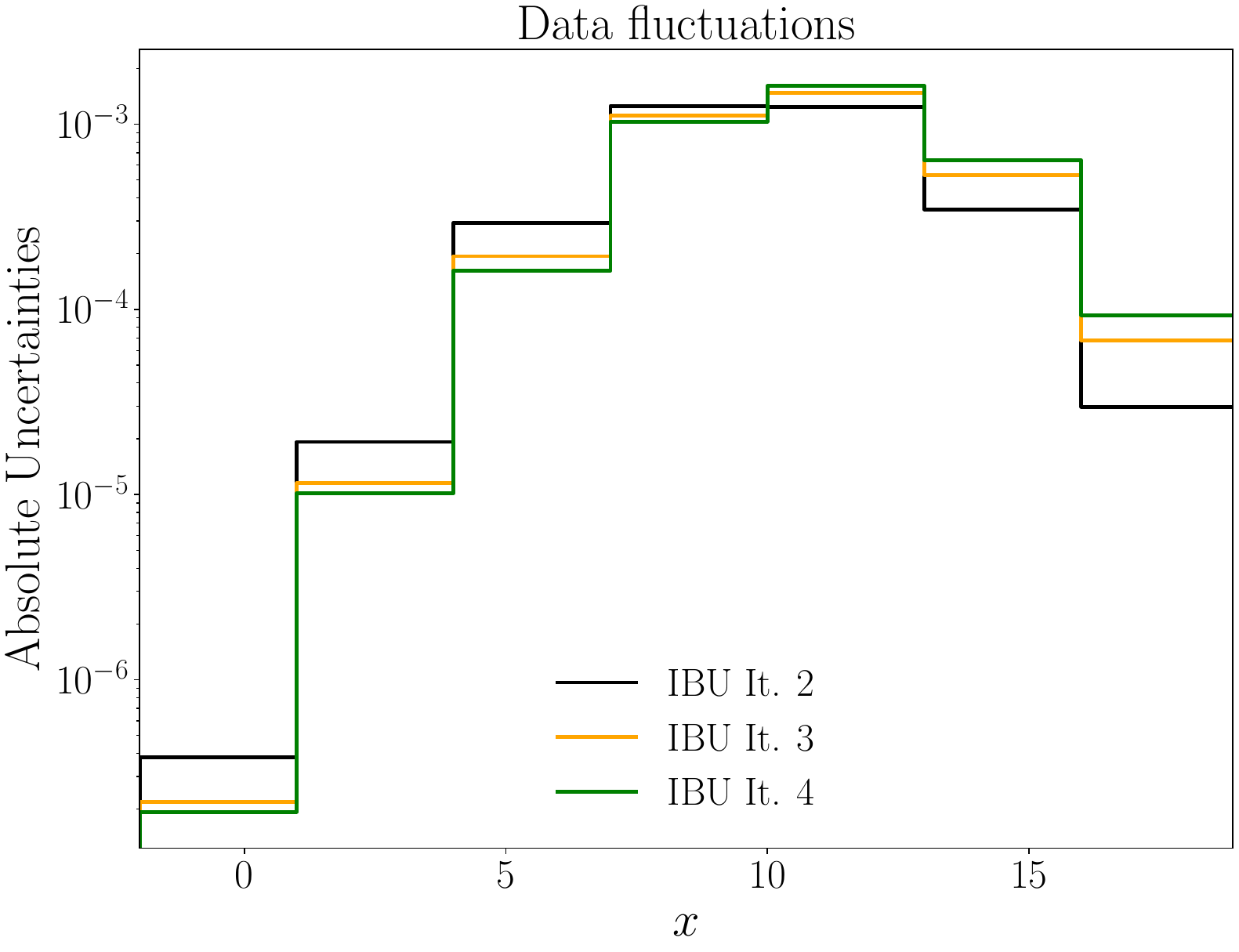}
    \caption{Correlations and amplitude of the statistical uncertainties for the unfolded distribution of a single event~(i.e. with $i_1 \equiv i_2$ in the relevant equations mentioned below). These are obtained for the analytic example using IBU~(see main text).
    The single event is set at $r_m=5$.
    For the first three left-hand plots fluctuations of the MC-based response matrix are implemented (according to $\mathrm{cov}^{\mathrm{MC}}$ in Eq.~\eqref{eq:single_covShape}), while the first three right-hand ones correspond to fluctuations of the detector-level data~(i.e. both the event that is unfolded and the rest of the data sample, with the normalized covariance matrix $\widetilde{\mathrm{cov}}^{\mathrm{Data}}$ evaluated according to  Eq.~\eqref{eq:single_cov_Data}).
    The first row shows the correlation matrices after one iteration, the second row after four iterations.
    The third row shows the evolution of the absolute uncertainties over several iterations.
    The fourth row shows the correlation matrix and absolute uncertainties for data fluctuations, with the covariances evaluated for the shape fluctuations of the unfolded distribution with four iterations, according to $\mathrm{cov}^{\mathrm{Data}}$ in Eq.~\eqref{eq:single_covShape}. The covariance after one iteration is in this latter case exactly zero~(as expected) and hence not displayed.
    }
\label{f:single_covariances_single}
\end{figure*}

Now the statistical covariance matrices for unfolded single events can also be derived.
The results are shown in Figure~\ref{f:single_covariances_single}, displayed again as a combination of correlation matrices and amplitudes of the uncertainties.
The covariance matrix of the full distribution is numerically equal to the sum of covariance matrices obtained for every pair of events (including the self-covariances).
The fluctuations of the response matrix induce features related to the ones observed above for the full distribution.
Indeed, there are significant (anti-)correlations among the bins in the region covered by the unfolded distribution, which is also shifted with the increasing number of iterations. 

At the first iteration, the fluctuations of the detector-level data event that is unfolded (and of the rest of the data sample), yield a correlation matrix containing only values which are exactly one~(based on the normalized covariance matrix $\widetilde{\mathrm{cov}}^{\mathrm{Data}}$ evaluated according to  Eq.~\eqref{eq:single_cov_Data}).
This is due to the fact that, at the first iteration, the unfolding matrix is not impacted by the fluctuation of the detector-level data.
Unfolding a single fluctuated bin will therefore yield the same output as the original bin, multiplied by a global factor~(i.e.\ a distribution with the same shape, but a different normalization).
Since this fluctuating factor is the same for all the bins of the single-event unfolded distribution, the bins are fully correlated.
For the same reason, in this case the covariance matrix associated to the shape of the single-event unfolded distribution is exactly zero.

With further iterations the correlations are reduced since the fluctuations of all the data events~(i.e. not just the one being unfolded) also impact the pseudo-inverted response matrix, as discussed above.
Furthermore, the correlation coefficients for the full unfolded distribution are typically smaller than the ones for the single-event correlations, when fluctuating simultaneously all the detector-level data.
This is due to the fact that the full covariance matrix additionally contains contributions from pairs of events that are not in the same bin, which dilute the correlations.
It was checked that closure between the per event- and full distribution-fluctuations is achieved at each iteration step.
Strong (anti-)correlations are observed for the shape of the single-event unfolded distribution, for which the integral is constraint to unity.

The absolute uncertainties of the single-event distribution~\footnote{Note that the statistical uncertainty originating from data fluctuations is scaled, following Eq.~\eqref{eq:single_cov_Data}, to account for the size of the sample of events available in the same reconstructed-level bin as the event that is unfolded, since all such events yield the same unfolded distribution.} are visualised in the third row of Figure~\ref{f:single_covariances_single}.
Their amplitudes increase somewhat with the increasing number of iterations and their distributions are progressively shifted towards the right.
This is due to the design of the toy model and of the unfolding procedure, where each iteration introduces such a shift of the unfolded distribution.
It can clearly be seen, that each event from the reconstructed-level data impacts multiple bins at the unfolded-level.
In addition, to reconstruct the uncertainty of the full distribution, the covariances of pairs of events from different bins need to be added, which further amplifies the enhancement of the diagonal uncertainties with the number of iterations.
The uncertainties associated to the shape of the single-event unfolded distribution (according to $\mathrm{cov}^{\mathrm{Data}}$ in Eq.~\eqref{eq:single_covShape}, displayed on the forth row of Figure~\ref{f:single_covariances_single}) are small compared to the ones from the covariance matrix $\widetilde{\mathrm{cov}}^{\mathrm{Data}}$, the latter being evaluated employing the normalization implemented in Eq.~\eqref{eq:single_cov_Data}.
The correlation matrices associated to data fluctuations, derived based on either Eq.~\eqref{eq:single_covShape} or Eq.~\eqref{eq:single_cov_Data}, are similar in this example due to the high statistics in the data distribution.

\subsection{Unfolding a Folded Truth Event}
\label{s:unfolding_folded_IcINN}

Instead of unfolding single events on detector-level, it is also interesting to smear a single truth-level event and unfold the resulting distribution.
Indeed, doing so enables a more thorough understanding of the properties of the event-by-event unfolding approaches.
For this exercise we consider one event out of the full truth distribution from Section~\ref{s:analytic}.
\begin{figure*}[t]
	\centering
	\includegraphics[width=0.45\linewidth]{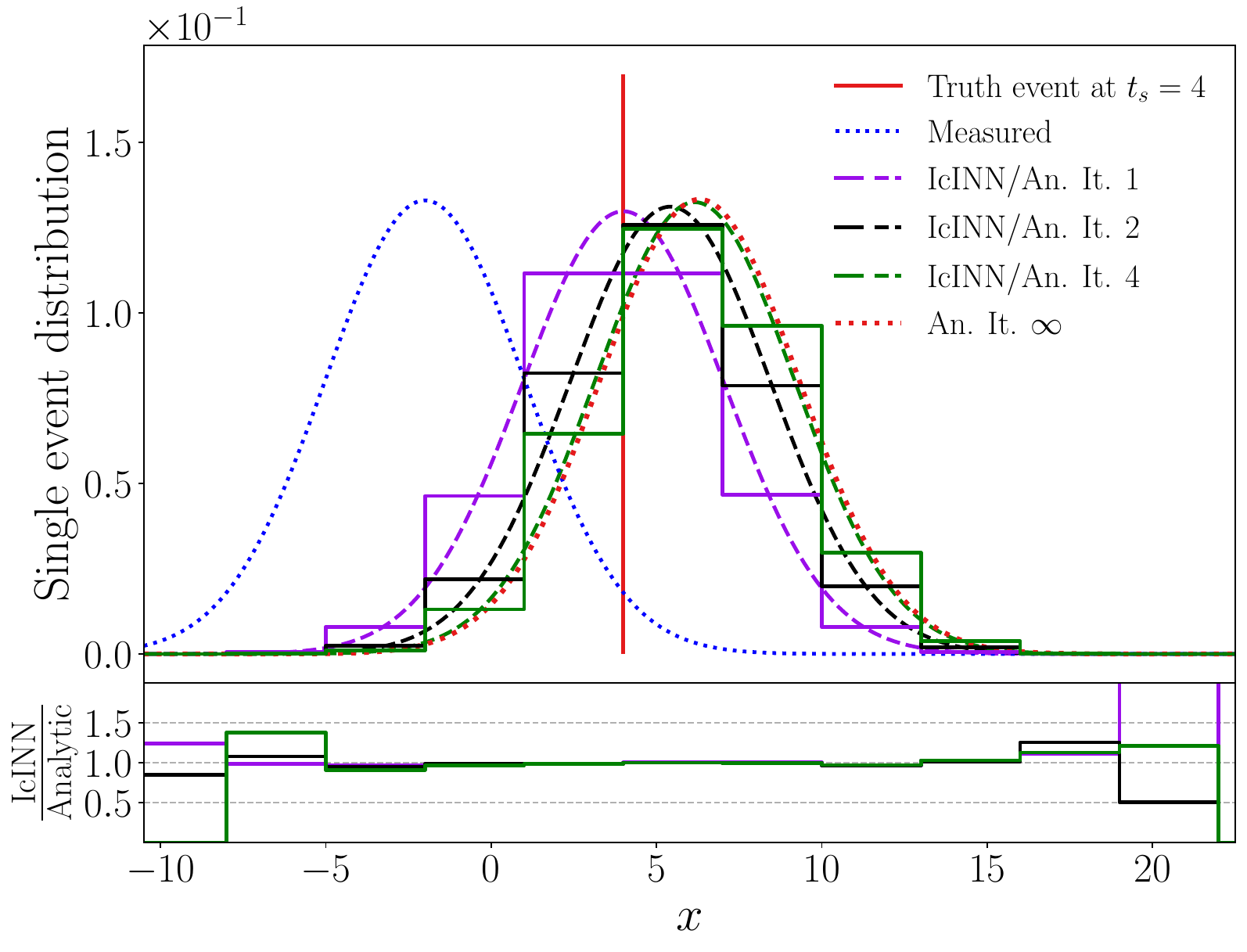}
    \hspace{0.5cm}
    \includegraphics[width=0.45\linewidth]{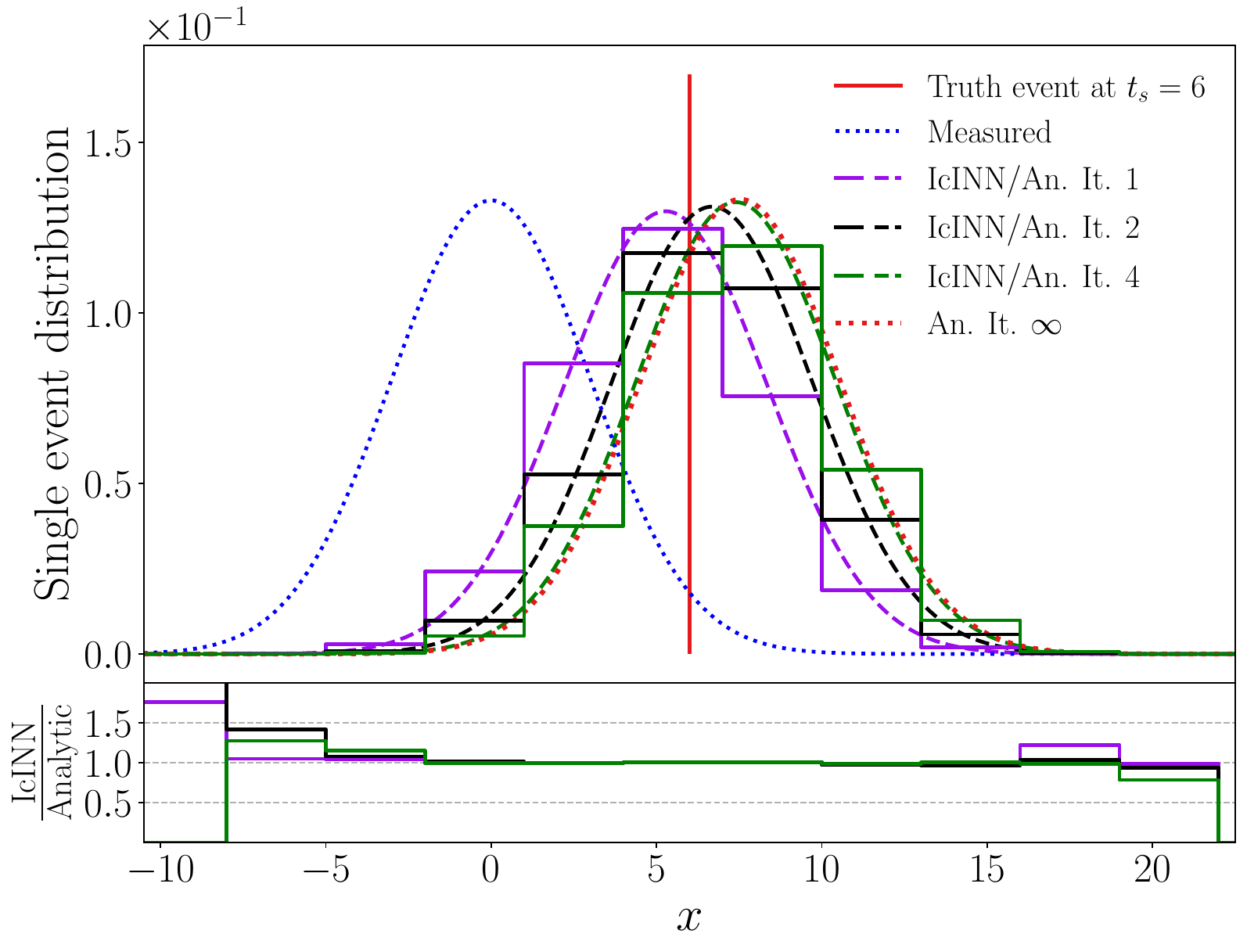}\\
    \includegraphics[width=0.45\linewidth]{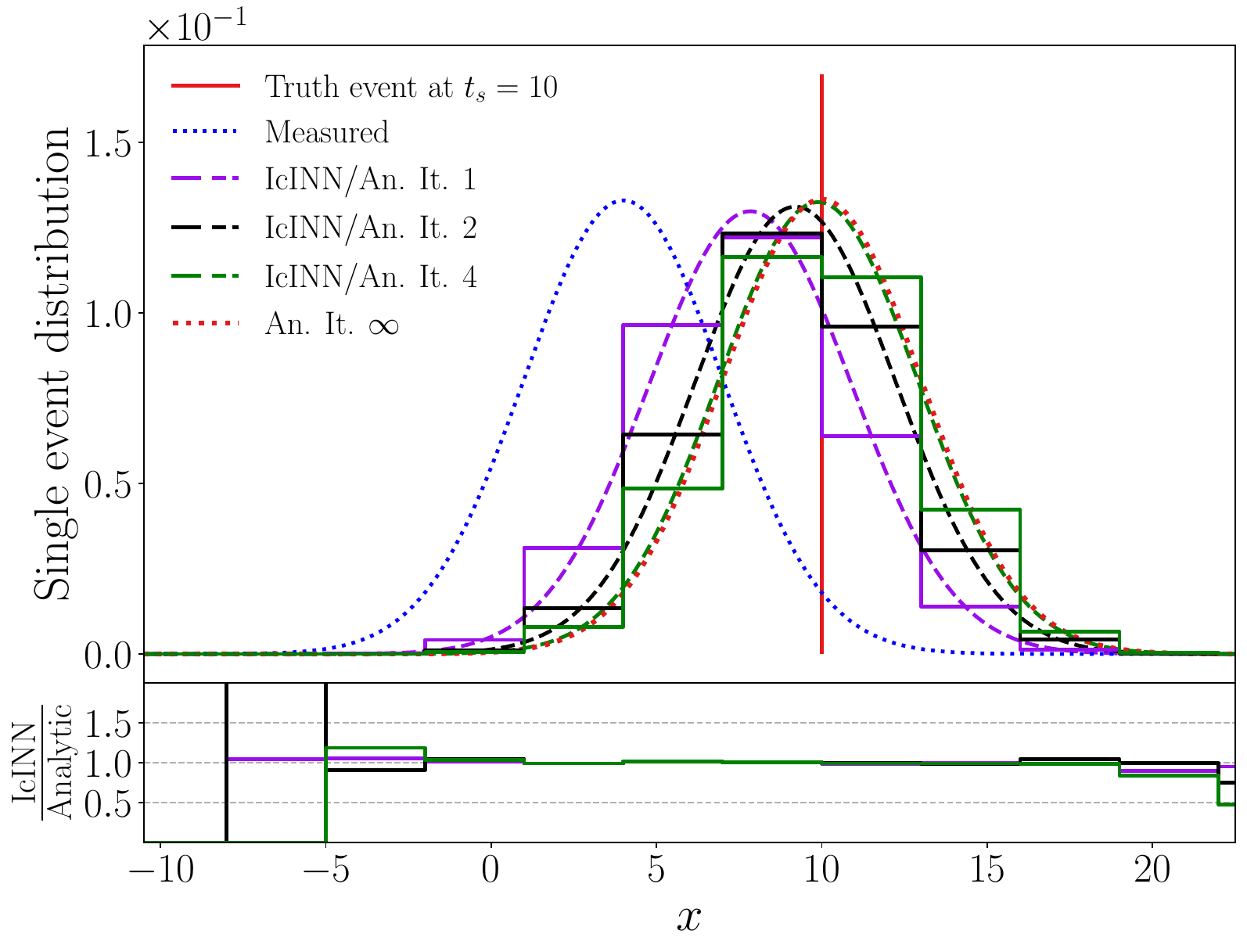}
    \hspace{0.5cm}
    \includegraphics[width=0.45\linewidth]{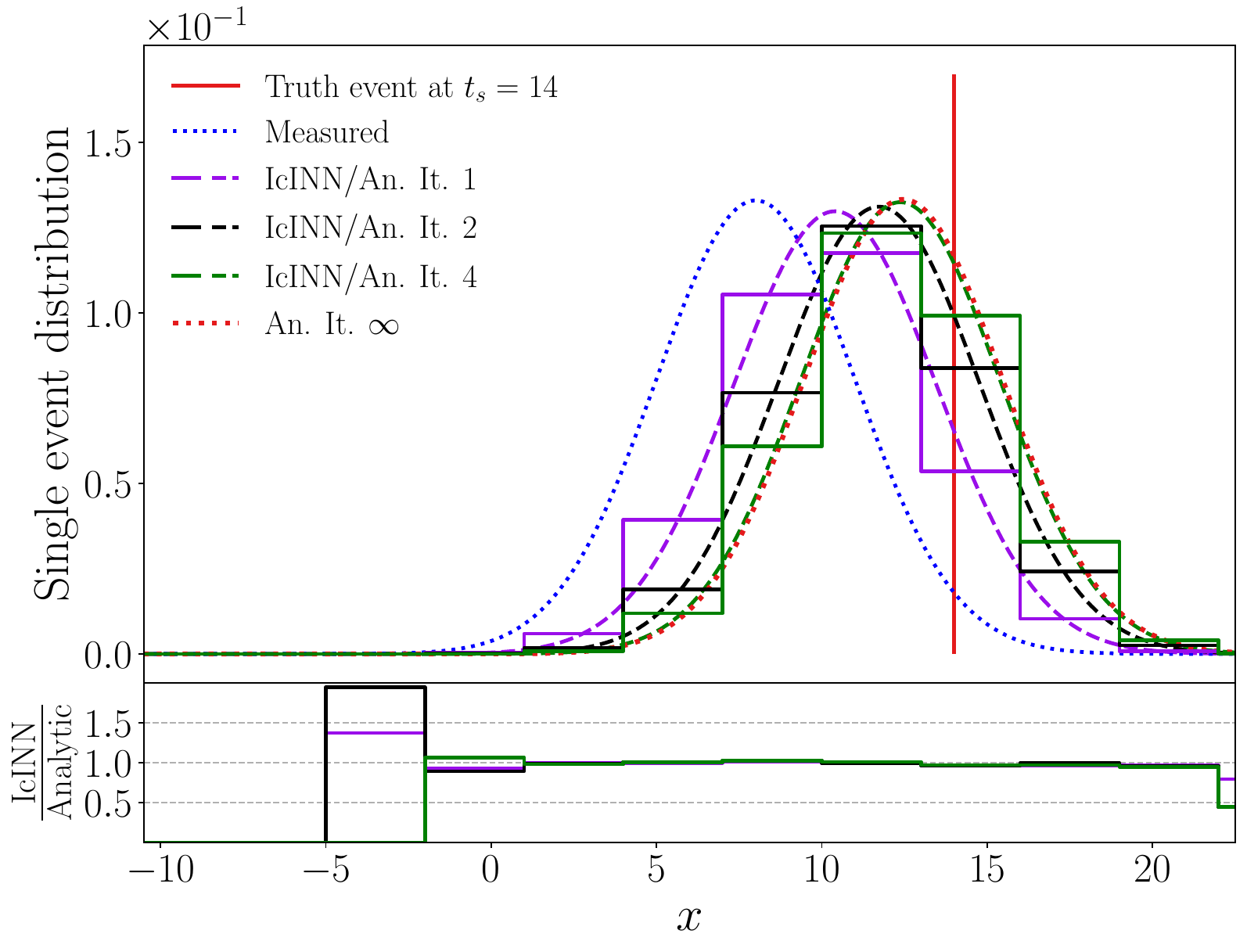}
	\caption{Examples of truth-level single events~(solid red) on which a detector smearing is applied to obtain their detector-level distributions~(dotted blue).
     These distributions are afterwards unfolded with the IcINN algorithm.
     The unfolding results after each iteration~(purple, black, green; solid) match their respective analytic predictions~(dashed lines).
     The analytic predictions for an infinite number of iterations are also shown~(dotted red).
     The truth-level single events are located at $t_s=4$ (top left), $t_s=6$ (top right), $t_s=10$ (bottom left) and $t_s=14$ (bottom right).}
     \label{f:truth_single_unfolded}
\end{figure*}
From an analytic point of view, the detector-level distribution can be predicted by applying the Gaussian smearing to a delta distribution of the single event $t_s$, which leads to a Gaussian distribution with $\mu_{\mathrm{single},m}=t_s+\mu_{\mathrm{smear}}$ and $\sigma_{\mathrm{single},m}=\sigma_{\mathrm{smear}}$.
This distribution can now be unfolded as in Section~\ref{s:analytic} to obtain as an unfolded distribution a Gaussian with parameters
\begin{align}
    \mu_{\mathrm{single,t}}&=\frac{t_s \, \sigma^2_{\mathrm{MC},t} + \mu_{\mathrm{MC},t}\sigma^2_{\mathrm{smear}}}{\sigma^2_{\mathrm{MC},t}+\sigma^2_{\mathrm{smear}}}, \\\
    \sigma_{\mathrm{single,t}}^2 &= \frac{2 \sigma^2_{\mathrm{smear}} \sigma^4_{\mathrm{MC},t} + \sigma^2_{\mathrm{MC},t} \sigma^4_{\mathrm{smear}}}{ (\sigma^2_{\mathrm{smear}} +\sigma^2_{\mathrm{MC},t})^2}.
\end{align}
To calculate the parameters of this distribution after any IcINN iteration $i$, the parameters $(\mu_{\mathrm{MC},t}, \sigma_{\mathrm{MC},t})$ have to be replaced by $(\mu_{u,i}, \sigma_{u,i})$ (in analogy to Section~\ref{s:analytic}).
In addition it is also possible to derive an analytic result that is equal to an "infinite" number of iterations by replacing the parameters $(\mu_{\mathrm{MC},t}, \sigma_{\mathrm{MC},t})$ with $(\mu_{\mathrm{Data},t}, \sigma_{\mathrm{Data},t})$, i.e. the parameters of the truth-level pseudo-data distribution from Section~\ref{s:analytic}.
This is due to the fact that $(\mu_{u,i}, \sigma_{u,i})$ converge for an infinite number of iterations to $(\mu_{\mathrm{Data},t}, \sigma_{\mathrm{Data},t})$.

The results for IcINN unfolding are shown in Figure~\ref{f:truth_single_unfolded}.~\footnote{IBU and IDS again produce similar results, as shown in~\ref{App:Unfolding_folded_matrix}.}
For this purpose the single truth-level event was smeared 50000 times and unfolded 10 times each.
It is clearly visible that not for all values of $t_s$ the mean values of the unfolded distributions converge towards the delta distribution of the single event $t_s$.
More specifically, for $t_s=4$ and $t_s=6$ the mean values converge numerically as well as analytically towards a higher value than $t_s$.
On the other hand, for $t_s=14$ the mean values converge towards a lower value.
Only for $t_s=10$ the mean values of the unfolded distribution seem to converge towards the delta distribution on truth level.
This behaviour is due to the fact, that the overall truth-level distribution (which is basically obtained with the IcINN) is located at $\mu_{\mathrm{Truth}}=10$.
The connection between the single-event unfolded distributions and the truth-level single-event $t_s$ is actually diluted by the smearing process.
The single-event unfolded distribution are therefore influenced by the overall unfolding result and are not entirely determined by the position of $t_s$.
Such influence, present for the single-event unfolding, vanishes when one sums the contributions from the full sample of events to get the total unfolded spectrum.
These observations indicate that for the event-by-event unfolding there can be subtleties~(i.e. extra potential sources of bias) beyond the ones present when unfolding the full distribution.
This calls for a very careful evaluation of the systematic uncertainties for an event-level unfolding result, a process that is typically dependent on the analysis scope.

\begin{figure*}[t]
    \centering
    \includegraphics[width=0.47\textwidth]{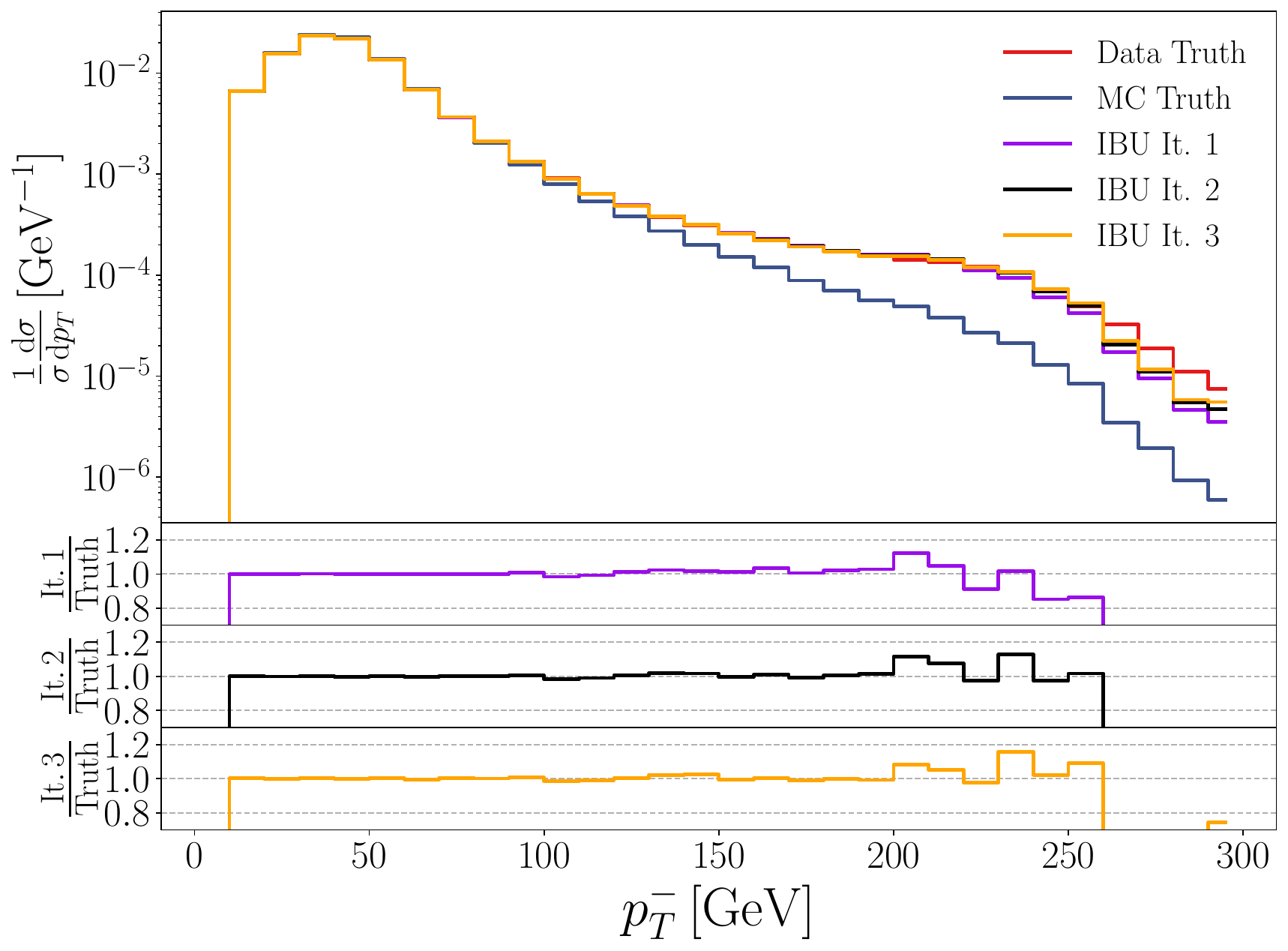}
    \includegraphics[width=0.47\textwidth]{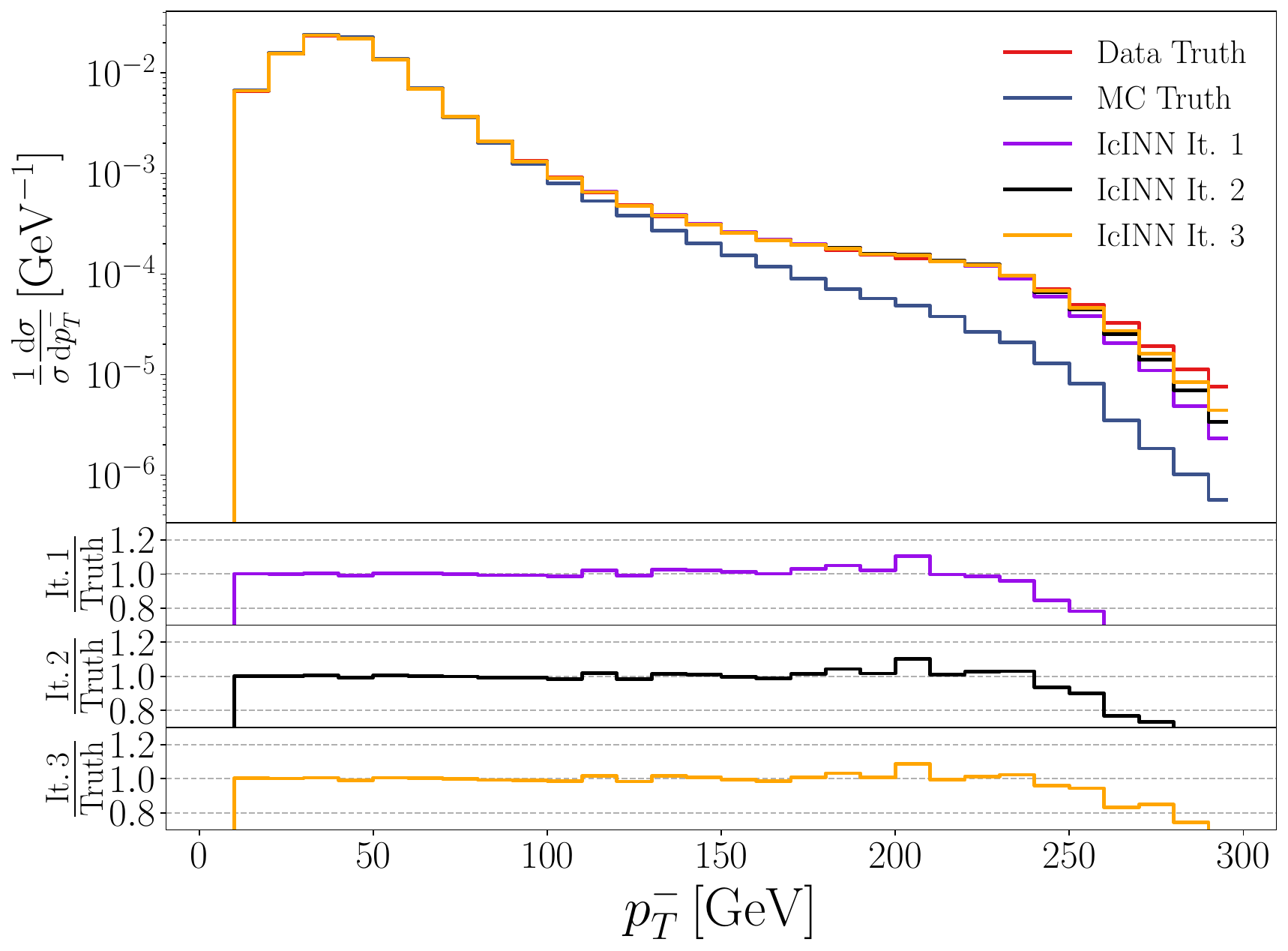}\\
    \vspace{0.5cm}
    \includegraphics[width=0.47\textwidth]{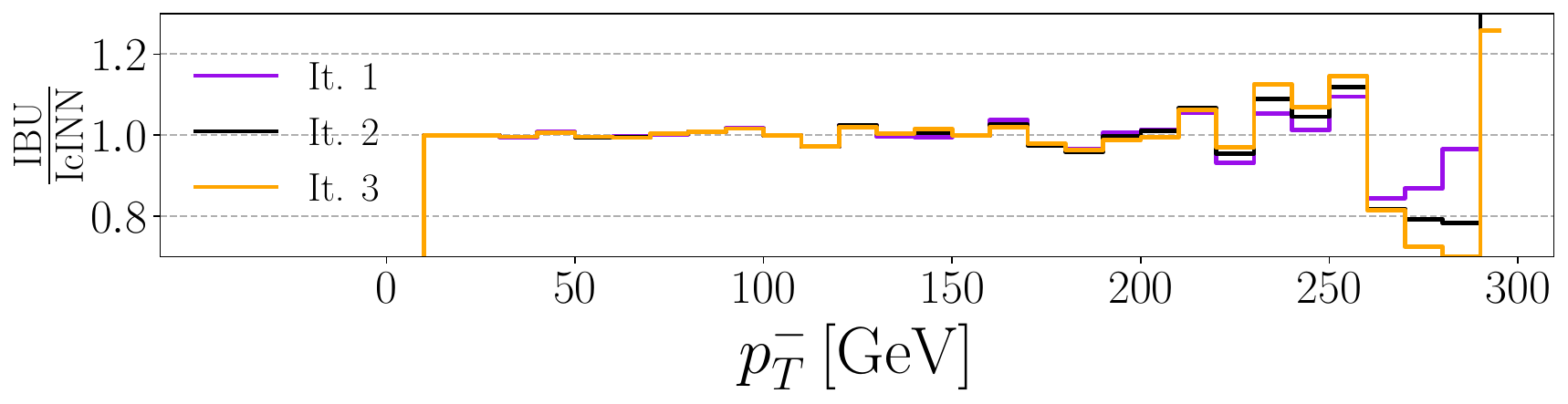}\\
    \caption{Unfolding result for Iterative Bayesian Unfolding~(top left) and IcINN unfolding~(top right, reproduced from Ref.~\cite{Backes:2022vmn}) for the transverse momentum of the muon $p_T^-$.
    Most of the distribution is unfolded correctly after a single iteration~(purple).
    The high-$p_T$ bins show a similar behavior as for the IcINN unfolding, i.e.\ they require more than one iteration~(black, orange). In addition, the figure shows the truth-level data (red) and the truth-level Monte Carlo (black).
    The panels below each of the top plots show the ratio between the unfolded distribution at each iteration step and the truth-level data.
    In the bottom plot the ratio of the results obtained with the different unfolding algorithms is shown.
    }
    \label{f:Zaa_IBU_result}
\end{figure*}
\section{Comparison for an EFT example}
\label{s:compare_IcINN_IBU}

At this point the single-event unfolded distributions of both the matrix--based Iterative Bayesian Unfolding as well as the ML--based Iterative cINN Unfolding have been validated to fulfill the analytic prediction of the toy model.
The next step is now to apply these ideas to a more realistic example, which is however not analytically solvable anymore.
Following again Ref.~\cite{Backes:2022vmn}, we use the process
\begin{align}
    pp \rightarrow Z \gamma \gamma, \qquad Z \rightarrow \mu^- \mu^+.
    \label{eq:mpp_Zgg}
\end{align}
The Monte Carlo events result from a pure Standard Model (SM) simulation of this process.
We use an Effective Field Theory (EFT) approach to add Beyond the Standard Model (BSM) contributions in the pseudo-experimental data.
This leads to significant differences between the distributions of the Monte Carlo simulation and of the pseudo-experimental data, hence the need for more than one unfolding iteration.
Further details on the simulation, like e.g.\ the choice and size of the Wilson coefficients, can be found in Ref.~\cite{Backes:2022vmn}.
MadGraph5~\cite{Alwall:2014hca} and Pythia8.3~\cite{Bierlich:2022pfr} are used for the event generation, while DELPHES~\cite{deFavereau:2013fsa} is used to simulate detector effects. 

The application of IBU to the $Z\gamma\gamma$ data is straight-forward.
The unfolding result is shown in Figure \ref{f:Zaa_IBU_result}. 
For simplicity, only one of the two muon momenta is unfolded.~\footnote{The IcINN unfolding is still two-dimensional: following Ref.~\cite{Backes:2022vmn}, a trivial second dimension is added.}
The unfolding result shows some structures which are very similar to the IcINN performance: the first iteration already predicts most of the truth-level data correctly.
Nevertheless, further iterations can improve the result in the high-$p_T$ region.~\footnote{Depending on the region of interest and the required precision in the study, one could question if three iterations are needed or if two are already sufficient.}

At this point it is possible to compare the unfolded distributions of a single measured event.
For illustrative purposes, a low-$p_T$ event at $45 \, \mathrm{GeV}$, as well as a high-$p_t$ one at $185 \, \mathrm{GeV}$, are chosen.
\begin{figure*}[p!]
    \centering
    \includegraphics[width=0.43\textwidth]{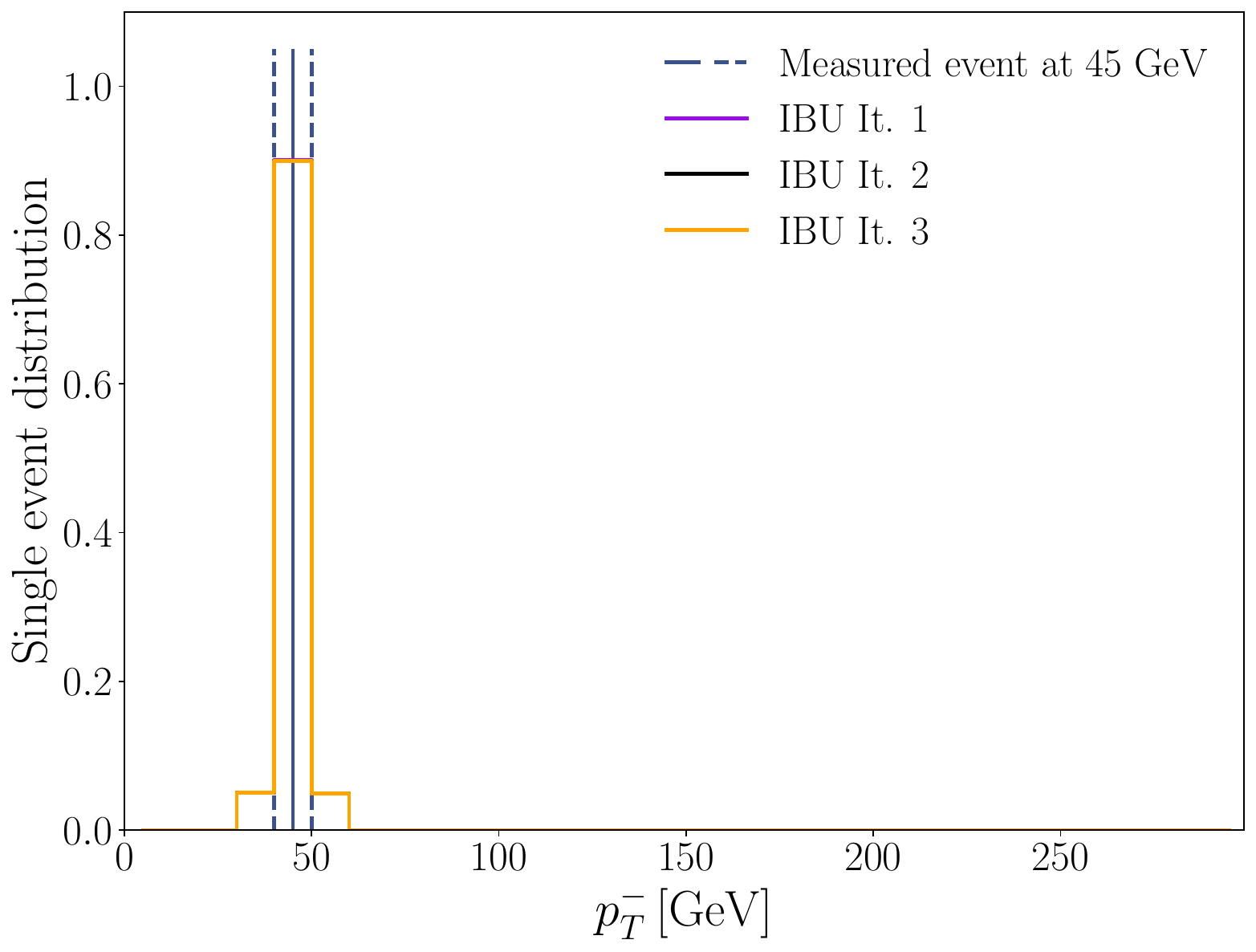} \hspace{0.2cm}
    \includegraphics[width=0.43\textwidth]{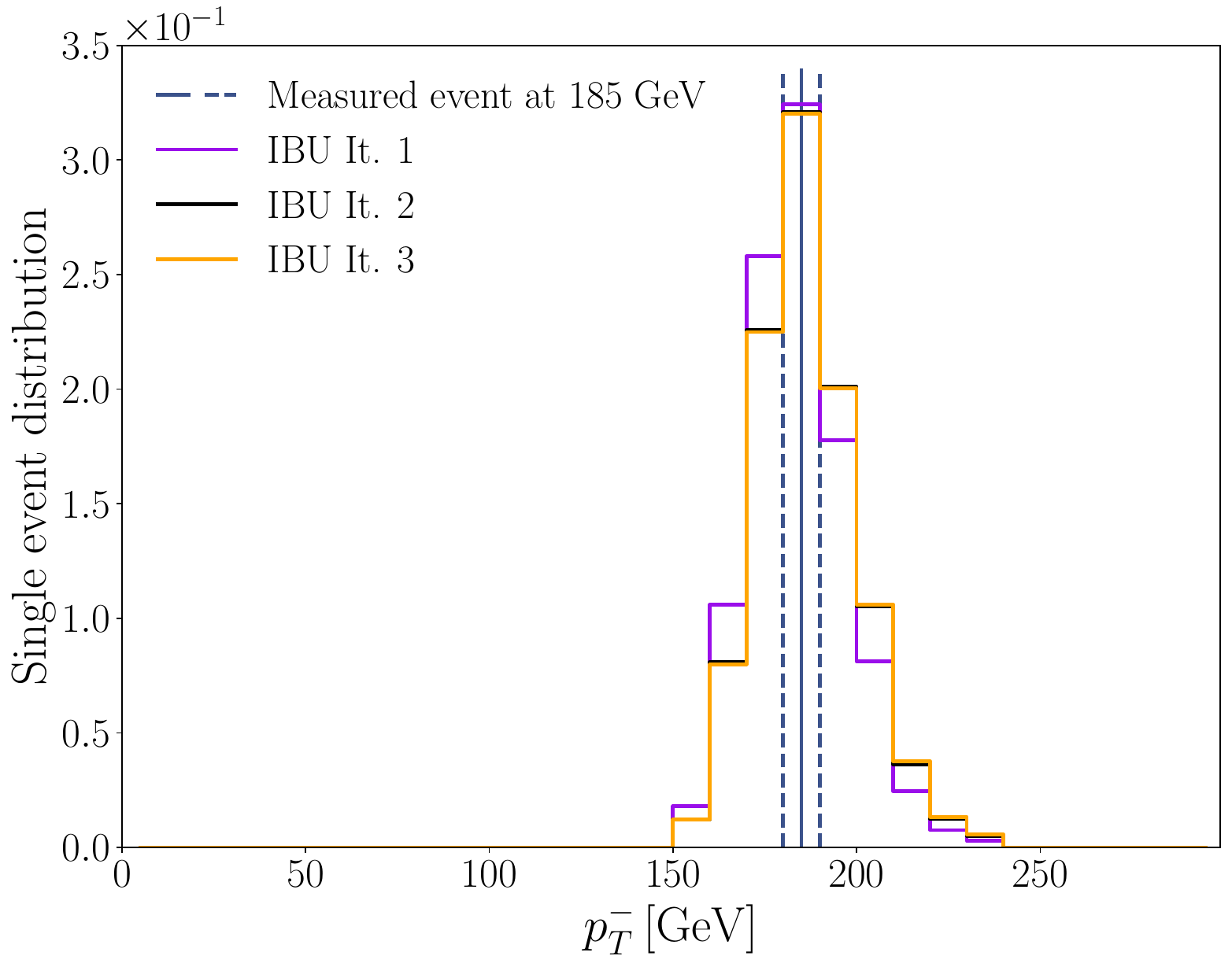} \\
    \includegraphics[width=0.43\textwidth]{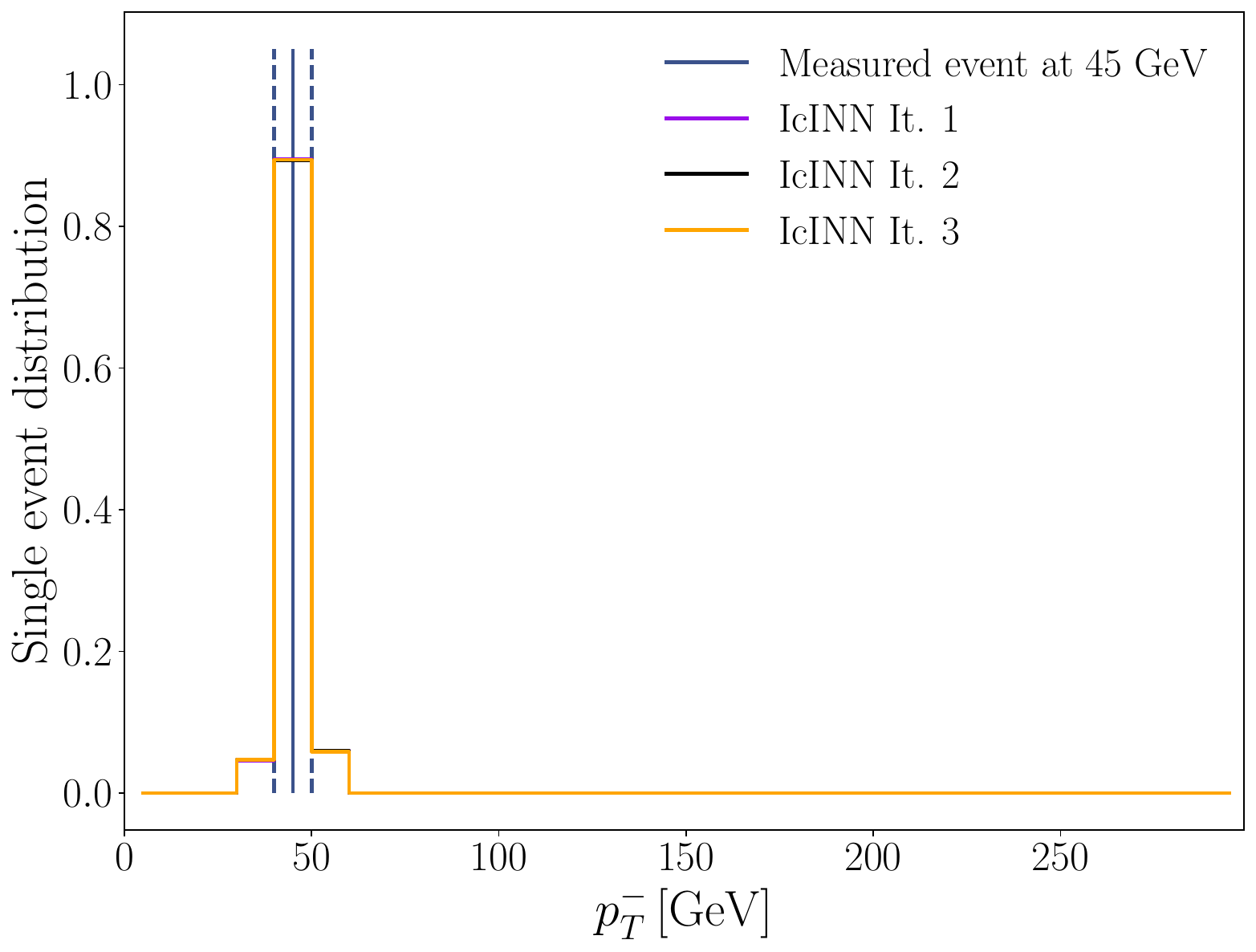} \hspace{0.2cm}
    \includegraphics[width=0.43\textwidth]{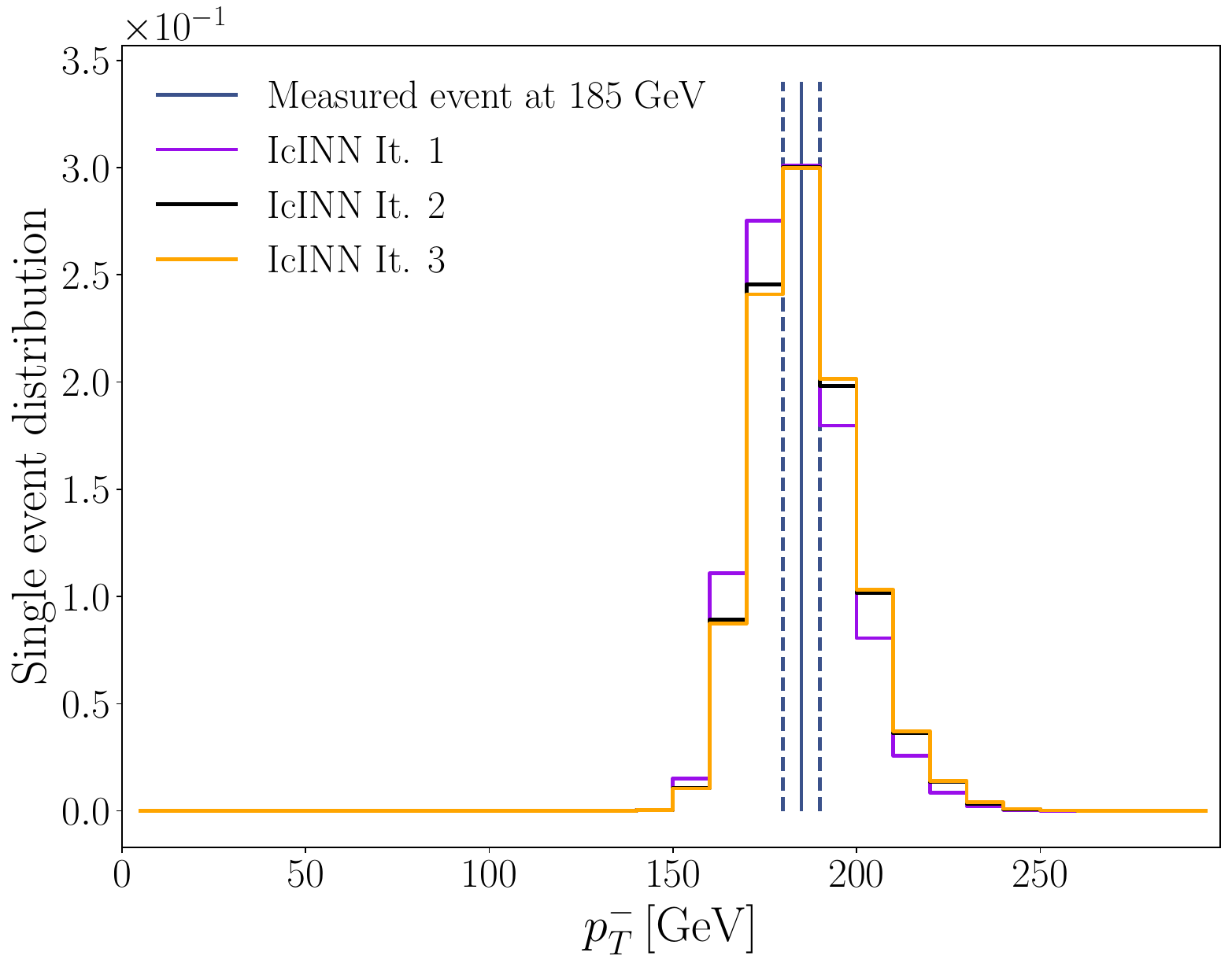} \\
    \includegraphics[width=0.43\textwidth]{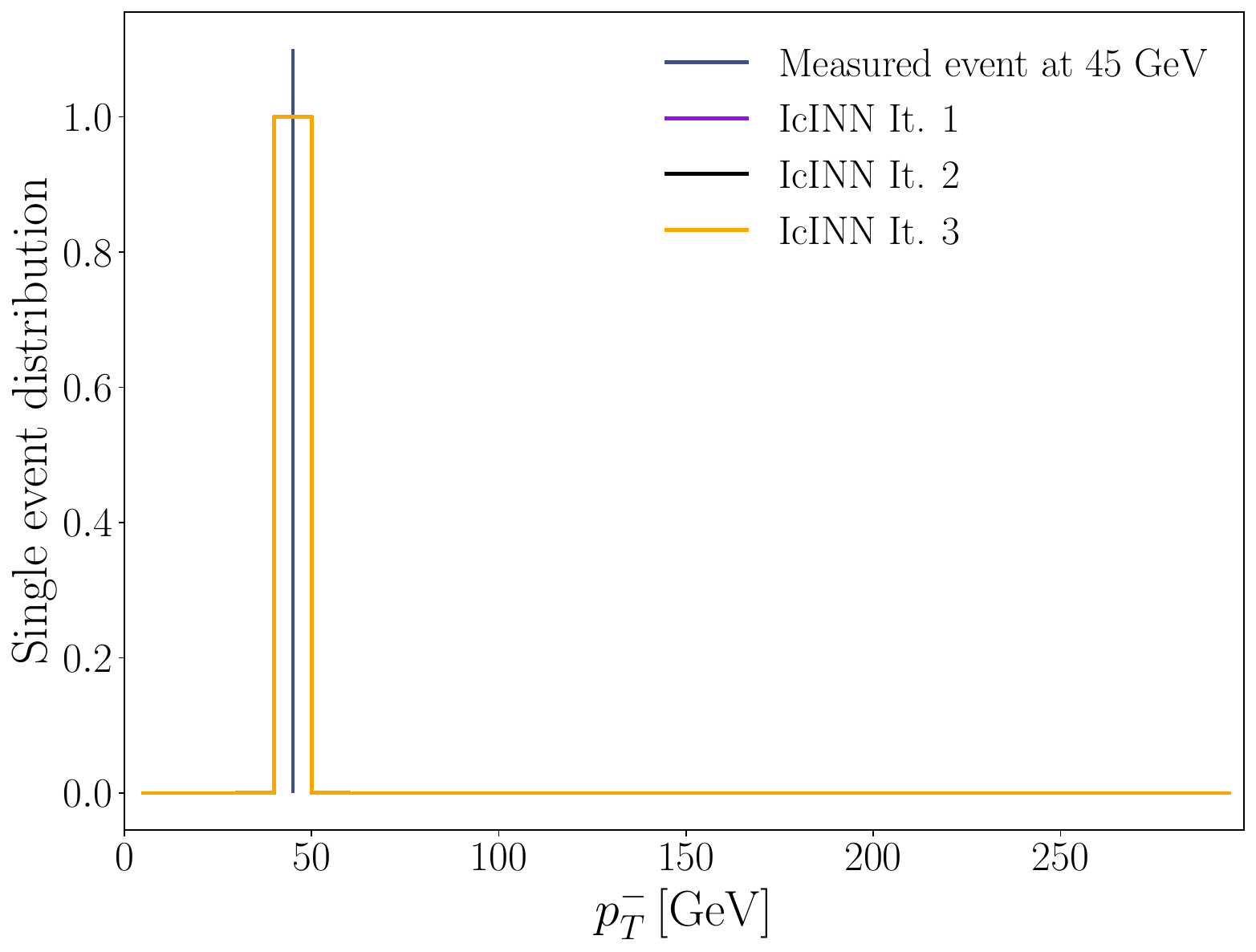} \hspace{0.2cm}
    \includegraphics[width=0.43\textwidth]{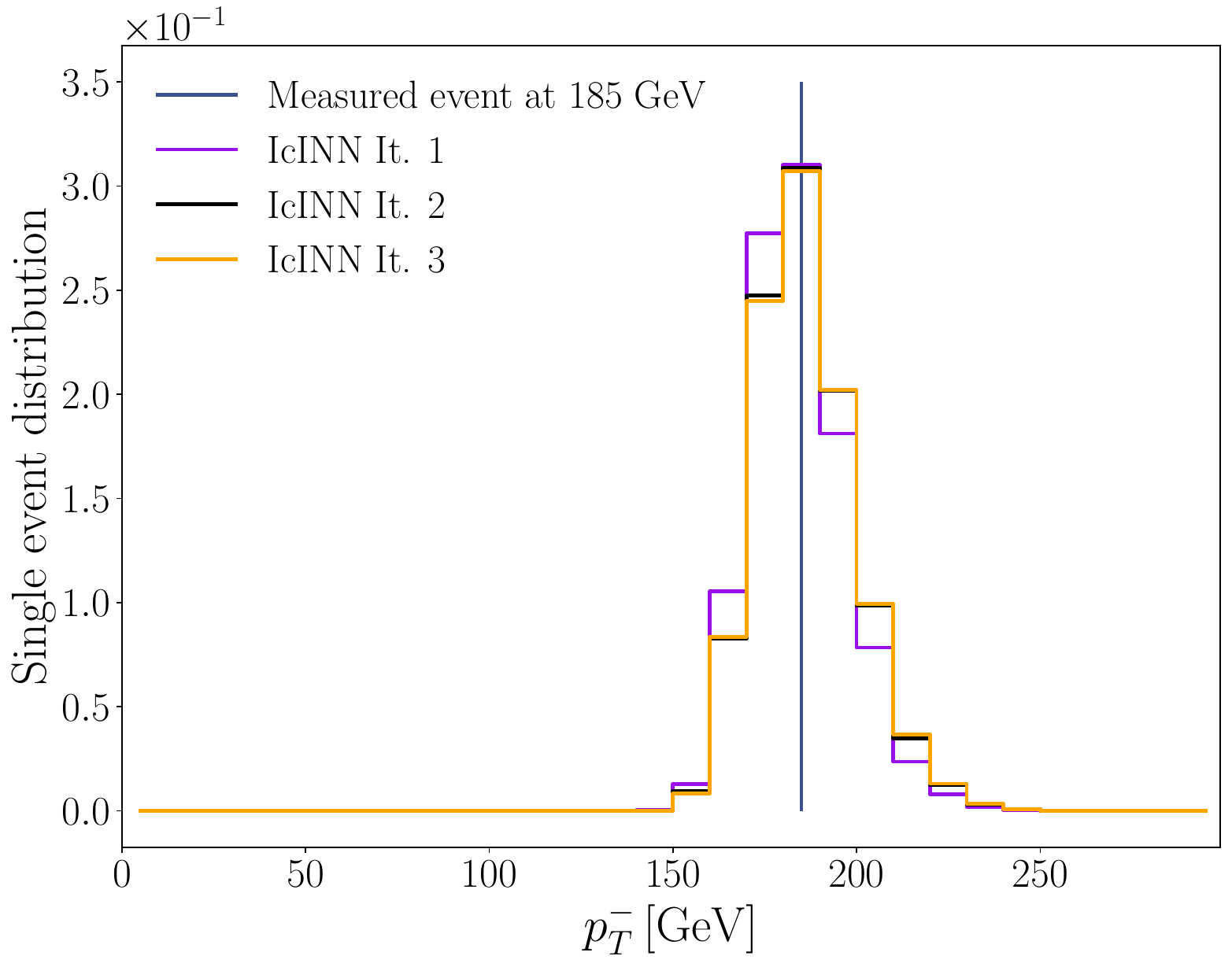} \\
    \includegraphics[width=0.43\textwidth]{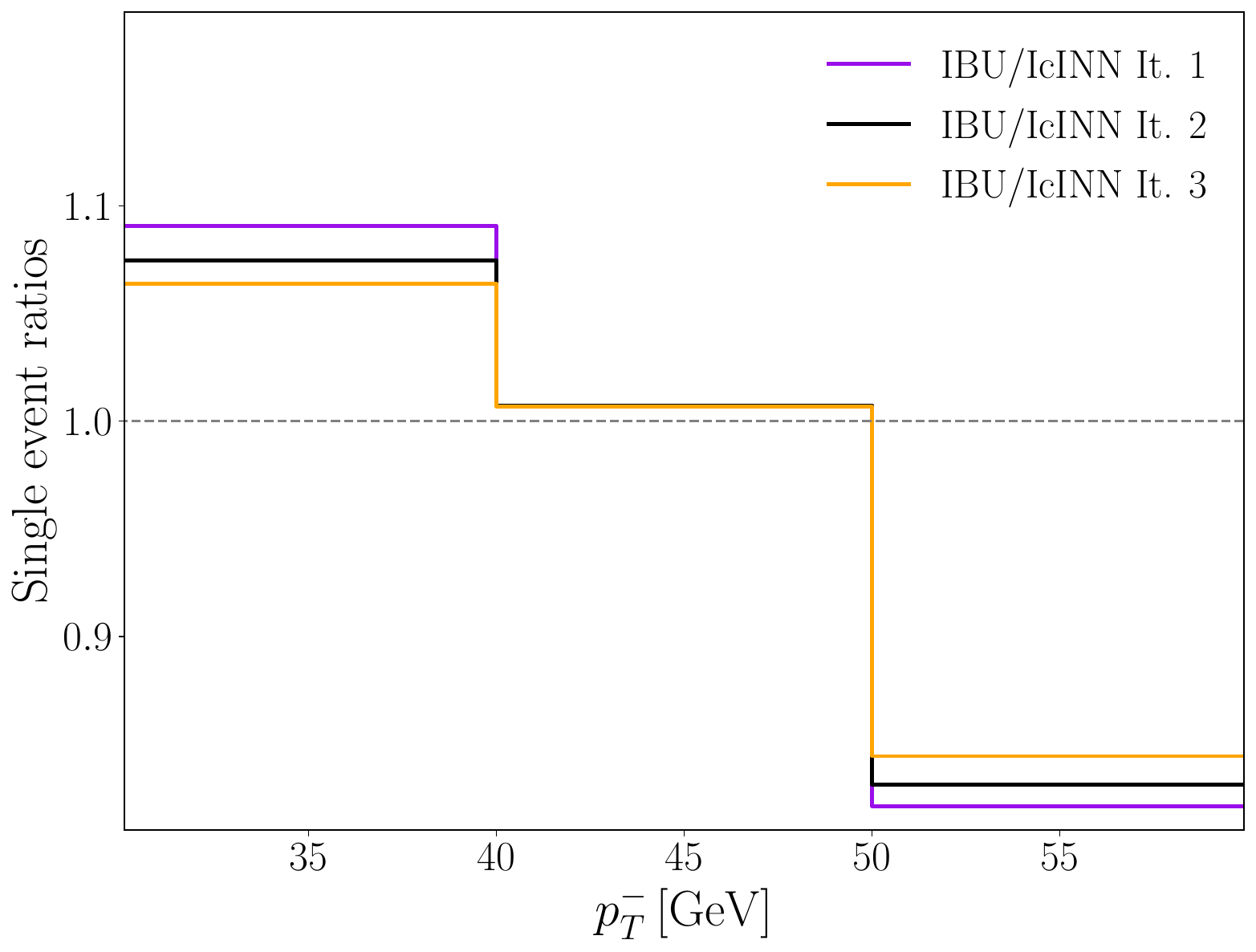} \hspace{0.2cm}
    \includegraphics[width=0.43\textwidth]{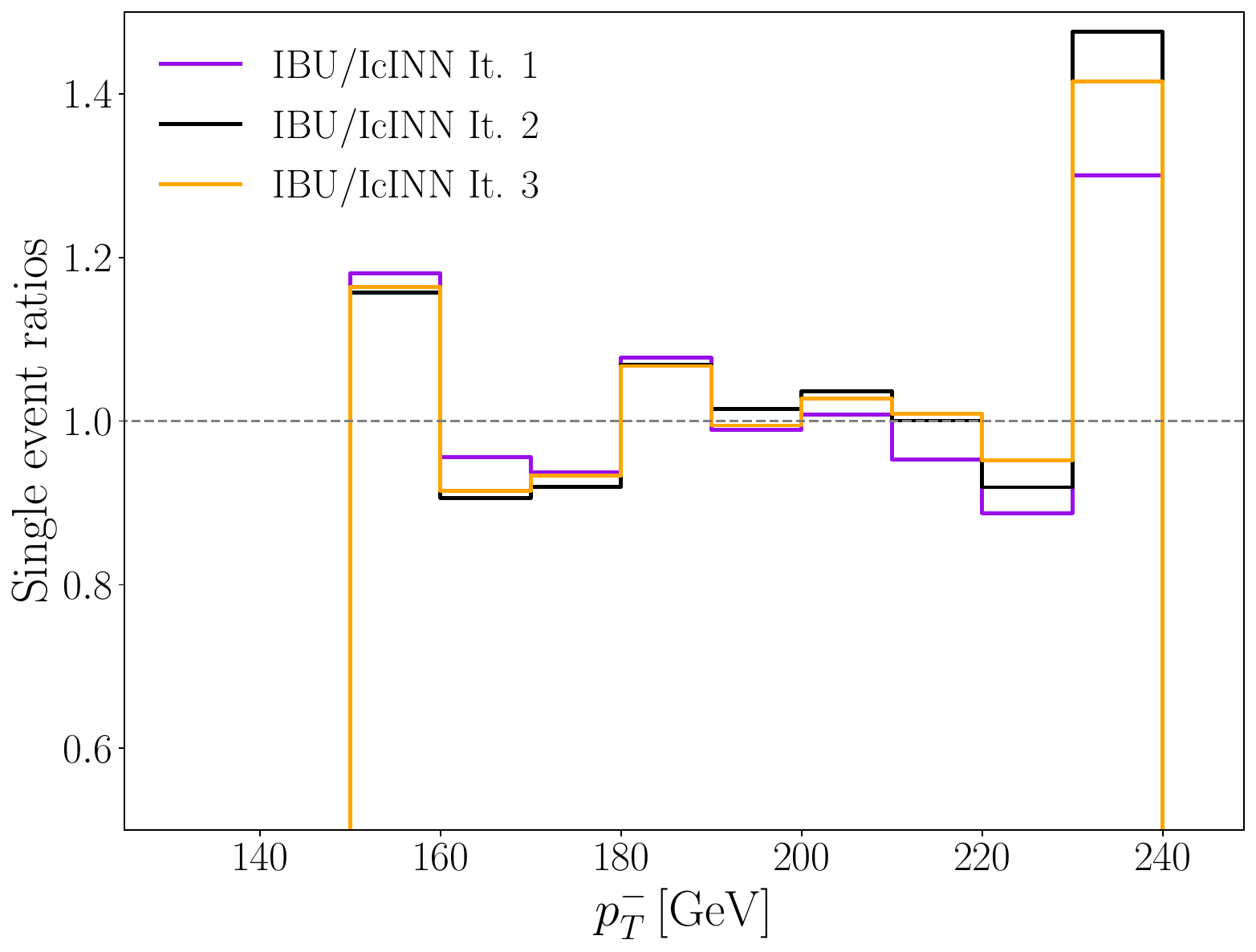} \\
    \caption{Single-event unfolded distributions after 1, 2 and 3 iterations (purple, black, orange).
    The left~(right) column shows the single-event unfolded distribution for an event at $45 \, \mathrm{GeV}$~($185 \, \mathrm{GeV}$).
    The results for the IBU are shown in the first row.
    In the second row the IcINN unfolds several events sampled from the full bin, while in the third row only the fixed event is unfolded with the IcINN. The fourth row shows the ratio of the IBU single-event distribution divided by the full-bin IcINN single-event distribution.} 
    \label{f:Zaa_single_events}
\end{figure*}
The single-event unfolded distributions are shown in Figure \ref{f:Zaa_single_events}.
For IcINN two different cases are shown: in the second row, the IcINN algorithm unfolded several events from the bin around $45\,\mathrm{GeV}$ to get a better comparison to the single-event unfolding of the matrix--based approaches, in the third row only the fixed events are unfolded.

In general, the single-event unfolded distributions produced by the IcINN unfolding and the IBU unfolding are very similar.
In the left column of Figure~\ref{f:Zaa_single_events} it is visible that the single-event unfolded distributions of low-$p_T$ events is not changing significantly with further iterations.
This is not surprising as the full unfolded distribution is in this region also nearly constant throughout the iterations.
It is visible, that the IBU is closer to the IcINN distribution if the IcINN unfolds events sampled in the full bin.
This is especially visible in the low-$p_T$ region, where the detector resolution is smaller than the bin size.

In the high-$p_T$ region, the single-event unfolded distributions are broader.
This is expected, since the detector resolution effects are chosen in this exercise to increase with higher $p_T$ \cite{Backes:2022vmn}.
In this $p_T$ region the IcINN results are very similar, not depending anymore on whether the event is fixed or sampled over the full bin.
The impact of the iterations is visible in the case of the IBU unfolding, as well as for the IcINN: the values in bins of the single-event unfolded distribution are progressively reduced on the low-$p_T$ side and increased on the high-$p_T$ side.
These observations are in agreement with the overall unfolding result, where multiple iterations also improve mainly the high-$p_T$ region of the unfolding problem.

Qualitatively the single-event unfolded distributions of IBU and IcINN match, but in a more quantitative comparison one expects to observe some differences.
This is due to two main reasons.
First, in the IcINN training and evaluation the truth level events are rescaled with a logarithmic function as a preprocessing to reduce the necessary amount of training.
For IBU no such transformation is applied here, although in general one can use variable binning adapted to the shape of the input distribution.
Second, the IBU single-event unfolding is limited to unfolding a full bin, while the IcINN is working on a non-binned phasespace.
It is simply not possible to completely overcome the information loss associated with the introduced binning, as even when using fine binning there is always some amount of information loss associated with it.
Nevertheless, the qualitative behaviour of the IcINN single-event unfolded distributions is confirmed with the matrix--based iterative unfolding algorithms.

\section{Summary and Conclusions}
\label{s:Conclusions}

In this paper we presented a novel approach to obtain single-event unfolded distributions for matrix--based algorithms by applying the unfolding matrix to one reconstructed event.
This approach enables a direct event-by-event comparison of matrix-- and ML--based unfolding methods.
The algorithm has been implemented for Iterative Bayesian unfolding (IBU) and Iterative Dynamic Stabilised (IDS) unfolding.
Both in the case of the Gaussian toy example as well as the $pp\rightarrow Z\gamma\gamma$ pseudo-data, the performance of the matrix--based algorithms was similar to the single-event unfolding of the Iterative cINN Unfolding (IcINN), with small differences in the bulk of the corresponding distributions.
It can hence be concluded that in these low-dimensional examples the matrix--based methods and IcINN agree.
Although systematic uncertainties still have to be evaluated carefully on the event-by-event basis, there is strong motivation to further use the IcINN single-event unfolded distributions in high-dimensional analyses, which are typically very challenging for matrix--based methods.

\section*{Acknowledgements}
AB would like to acknowledge support by the BMBF for the AI junior group 01IS22079.
AB and BM gratefully acknowledge the continuous support from LPNHE, CNRS/IN2P3, Sorbonne Universit\'e and Universit\'e de Paris Cit\'e.
MB acknowledges support by the IMPRS-PTFS.

\appendix

\section{Implementing a Weighting Approach}
\label{App:WeightingApproach}

An alternative approach to obtain a single-event unfolded distribution, applicable for SVD too, can be implemented by introducing a small perturbation in the detector-level data histogram.
Doing so, the complete matrix--based procedure is performed once with the original distribution, once with a slightly changed detector-level distribution $r_i$.
A single bin $i_s$ of the measured distribution~(i.e.\ the one containing the event to be unfolded) is modified by multiplying it with a weighting factor $w$.
The single-event unfolded histogram $e_j(i_s)$ is defined as the normalized difference between the unfolded distributions
\begin{align}
    e_j(i_s) = \frac{1}{(w-1)r_{i_s}} \left( u_j' - u_j \right),
\end{align}
using the original unfolding result $u_j$ of the original detector-level distribution, as well as the unfolding result $u_j'$ derived from the reweighted detector-level data.

\begin{figure*}[p!]
	\centering
	\includegraphics[width=0.48\linewidth]{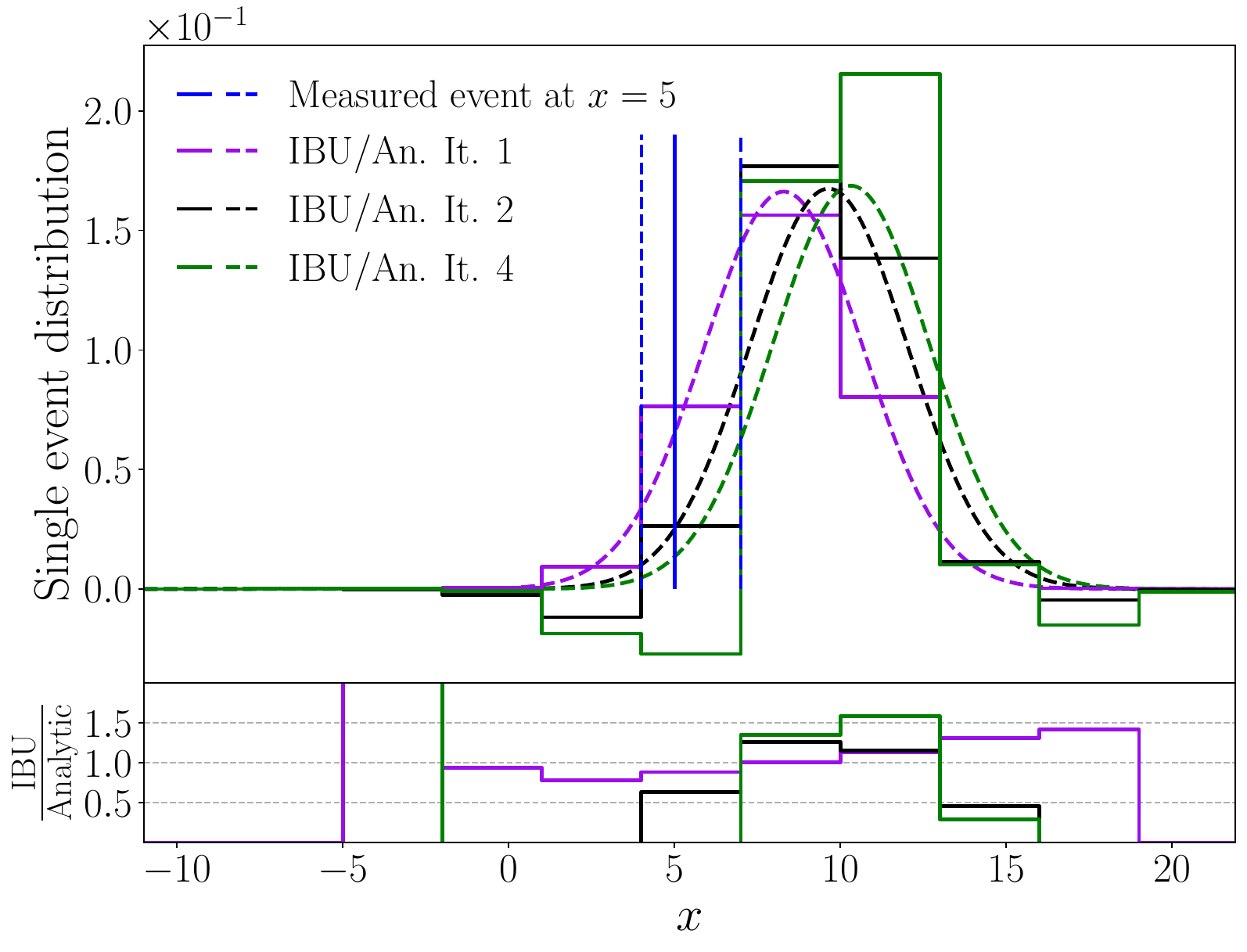}
    \hspace{0.1cm}
    \includegraphics[width=0.48\linewidth]{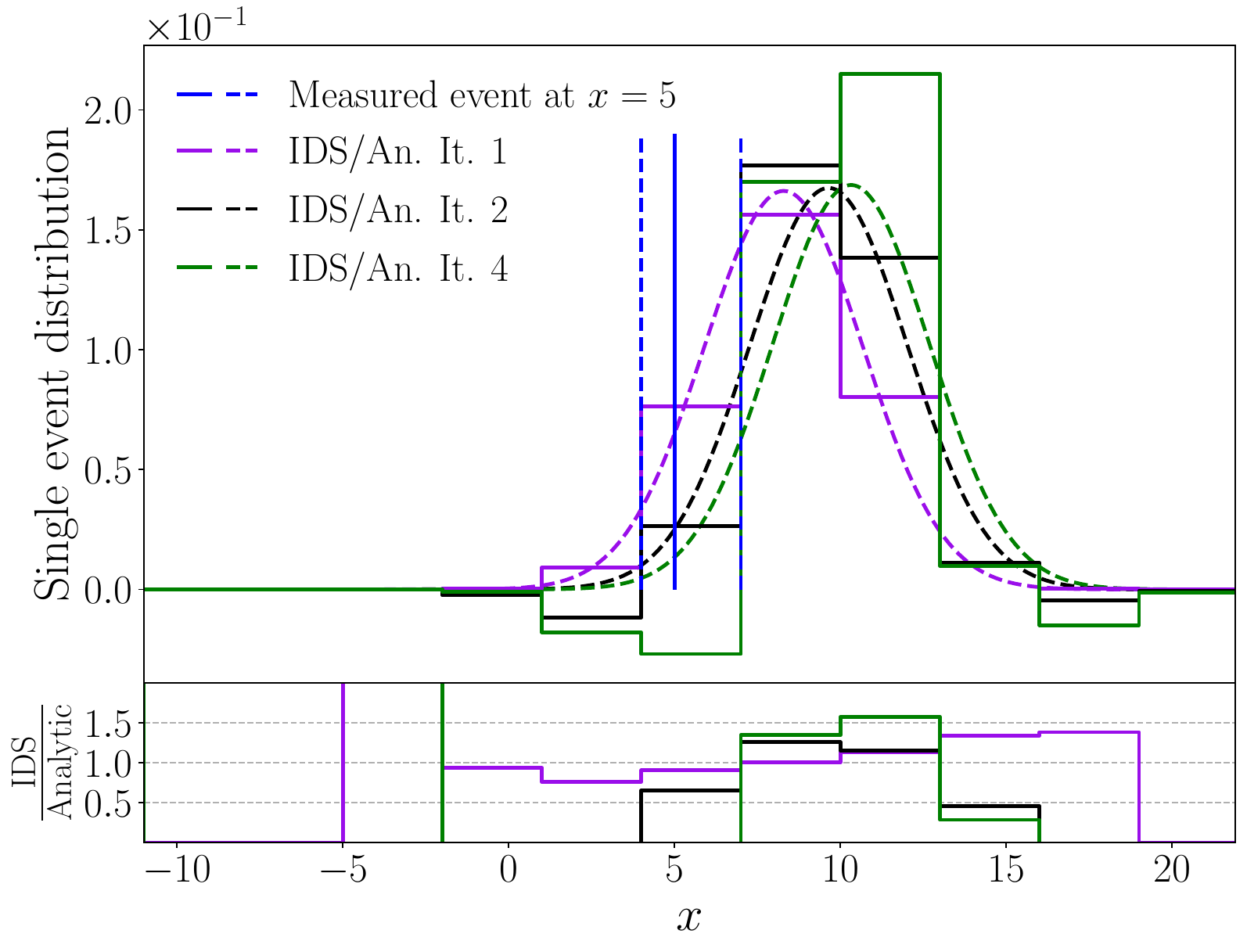} \\
    \includegraphics[width=0.48\linewidth]{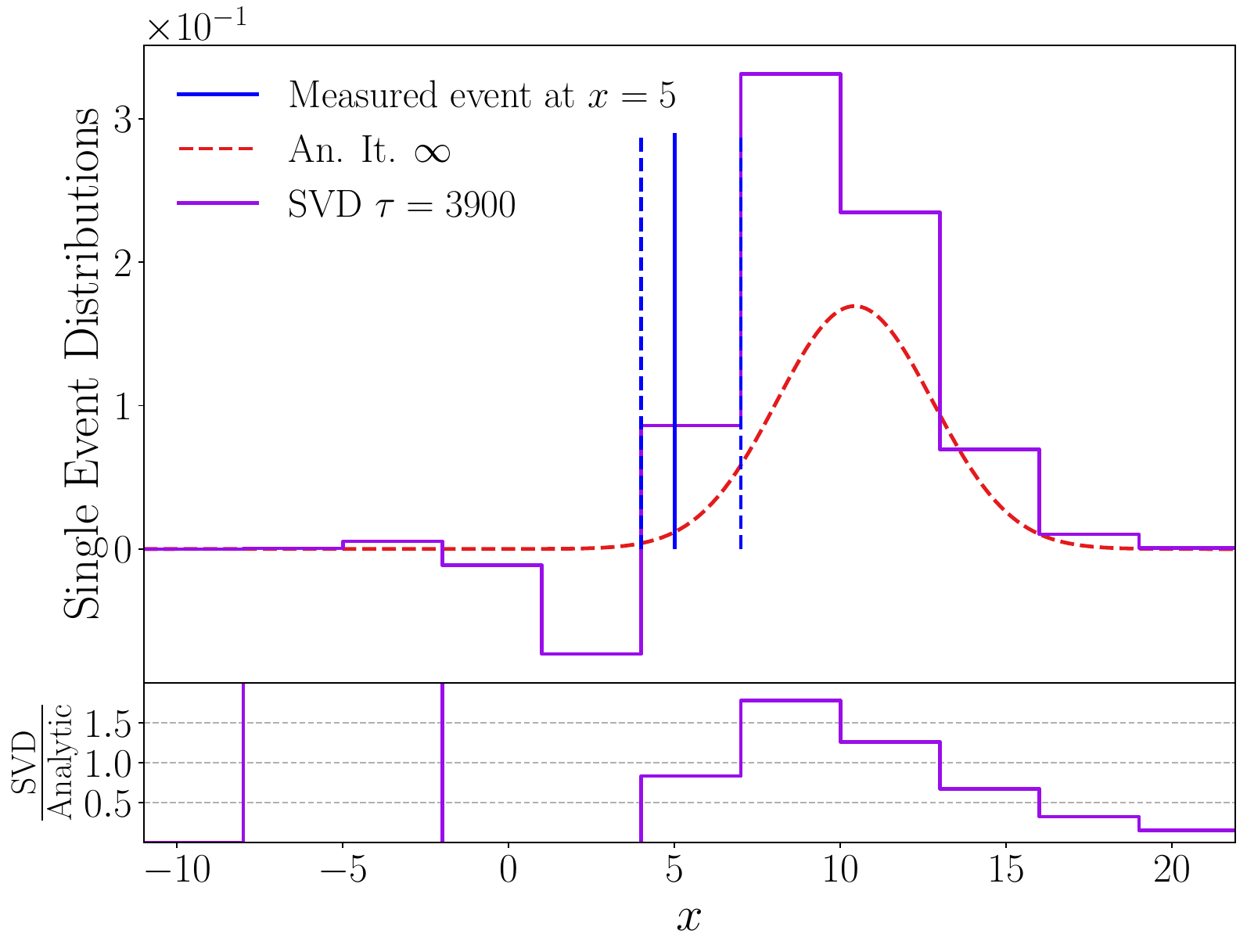}
    \includegraphics[width=0.48\linewidth]{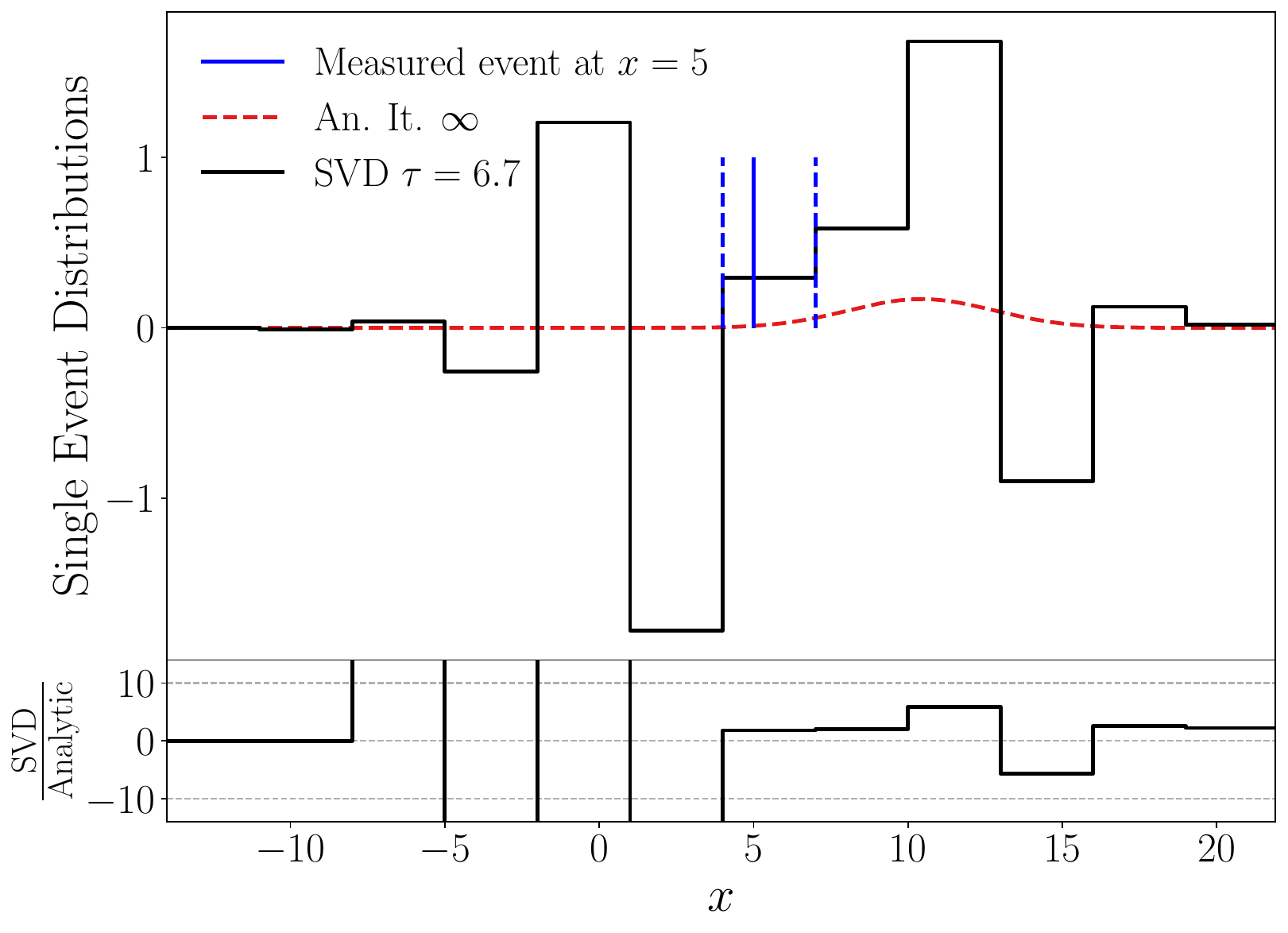}\\
    \includegraphics[width=0.48\linewidth]{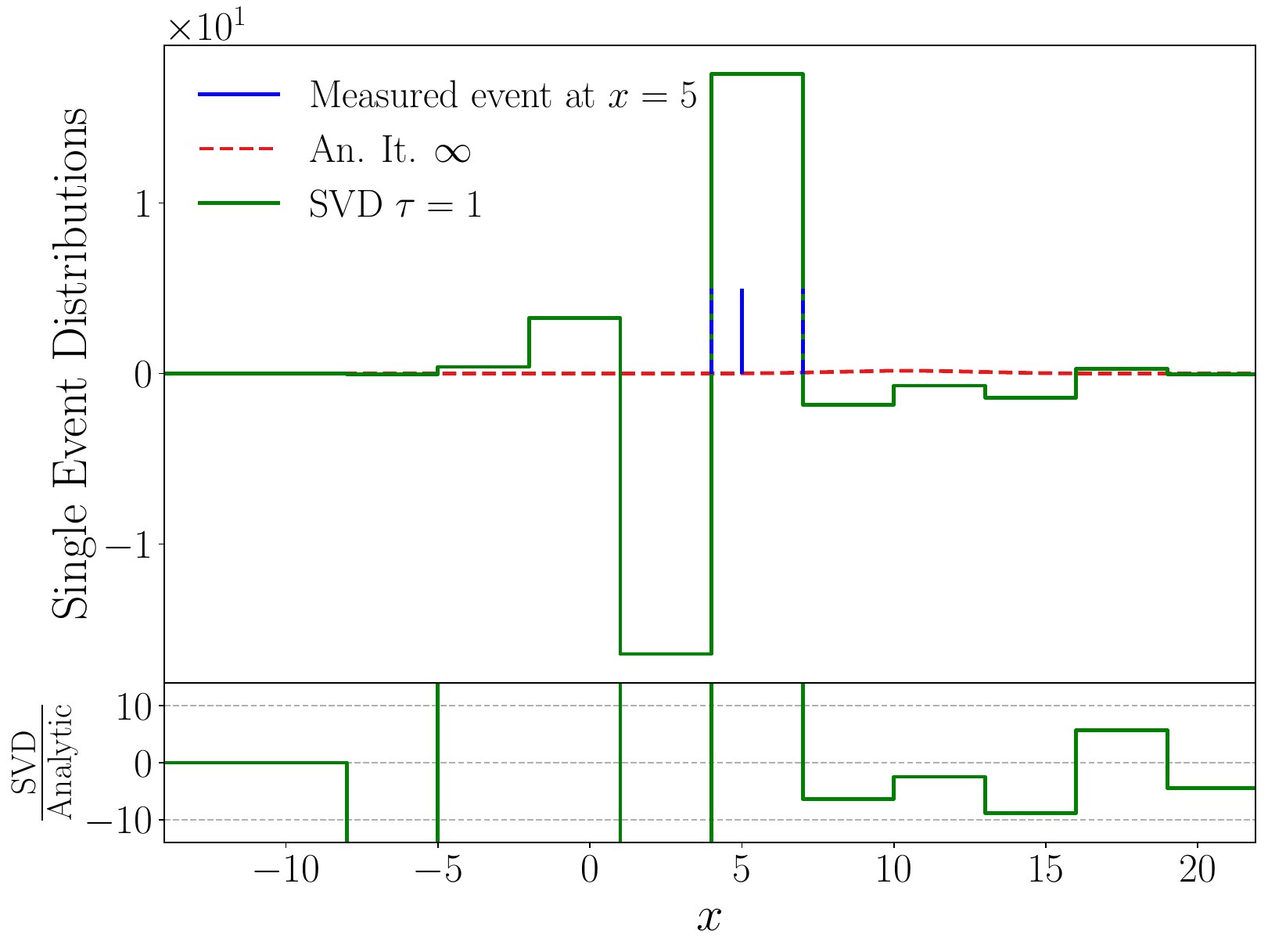}
	\caption{Single-event unfolding using a weighting approach.
     In the upper left the result for IBU is shown, on the upper right the result for IDS.
     The blue bin is unfolded.
     The dashed lines (purple, black, green) are the analytic predictions for each iteration, the solid lines (purple, black, green) are the unfolding results of the single-event distributions after each iteration.
     In the first iteration the single-event unfolded distributions match their analytic predictions.
     For higher iterations, this is not the case.
     The lower three plots show the results for SVD with several different regularisation values $\tau$.
     The analytic prediction (dashed; red) is derived by taking the limit of infinite iterations.
     The single-event distribution resulting from the strongest regularization (middle left plot; purple) is closest to the analytic prediction, while the other two examples feature larger fluctuations.
     A weighting factor $w=1.001$ is used for these examples.}
	\label{f:single_all_algorithms}
\end{figure*}

The results of this procedure for the analytic toy model are shown in Figure~\ref{f:single_all_algorithms}.
The parameters for IDS were chosen as in Section~\ref{s:analytic}.
Due to the weak level of regularisation these parameters introduce it is not surprising that IBU and IDS behave very similarly (upper left and upper right plot).
In the first iteration the single-event unfolded distribution matches the prediction of the analytic approach.
The single-event distributions of the second and fourth iteration do not match the analytic predictions, instead there is a small shift as well as negative entries.
It is not surprising to see such a behaviour, since this weighting approach is fundamentally different to the analytic unfolding of a delta distribution.

This behavior can be explained by looking at the algorithms more closely: the first iteration is the pseudo-inversion of the response matrix using Bayes' theorem.
This pseudo-inversion is based solely upon the original Monte Carlo simulation, which is the same regardless whether the detector-level data distribution is weighted or not.
The pseudo-inverse, i.e.\ the unfolding matrix, is therefore not impacted by the changed detector-level data and the extracted single-event unfolded distribution is exactly a column of the unfolding matrix (for a fixed index $i_s$, as it is used in Eq.~\eqref{eq:IBU_unfold}).
This procedure agrees well with the analytic prediction. 

In the examples of Figure \ref{f:single_all_algorithms} a factor of $w=1.01$ was used, thus the height of one bin of the detector-level data was increased by one percent.
After the first iteration, the unfolded weighted data is set to be the new prior for the pseudo-inversion, which impacts in a coherent way several truth-level bins.
This then implies that the subsequent unfolding matrix is impacted by the change applied to the detector-level data distribution.
That effectively enhances, in the unfolding correction, the amount of migrations towards the corresponding increased bin~(keeping in mind that the toy model considered here also applies a shift between truth and reconstructed quantities, in addition to the Gaussian smearing).
Due to the Gaussian nature of the resolution smearing, the bins next to the reweighted one are impacted stronger than bins that are farther away.

The shape of the sum of all the individual unfolded event distributions obtained here is somewhat different than the one of the full unfolded detector-level data distribution~(while the integral is preserved).
The differences are of the order of $1\%$ in the peaks and $5\%$ in the tails, also depending on the unfolding algorithm and the value of $w$.
This mismatch is caused by the fact that this approach actually yields an approximation of the single-event unfolded distribution, which can be improved with $w \to 1$.
This behavior, as well as the result of the first iteration, shows that it might be preferable to instead use a procedure based on extracting a particular column of the unfolding matrix after each iteration (see Section~\ref{Sec:OneEvUnfolding}).

The single-event distributions of the SVD algorithm are shown for several regularization values $\tau$ in the lower three plots of Figure~\ref{f:single_all_algorithms}.
The analytic reference function (dashed; red) was obtained by calculating the limit of the single-event unfolded distribution for an infinite number of iterations. 
For a strong regularization of $\tau=3900$ the single-event unfolded distribution is closest to the analytic prediction (purple; middle-left plot).
For smaller $\tau \in \lbrace 6.7, 1 \rbrace$ the single-event unfolded distributions are clearly impacted by the high frequency problem (black, green; middle-right and lower plot).
For the unfolding of the full distribution, $\tau=6.7$ was the regularization factor which yielded the best results.
This regularization appears not to be strong enough to ensure the smoothness of the single-event unfolded distribution.
In this case it is hard to find a good working point for SVD, which would ensure both a good unfolding result for the full distribution and an appropriate regularization for the single-event unfolded distributions.

\begin{figure*}[p!]

	\centering
	\includegraphics[width=0.43\linewidth]{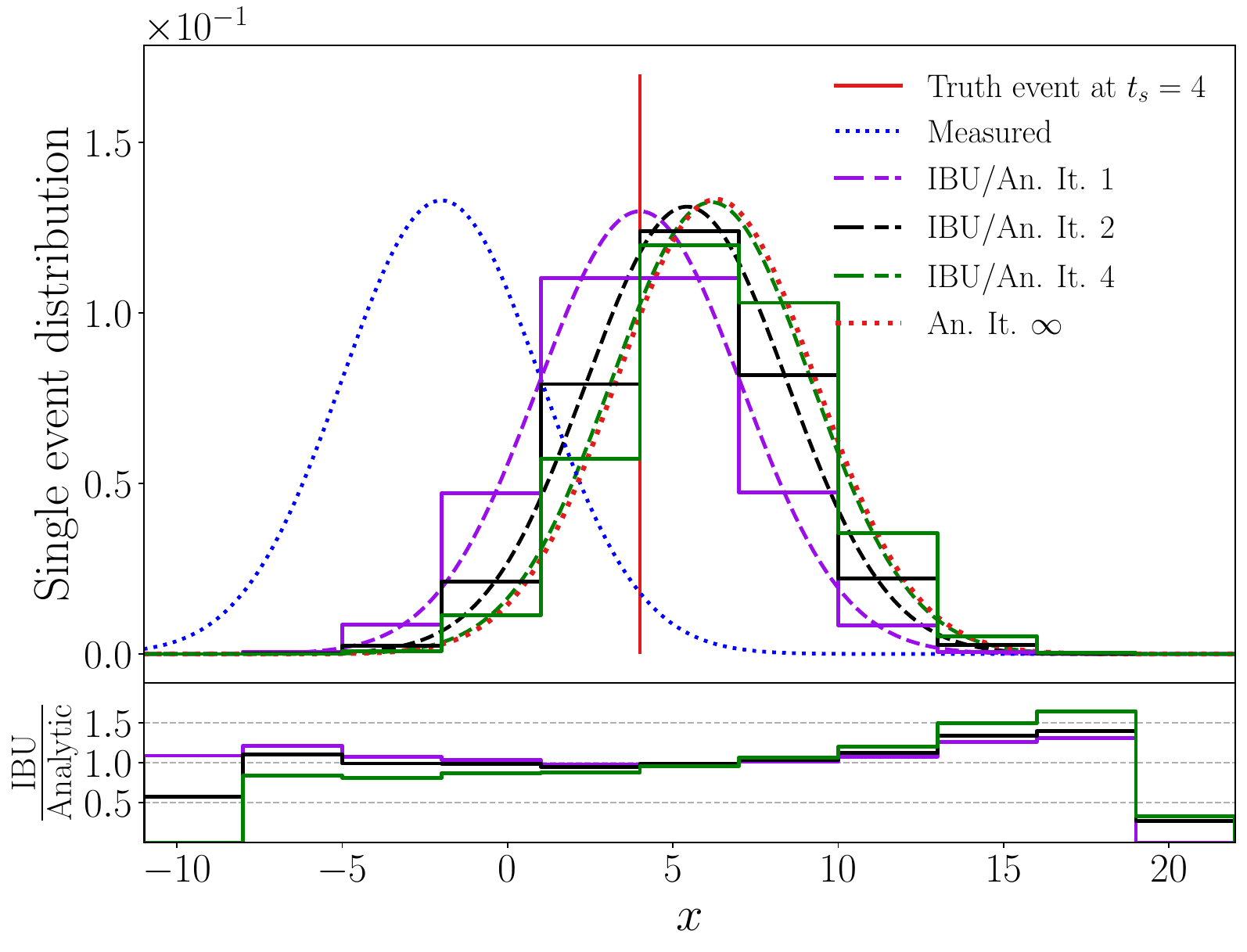}
    \hspace{0.5cm}
    \includegraphics[width=0.43\linewidth]{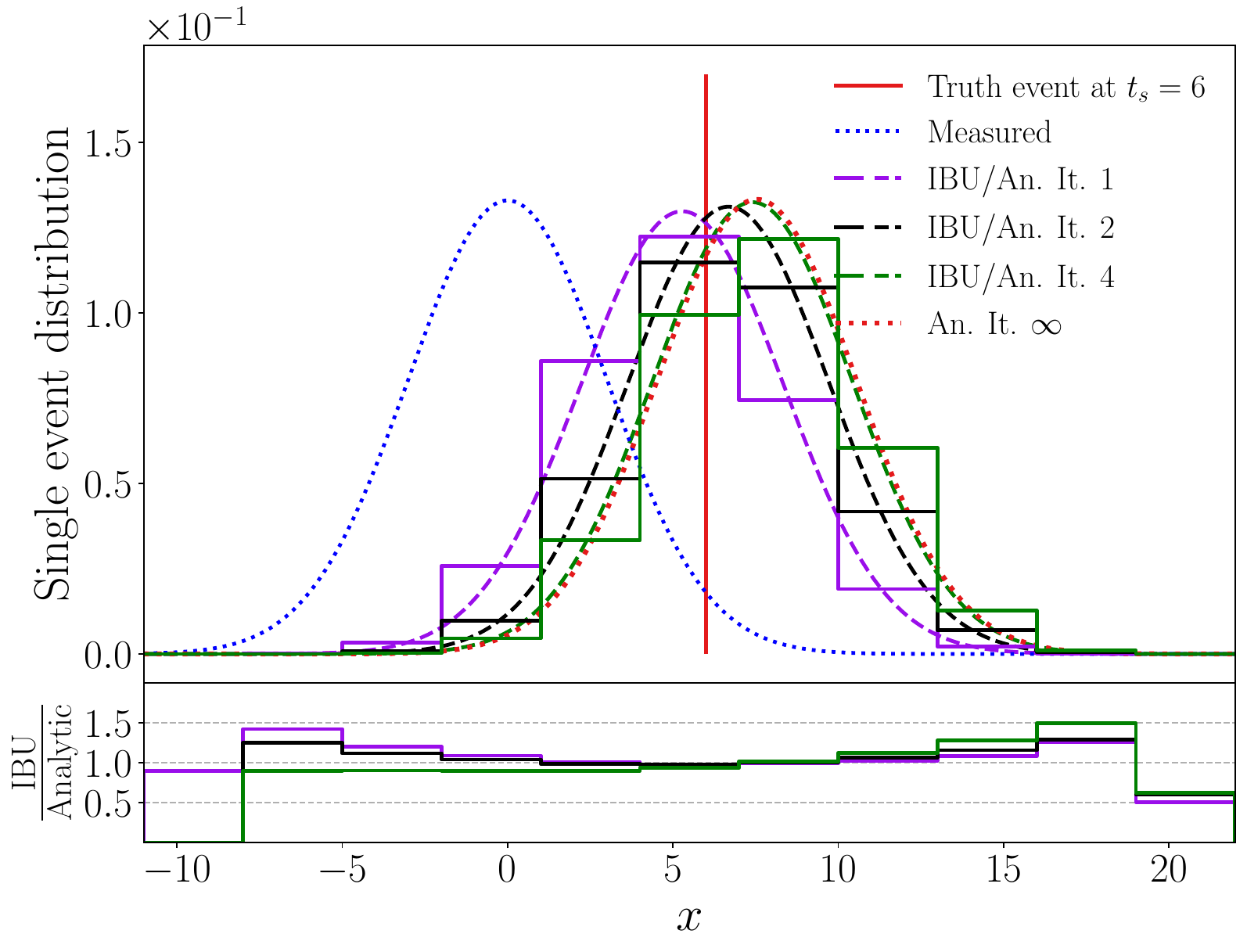}\\
    \includegraphics[width=0.43\linewidth]{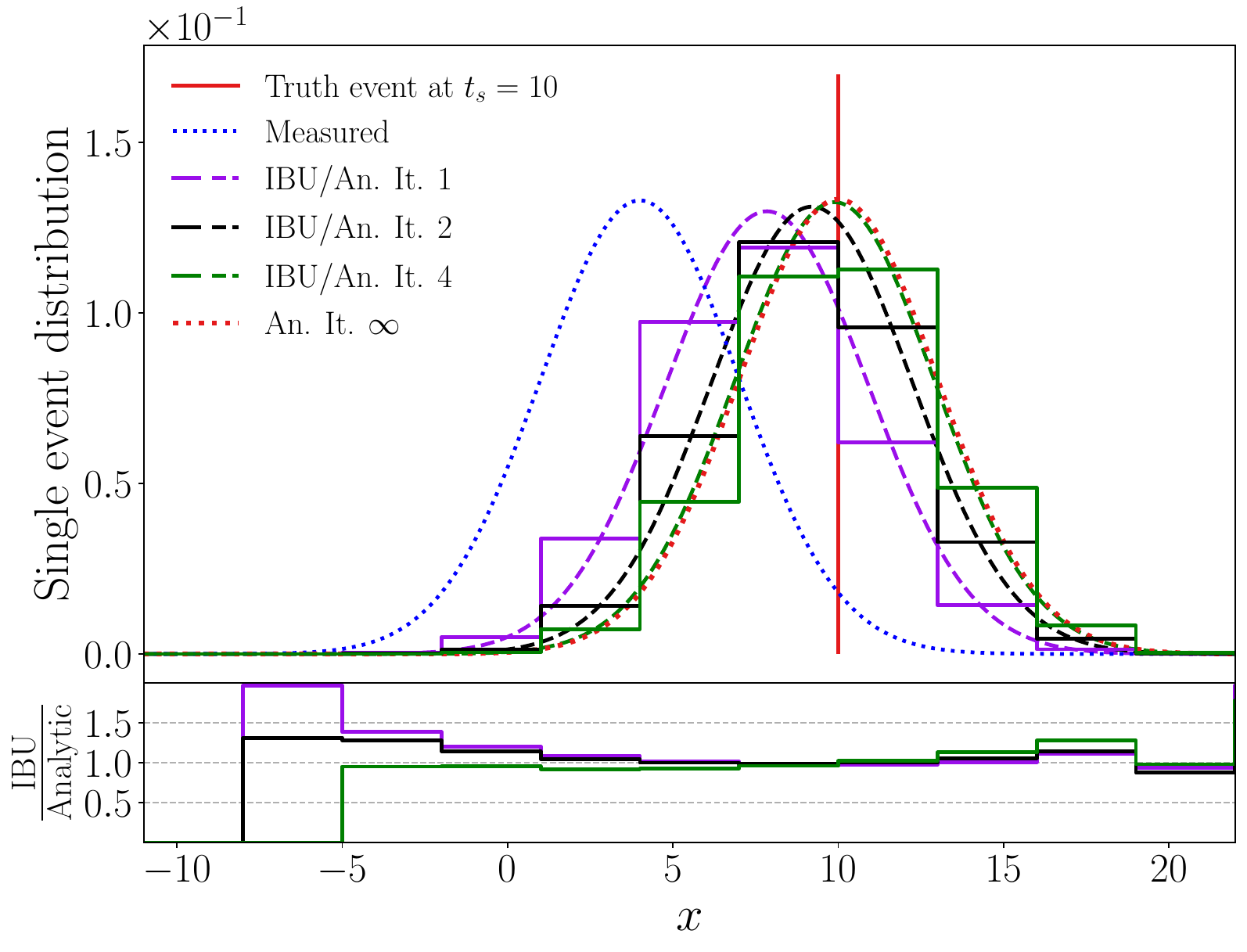}
    \hspace{0.5cm}
    \includegraphics[width=0.43\linewidth]{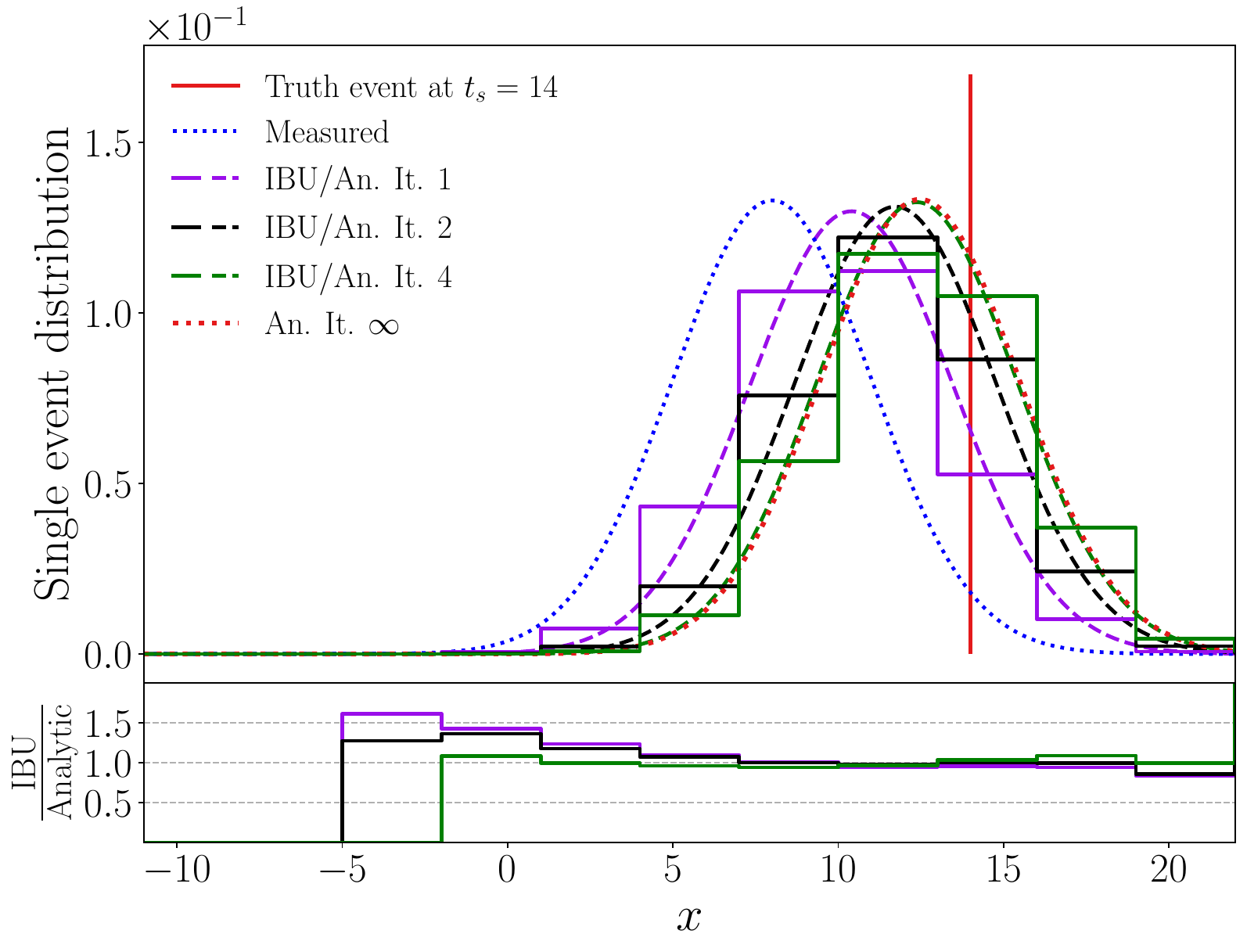}\\
    \vspace{0.1cm}
	\includegraphics[width=0.43\linewidth]{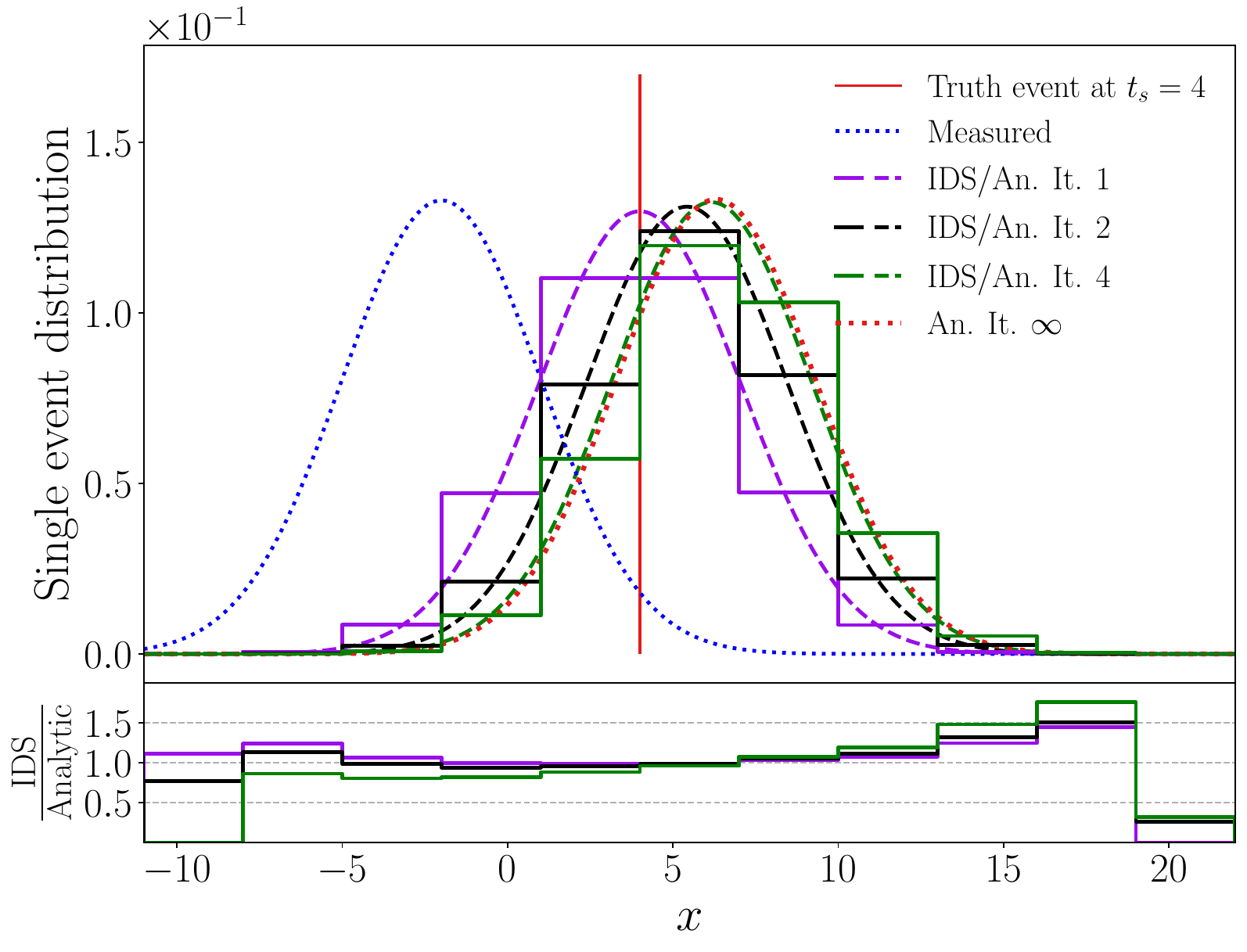}
    \hspace{0.5cm}
    \includegraphics[width=0.43\linewidth]{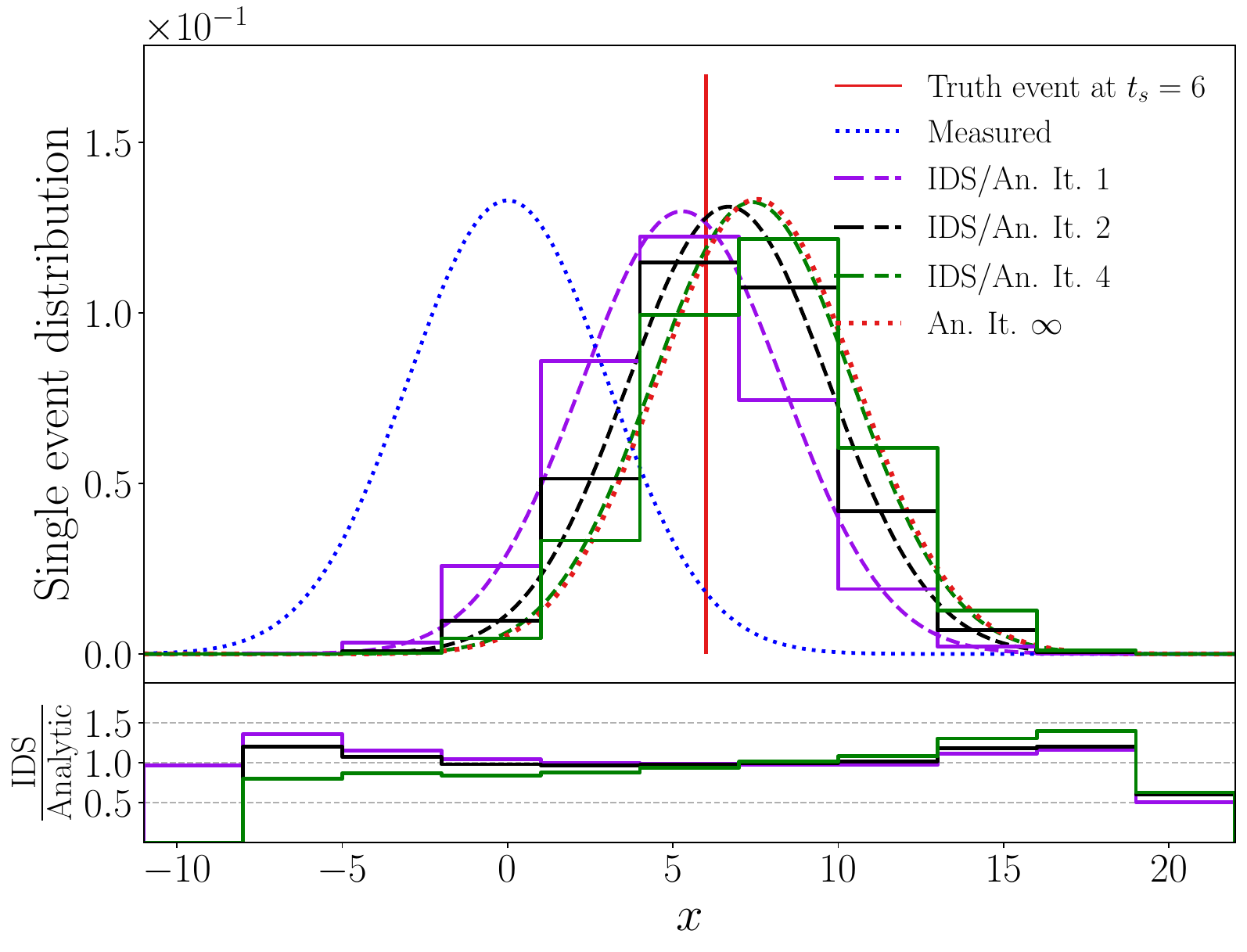}\\
    \includegraphics[width=0.43\linewidth]{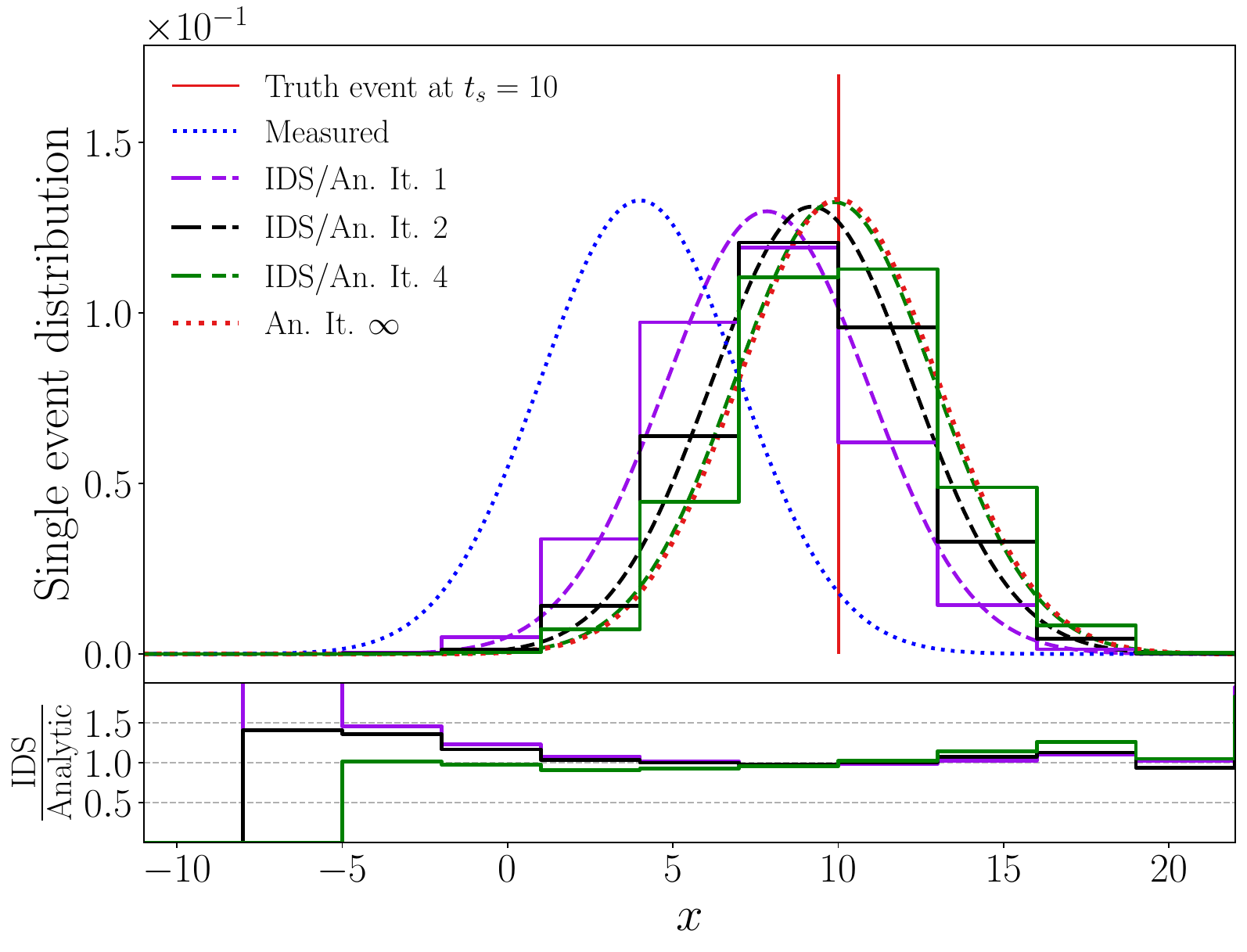}
    \hspace{0.5cm}
    \includegraphics[width=0.43\linewidth]{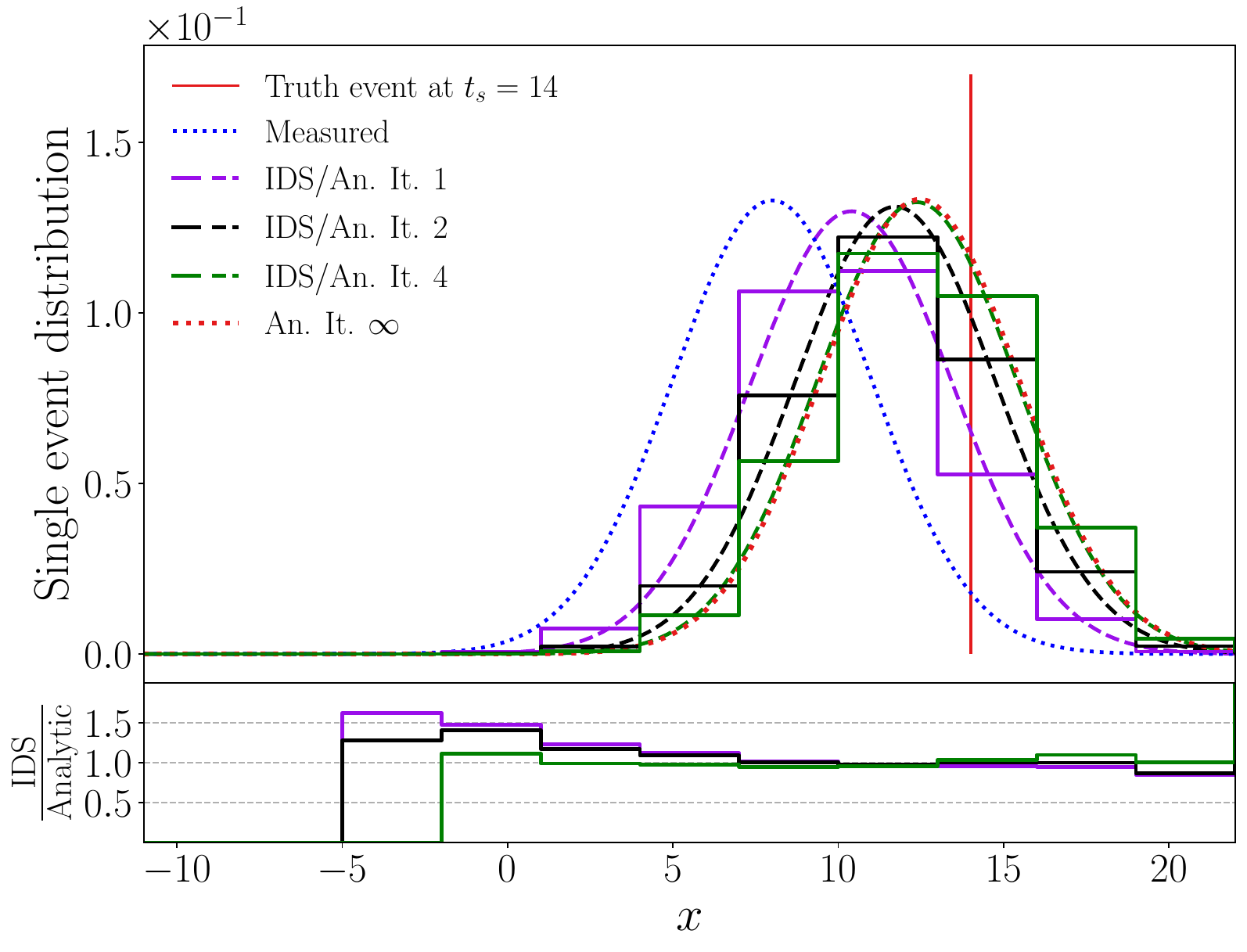}
    \caption{Unfolding of a folded distribution~(dotted blue) of a truth-level event~(solid red) at $t_s \in \lbrace 4,6,10,14 \rbrace$ with IBU~(first two rows) and IDS~(last two rows). The results after each iteration~(purple, black, blue solid lines) as well as the analytic prediction~(purple, black, blue dashed lines) are given. Finally, the analytic prediction for infinite iterations~(dotted red) is also given. }
	\label{f:truth_single_unfolded_matrix}
\end{figure*}

\begin{figure*}[h!]

	\centering
	\includegraphics[width=0.43\linewidth]{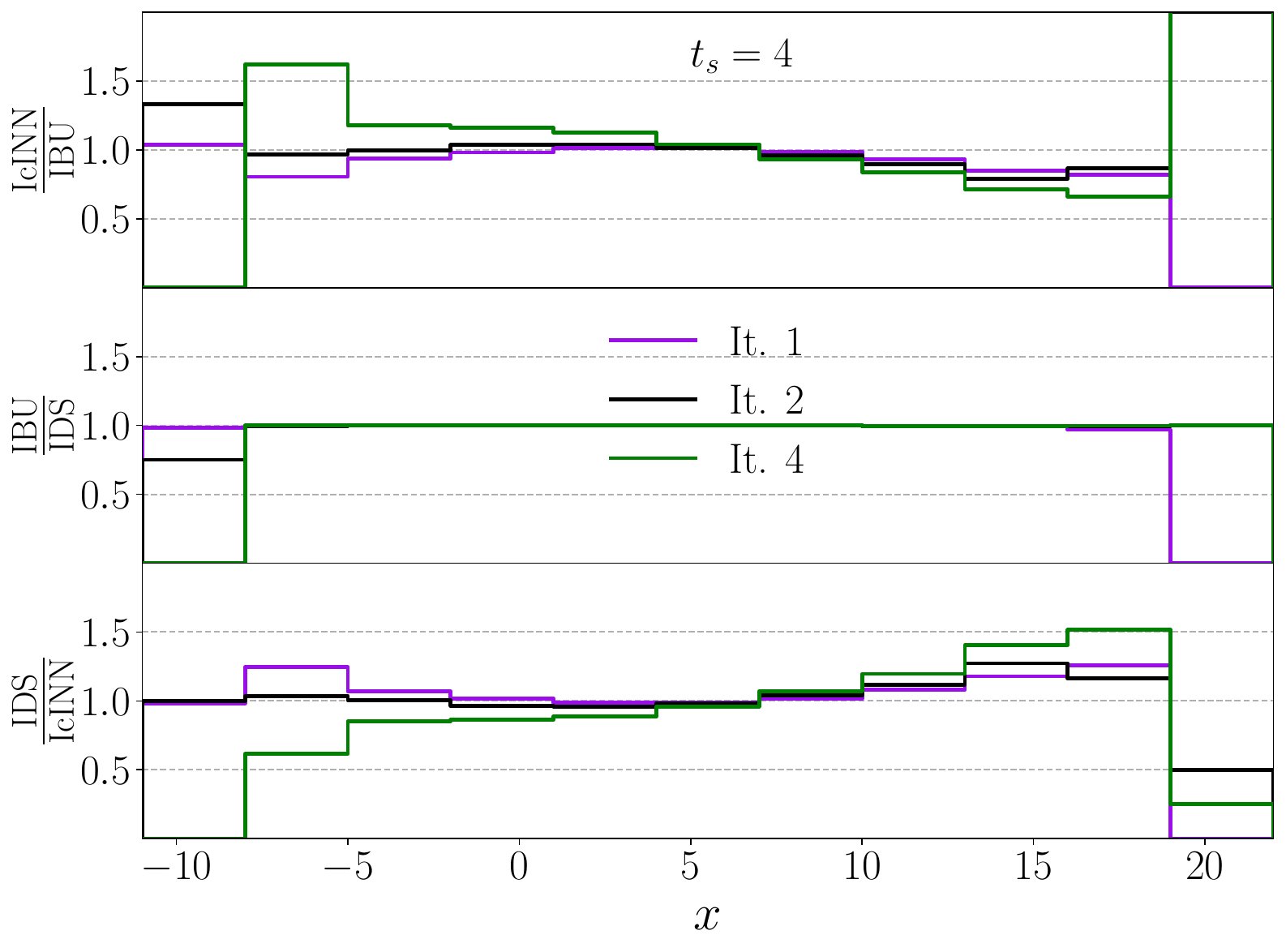}
    \hspace{0.5cm}
    \includegraphics[width=0.43\linewidth]{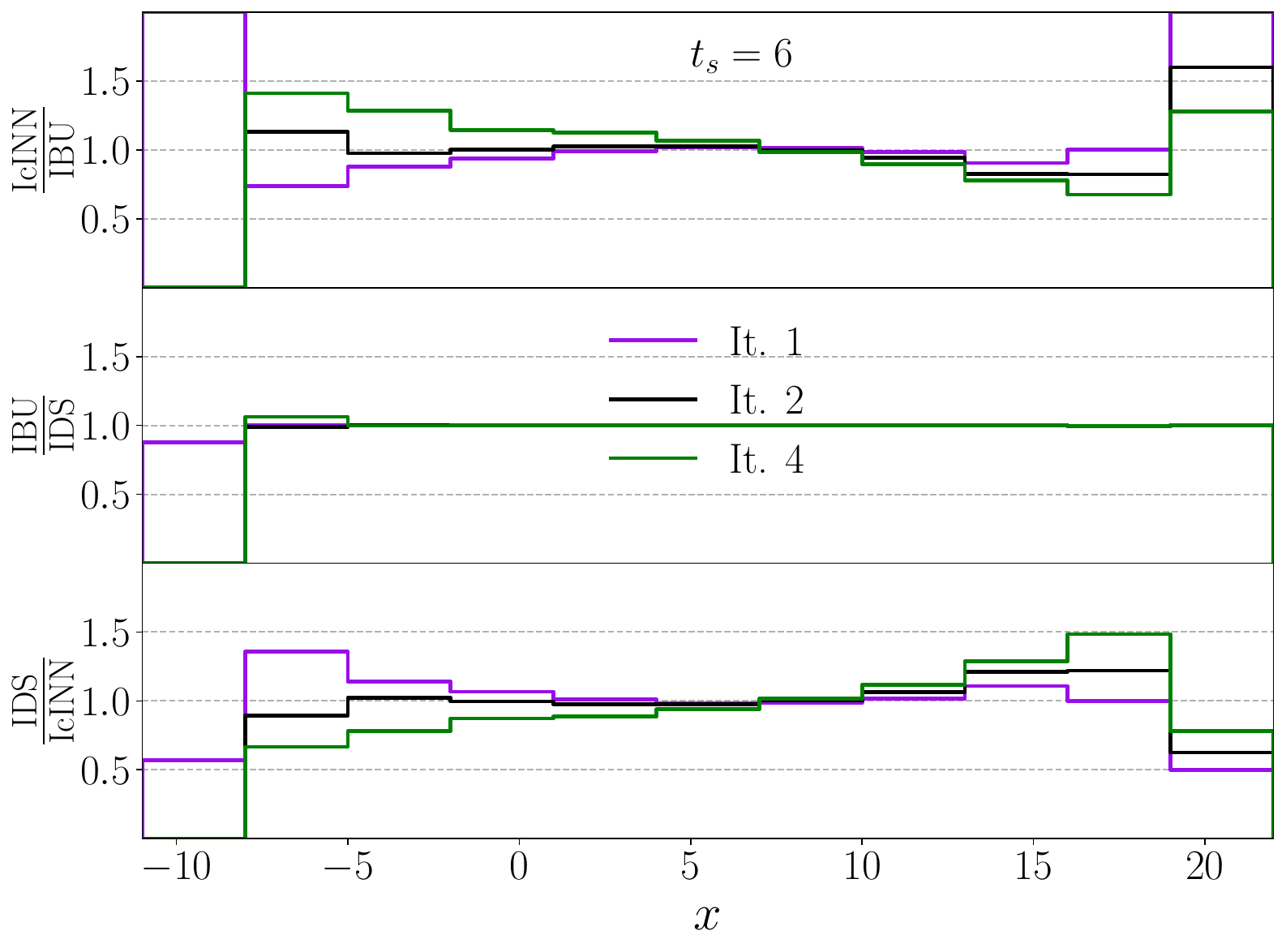}\\
    \includegraphics[width=0.43\linewidth]{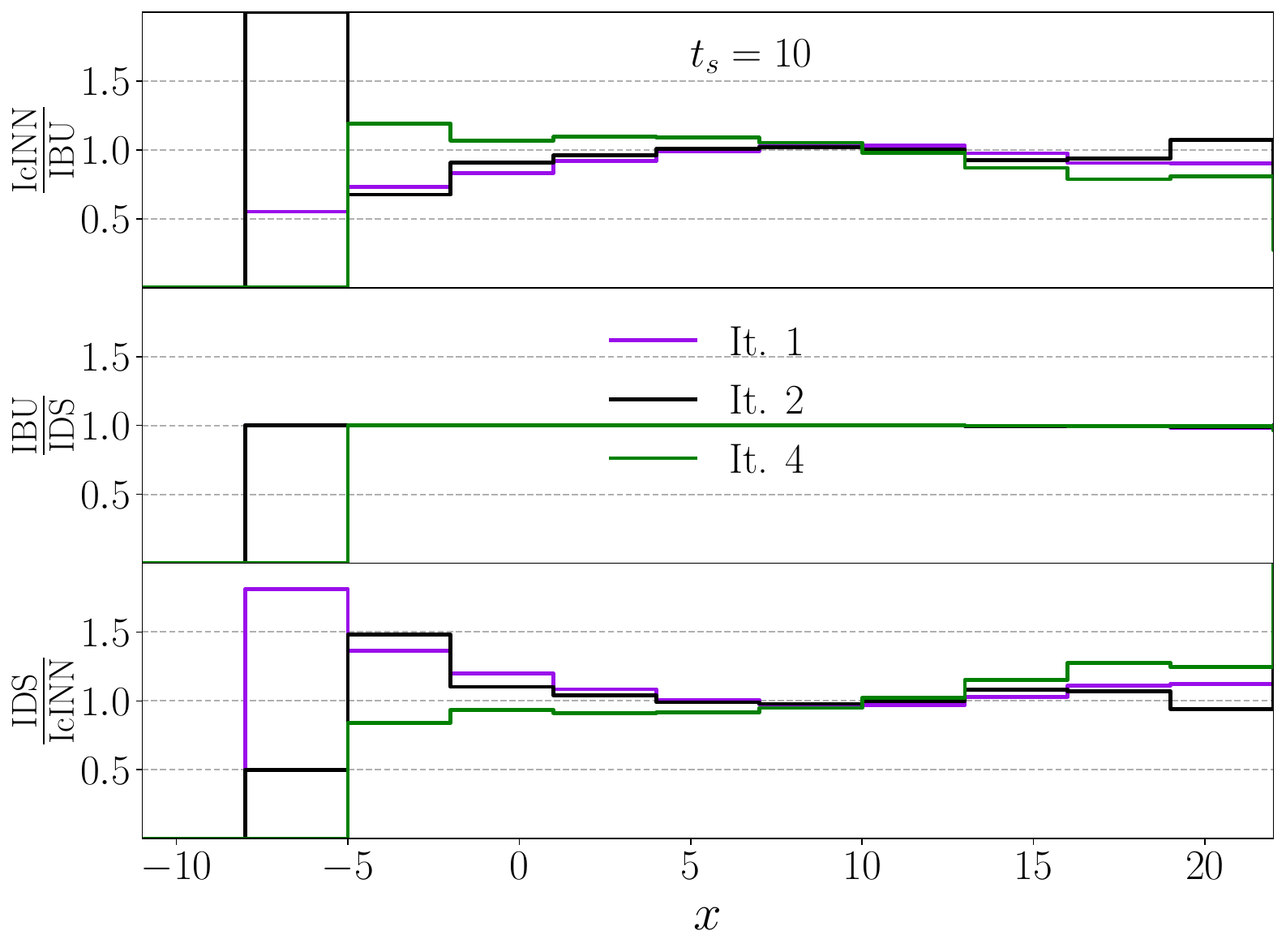}
    \hspace{0.5cm}
    \includegraphics[width=0.43\linewidth]{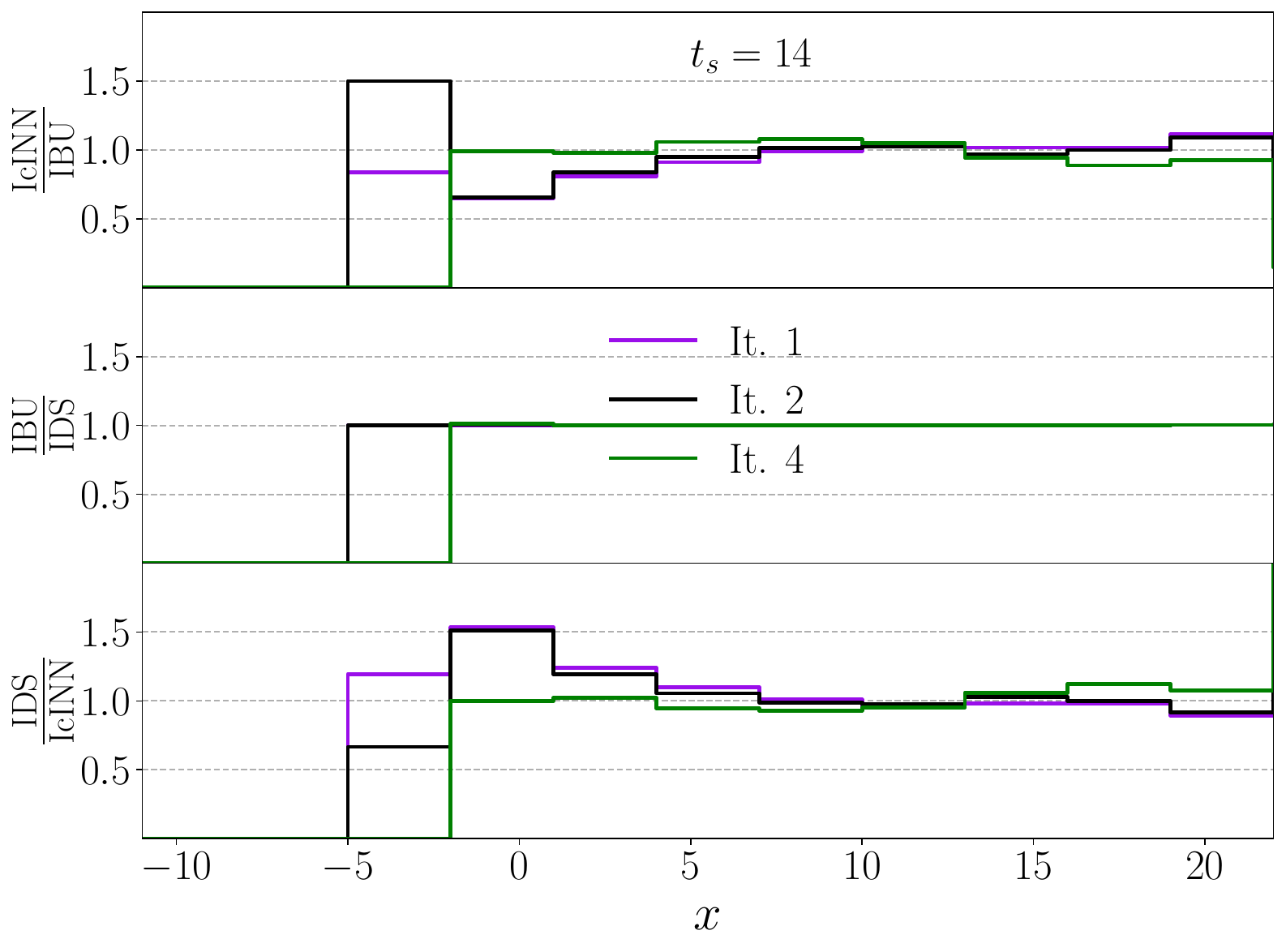}\\
    
    \caption{Ratio plots of the obtained unfolded distributions from Figure~\ref{f:truth_single_unfolded}~(IcINN) and Figure~\ref{f:truth_single_unfolded_matrix}~(IBU and IDS). The ratios are plotted for each pair of algorithms from (IcINN, IBU, IDS). The truth level event is at $t_s=4$ in the top left plot, $t_s=6$ in the top right plot, at $t_s=10$ in the bottom left plot and at $t_s=14$ in the bottom right plot.} 
	\label{f:truth_single_unfolded_ratios}
\end{figure*}

\section{Comments on Single-Event Uncertainties}
\label{a:single_event_uncertainty}

The full unfolded distribution $u_j^{(n)}$ of one fluctuated pseudo-experiment $(n)$ can be expressed in terms of the single-event unfolded distributions $e_{j}^{(n)}(i_s)$ as
\begin{align}
    u_j^{(n)} = \sum_{i_s=1}^{N_\mathrm{bins}} B^{(n)}(i_s) \cdot e_{j}^{(n)}(i_s),
    \label{eq:single_represent_u}
\end{align}
with $N_\mathrm{bins}$ being the number of bins on detector-level and $B^{(n)}(i_s)$ being the number of events in the bin $i_s$ of the experimental detector-level data distribution that is unfolded.
This equation also indicates that every single-event unfolded distribution can contribute to every bin of the full unfolded distribution.
Eq.~\eqref{eq:single_represent_u} can be plugged into Eq.~\eqref{eq:full_covariance} to calculate the covariance of the full distribution as
\begin{align}
  \mathrm{cov}_{kl} &= \frac{1}{N_{\mathrm{f}}-1} \sum_{n=1}^{N_{\mathrm{f}}} \nonumber \\
  & \qquad \, \, \, \sum_{i_1=1}^{N_\mathrm{bins}} \Bigl(B^{(n)}(i_1) \cdot e_{k}^{(n)}(i_1)-\overline{B(i_1) \cdot e_{k}(i_1)}\Bigr)   \nonumber \\
  & \qquad \cdot  \sum_{i_2=1}^{N_\mathrm{bins}} \Bigl(B^{(n)}(i_2)\cdot e_{l}^{(n)}(i_2) -\overline{B(i_2) \cdot e_{l}(i_2)}\Bigr), \label{eq:single_cov_long}
\end{align}
where we define
\begin{align}
    \overline{B(i_s) \cdot e_k(i_s)} = \frac{1}{N_f}\sum_{n'=1}^{N_f} B^{(n')}(i_s) \cdot e_k^{(n')}(i_s).
    \label{eq:single_meanNorm}
\end{align}

The two different sources of Poissonian fluctuations, for the data and respectively the MC-based response matrix, need to be treated differently because of their impact on the number of entries in the detector-level measured distribution $B^{(n)}(i_s)$.
In both cases, the covariance matrix among the shapes of two single-event unfolded distributions~(each of which is normalized to unit integral) can be evaluated according to Eq.~\eqref{eq:single_covShape}.
When a fluctuation of the response matrix is applied, the entries of the detector-level data $r_i$ and in consequence $B(i_s)$ are constant, as it is also the case for the integral of the corresponding unfolded distribution.
This makes it possible to evaluate the corresponding contribution to the covariance matrix of the full unfolded distribution, based on the covariances of the shapes of the single-event unfolded distributions, as
\begin{align}
    \mathrm{cov}_{kl}^{\mathrm{MC}} = \sum_{i_1=1}^{N_\mathrm{bins}} \sum_{i_2=1}^{N_\mathrm{bins}} B(i_1) B(i_2) \,\mathrm{cov}_{kl}^{\mathrm{MC}}(e(i_1), e(i_2)),
    \label{eq:single_cov_MC}
\end{align}
This relation allows for a closure test.
The interpretation of Eq.~\eqref{eq:single_cov_MC} is straight-forward: the double-sum gives one contribution for each pair of single events, while multiple events in the same bin deliver the same contribution.
Eq.~\eqref{eq:single_cov_MC} is thus a summation of all possible combinations of covariances between two single-event distributions.

The second case consists in applying the Poissonian fluctuation to the detector-level experimental data that need to be unfolded.
By definition this leads to a variation of $B^{(n)}(i_s)$, making it dependent on the pseudo-experiment index.
Therefore, the $B^{(n)}(i_1)$ and $B^{(n)}(i_2)$ in Eq.~\eqref{eq:single_cov_long} do not factor out of the sum over the $N_{\mathrm{f}}$ fluctuations.
While the covariance matrix among the shapes of two single-event unfolded distributions is provided by Eq.~\eqref{eq:single_covShape}, one can also evaluate a normalized covariance of the unfolded distributions, accounting in addition for fluctuations of the number of entries in a pair of bins, as
\begin{align}
    \widetilde{\mathrm{cov}}_{kl}^{\mathrm{Data}}(B(i_1) \cdot e(&i_1),B(i_2) \cdot e(i_2)) = \frac{1}{(N_{\mathrm{f}}-1)} \cdot \nonumber \\
    \frac{1}{ \sqrt{B(i_1) B(i_2)}} \sum_{n=1}^{N_{\mathrm{f}}} 
    &\Bigl(B^{(n)}(i_1) \cdot e_{k}^{(n)}(i_1)-\overline{B(i_1) \cdot e_{k}(i_1)}\Bigr) \nonumber \\
    \cdot&\Bigl(B^{(n)}(i_2)\cdot e_{l}^{(n)}(i_2) -\overline{B(i_2) \cdot e_{l}(i_2)}\Bigr).
    \label{eq:single_cov_Data}
\end{align}
This formula can be understood in terms of fluctuations of the normalization and shape of the distributions, which can be factorised.
Since $B^{(n)}(i_s)$ is drawn from a Poissonian with expectation value $B(i_s)$, provided that the corresponding detector-level bin contains enough events, these fluctuations will have little impact on the majority of the single-event unfolded distributions.
Summing over the bin indices and compensating for the normalization factor allows to obtain closure for the total covariance matrix, as 
\begin{align}
    \mathrm{cov}_{kl}^{\mathrm{Data}} = & \sum_{i_1=1}^{N_\mathrm{bins}} \sum_{i_2=1}^{N_\mathrm{bins}} \sqrt{B(i_1) B(i_2)} \cdot \nonumber \\ &\widetilde{\mathrm{cov}}_{kl}^{\mathrm{Data}}(B(i_1) \cdot e(i_1),B(i_2) \cdot e(i_2)),
    \label{eq:single_cov_DataClosure}
\end{align}
which has also been checked numerically.
While several normalization options are possible for $\widetilde{\mathrm{cov}}_{kl}^{\mathrm{Data}}$, the choice made in Eq.~\eqref{eq:single_cov_Data} enables an interpretation of Eq.~\eqref{eq:single_cov_DataClosure} in terms of sums of contributions from individual events, in simple examples where one has little or no bin-to-bin migrations, or when applying this same approach to the reconstructed-level distribution.~\footnote{This feature is due to the fact that a Poisson distribution of $N$ events is equivalent to $N$ independent Poisson distributions for one event each, which is at the basis of the Bootstrap approach~(see e.g.\ Refs.~\cite{Efron1992,ATLAS:2021kho} and references therein). It is also due to the fact that in the current approach the shape of the unfolded distribution for a given event is determined by the reconstructed-level bin that it corresponds to, while in the simple examples considered here the $e(i_s)$ distribution has non-zero values only in the bin $i_s$.}
At the same time, this approach does take into account the impact of the fluctuations applied to the entries of a given reconstructed bin on the unfolded distribution obtained for another bin, evaluating the corresponding induced cross-correlations that are especially relevant for iterative unfolding methods.

\section{Unfolding a Folded Truth Event with IBU and IDS}
\label{App:Unfolding_folded_matrix}

Similarly to the studies discussed in Section~\ref{s:unfolding_folded_IcINN}, an unfolding of a folded truth-level event can be implemented also with the IBU and IDS unfolding methods.
The results for both algorithms and $t_s \in \lbrace 4,6,10,14 \rbrace$ are shown in Figure~\ref{f:truth_single_unfolded_matrix}.
In the ratio plots of Figure~\ref{f:truth_single_unfolded_ratios} it is shown that the resulting distributions are very similar to the ones obtained for IcINN. Hence the same interpretations are also applicable here.

\section{Normalized correlation distributions for different unfolding algorithms}
\label{App:Normalized_correlation_distribution}

\begin{figure*}[t]
    \centering
    \includegraphics[width=0.47\textwidth]{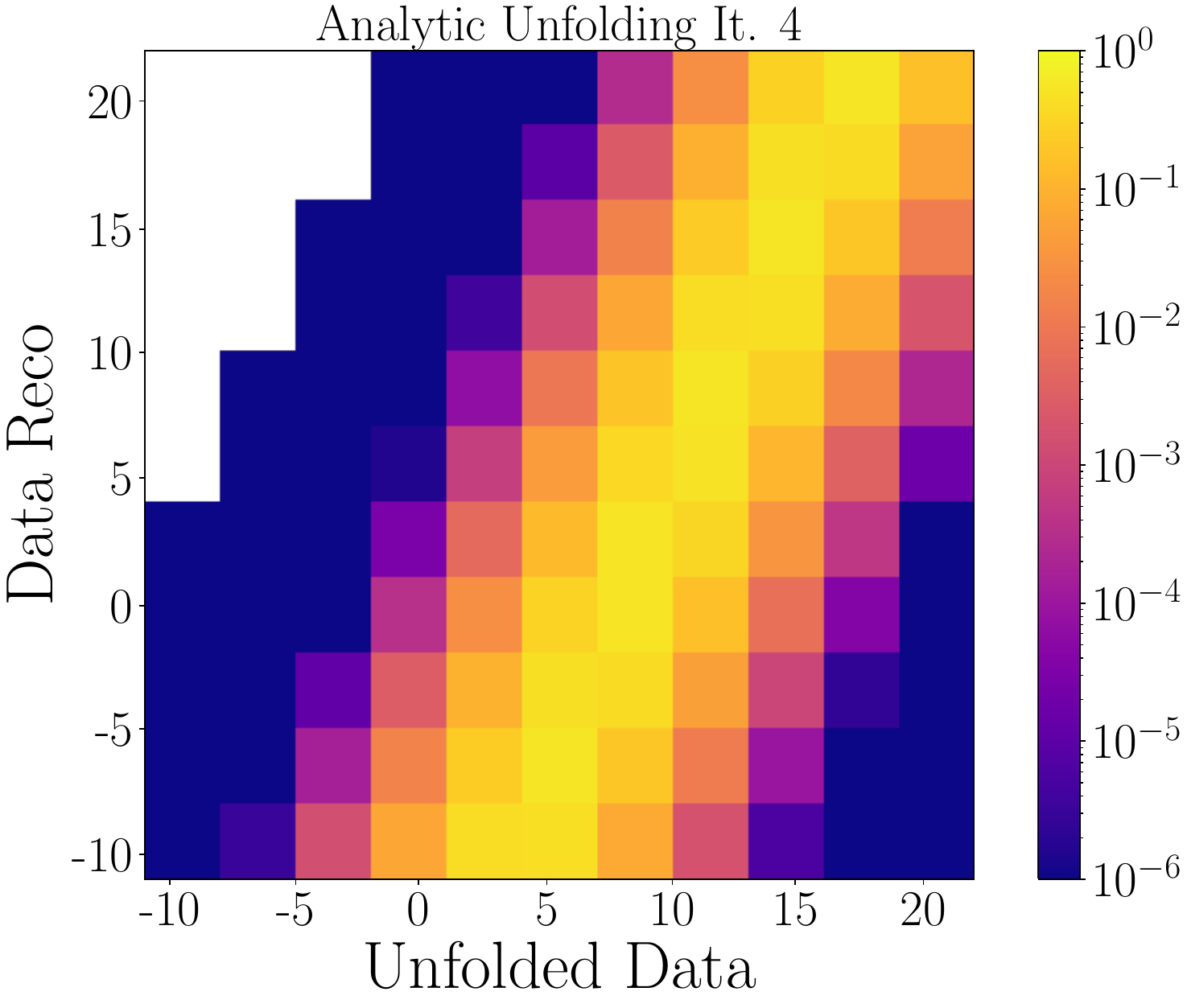}
    \hspace{0.1cm}
    \includegraphics[width=0.47\textwidth]{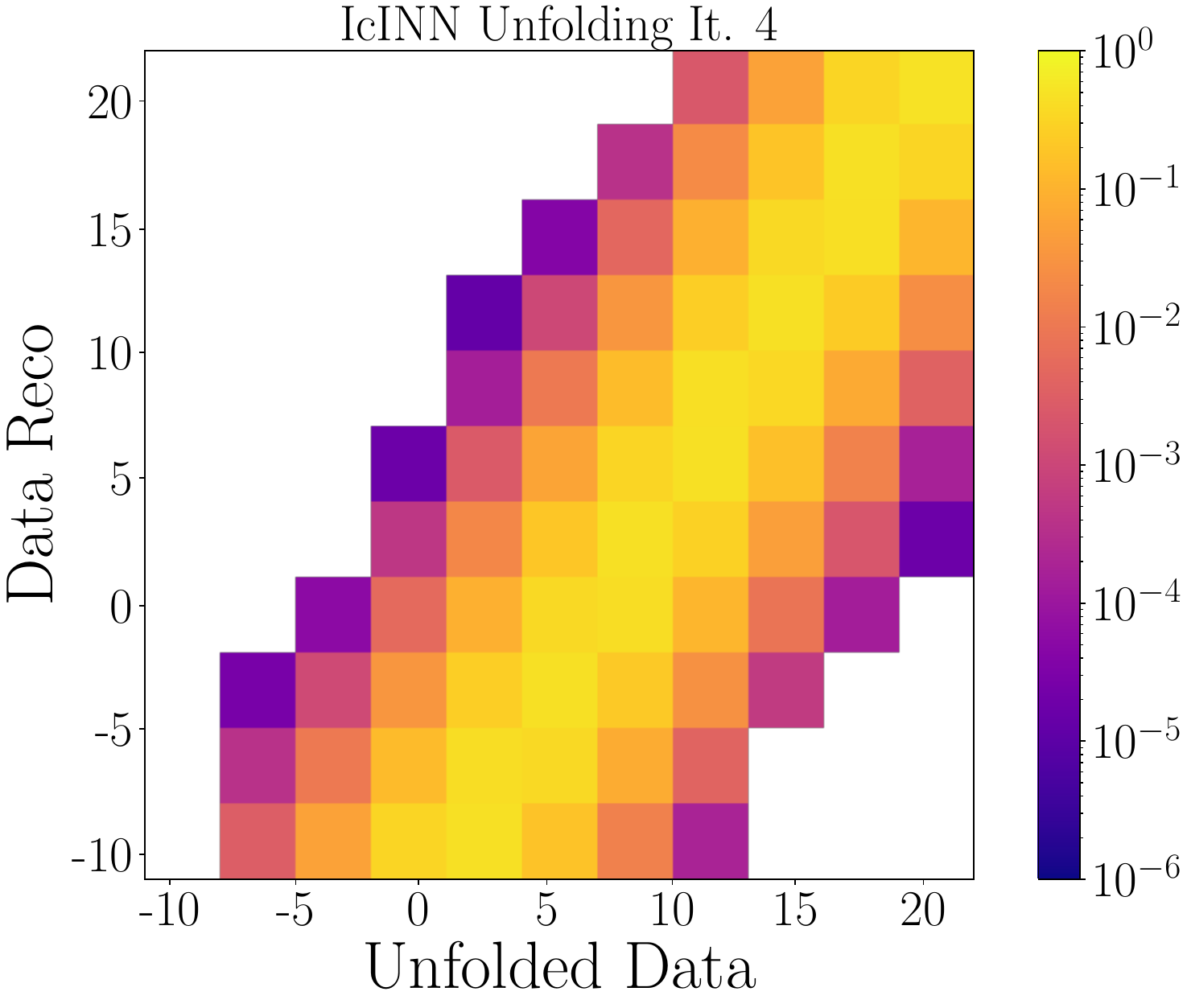}\\
    \vspace{0.3cm}
    \includegraphics[width=0.47\textwidth]{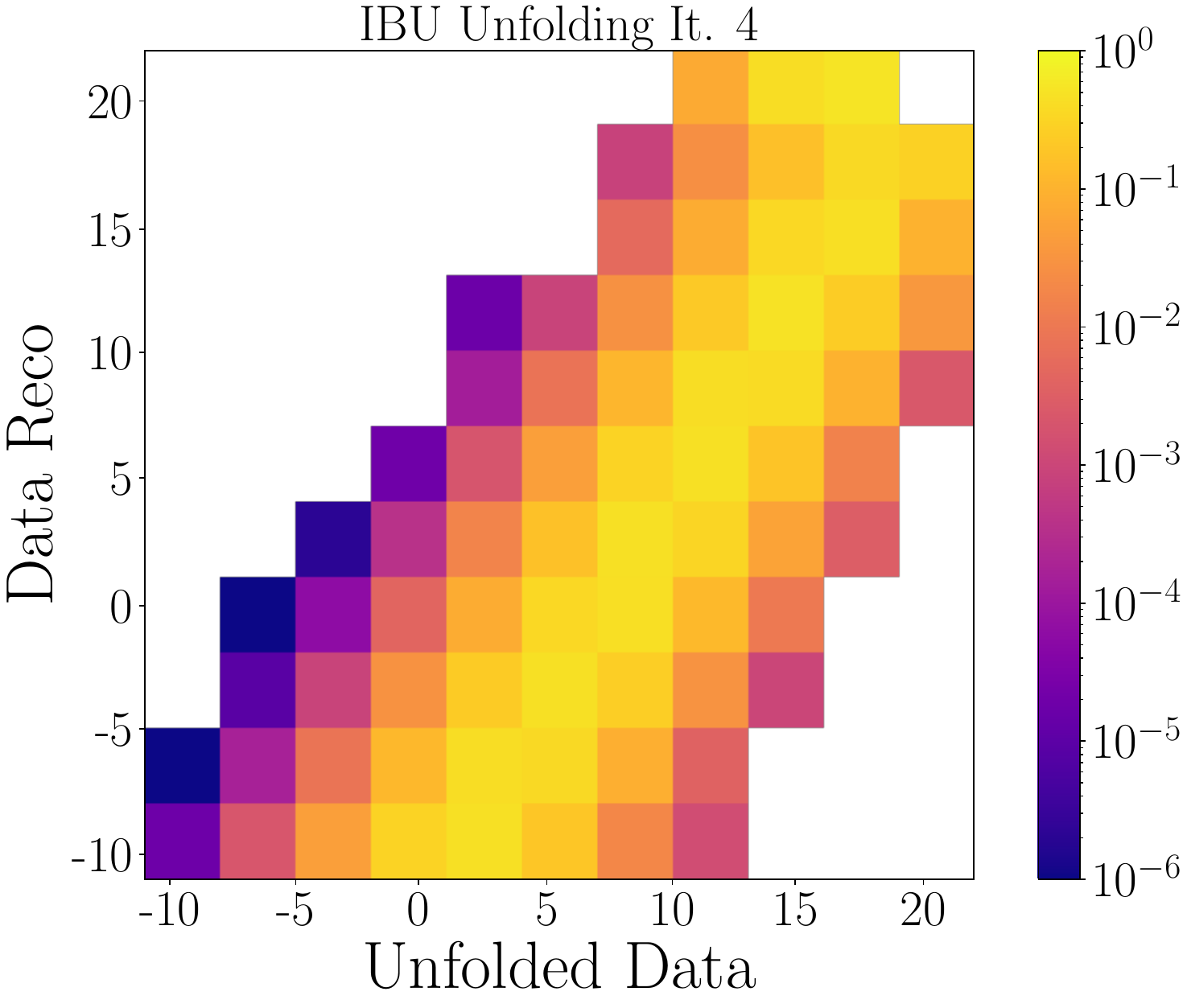}
    \hspace{0.1cm}
    \includegraphics[width=0.47\textwidth]{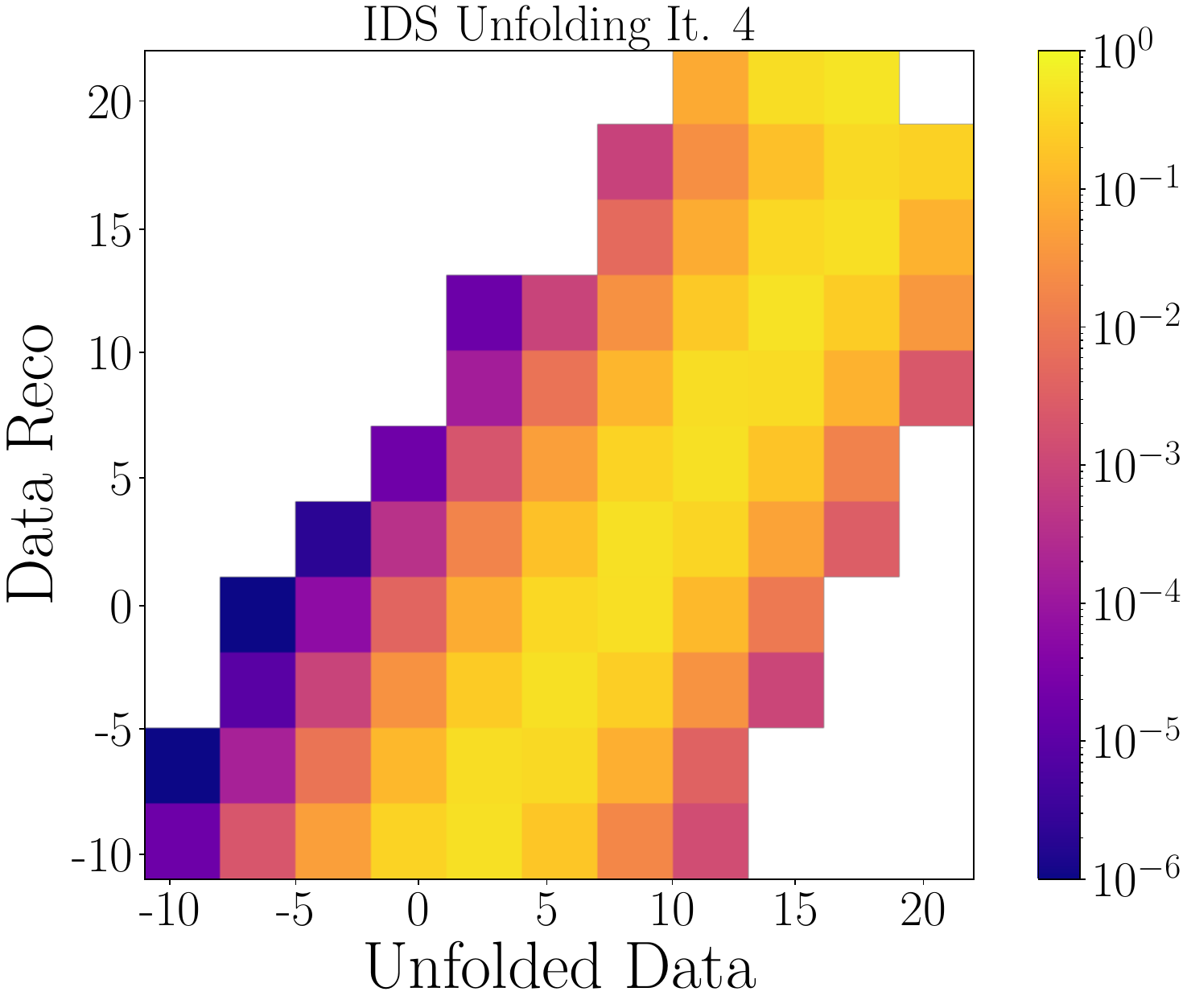}
    \caption{Correlation distributions of Figure~\ref{f:2D_measured_unfolded} normalized according to Eq.~\ref{eq:R'formula} (i.e.\ the posterior distribution $R'_{ji}$). Again the results for an analytic prediction for the toy model (top left) and the results for IcINN (top right), IBU (bottom left) and IDS (bottom right) are shown for 4 iterations.}
    \label{f:2D_measured_unfolded_norm}
\end{figure*}

In order to visualize the probabilistic unfolding matrix for each algorithm it is instructive to normalize the plots of Figure~\ref{f:2D_measured_unfolded} according to Eq.~\ref{eq:R'formula}.
The resulting plots are shown in Figure~\ref{f:2D_measured_unfolded_norm}.
%


%
\end{document}